\documentclass[fleqn,usenatbib]{mnras}
\usepackage{newtxtext,newtxmath}
\usepackage[T1]{fontenc}
\usepackage{ae,aecompl}
\usepackage{graphicx}	% Including figure files
\usepackage{amsmath}	% Advanced maths commands
\usepackage{threeparttable}
\usepackage{multirow}
\usepackage{romannum}

\usepackage{appendix}

\title[PAH emission model]{Polycyclic Aromatic Hydrocarbon emission model in photodissociation regions\\
\begin{LARGE} 
  \Romannum{1}: Application to the 3.3, 6.2, and 11.2 $\mu$m bands
\end{LARGE} }

\author[A. Sidhu et al.]{
Ameek Sidhu,$^{1,2}$\thanks{E-mail: asidhu92@uwo.ca}
A.G.G.M. Tielens,$^{3,4}$
Els Peeters$^{1,2,5}$
and Jan Cami$^{1,2,5}$
\\
$^{1}$Department of Physics \& Astronomy, University of Western Ontario, London, ON, N6A 3K7, Canada\\
$^{2}$Institute for Earth and Space Exploration, University of Western Ontario, London, ON, N6A 3K7, Canada\\
$^{3}$Leiden Observatory, Leiden University, Niels Bohrweg 2, 2333 CA Leiden, Netherlands \\
$^{4}$ Department of Astronomy, University of Maryland, College Park, MD 20742, USA \\
$^{5}$ SETI Institute, 189 Bernardo Avenue, Suite 100, Mountain View, CA 94043, USA
}

\begin{document}
\label{firstpage}
\pagerange{\pageref{firstpage}--\pageref{lastpage}}
\maketitle

\begin{abstract}
We present a charge distribution based model that computes the infrared spectrum of polycyclic aromatic hydrocarbon (PAH) molecules using recent measurements or quantum chemical calculations of specific PAHs. The model is applied to a sample of well-studied  photodissociation regions (PDRs) with well-determined physical conditions (the radiation field strength, $G_{0}$, electron density $n_{e}$, and the gas temperature, $T_{\rm gas}$). Specifically, we modelled the emission of five PAHs ranging in size from 18 to 96 carbon atoms, over a range of physical conditions characterized by the ionization parameter $\gamma = G_{0}\times T_{\rm gas}^{1/2}/n_{e}$. The anions emerge as the dominant charge carriers in low $\gamma $ ($< 2\times 10^{2}$) environments, neutrals in the intermediate $\gamma$ ($10^{3} - 10^{4}$) environments, and cations in the high $\gamma$ ($ > 10^{5}$) environments. Furthermore, the PAH anions and cations exhibit similar spectral characteristics. The similarity in the cationic and anionic spectra translates into the interpretation of the 6.2/(11.0+11.2) band ratio, with high values of this ratio associated with large contributions from either cations or anions. The model's predicted values of 6.2/(11.0+11.2) and 3.3/6.2 compared well to the observations in the PDRs NGC~7023, NGC~2023, the horsehead nebula, the Orion bar, and the diffuse ISM, demonstrating that changes in the charge state can account for the variations in the observed PAH emission. We also reassess the diagnostic potential of the 6.2/(11.0+11.2) vs 3.3/(11.0+11.2) ratios and show that without any prior knowledge about $\gamma$, the 3.3/(11.0+11.2) can predict the PAH size, but the 6.2/(11.0+11.2) cannot predict the $\gamma$ of the astrophysical environment.
\end{abstract}


%%%%%%%%%%%%%%%%%%%%%%%%%%%%%%%%%%%%%%%%%%%%%%%%%%

%%%%%%%%%%%%%%%%% BODY OF PAPER %%%%%%%%%%%%%%%%%%
\section{Introduction}
The strong emission features at 3.3, 6.2, 7.7, 8.6, 11.2, and 12.7 $\mu$m in a typical mid-infrared (mid-IR) spectrum are characteristic of a family of complex organic molecules known as Polycyclic Aromatic Hydrocarbons \citep[PAHs;][]{Sellgren:1983, Leger:1984, Allamandola:1985, Allamandola:1989}. PAH molecules account for $\sim$ 15\% of the total cosmic carbon \citep{Allamandola:1989} and $\sim$ 20 \% of the total IR power of the Milky Way and star-forming galaxies \citep{Madden:2006, Smith:2007}. Emission from PAHs typically originates from photo-dissociation regions (PDRs), where the physics and chemistry of the gas are driven by far-ultraviolet (FUV; 6--13.6 eV) photons \citep{Hollenbach:97}. Gas within a PDR is stratified, with regions of atomic gas found closer to the star and molecular gas further away. The atomic gas consists largely of hydrogen (H) and ionised carbon (C), and the molecular gas of molecular hydrogen (H$_{2}$) and carbon monoxide (CO). PAH emission typically outlines the PDR surface, decreasing into the PDR as the UV radiation field is attentuated.

The physical conditions (radiation field strength $G_{0}$, gas density $n_{\rm gas}$, and gas temperature $T_{\rm gas}$) within the PDRs regulate the molecular properties of the PAH population such as their charge, size, and molecular structure \citep[e.g.][]{Hony:2001, Peeters:prof6:02, Galliano:2008}. Several experimental and theoretical studies have shown that changes in the PAH properties affect the spectral characteristics of the PAH population's emission features. For example, changes in the charge state of PAHs influence the intensity of the 6.2, 7.7, and 8.6 $\mu$m features relative to the 11.2 $\mu$m feature, changes in the size distribution influence the relative intensity of the 3.3 $\mu$m to 11.2 $\mu$m feature, and changes in the molecular structure influence the relative intensities of the features in the 11-14 $\mu$m region \citep[e.g.][]{Hudgins:1999, Allamandola:1999, Hony:2001, Peeters:prof6:02, Bauschlicher:2008, Bauschlicher:2009, Ricca:2012, Candian:14}. The PAH emission features observed in PDRs show spectral variations, the most prominent of which are due to changes in the charge state of the PAH population \citep[e.g.][]{Joblin:1996, Sloan:99, Allamandola:1999, Bregman:2005, Compiegne:2007, Galliano:2008, Rosenberg:11, Peeters:2017, Sidhu:2021}.

\citet{Bakes:2001} presented a PAH emission model that calculates the charge distribution of PAHs and uses these results to determine the PAH emission in astrophysical environments. In this paper, we revisit the emission model adopting recent experimentally or quantum chemically determined PAH characteristics such as the ionization potentials, photo-absorption cross-sections, IR cross-sections, for individual PAHs. The aim of this paper is to demonstrate that the PAH charge distribution can account for the observed spectral characteristics of PAHs by comparing the model results with observations of PDRs exhibiting a wide range of physical conditions. This paper is organized as follows. In Section~\ref{sec:emission_model}, we describe the PAH emission model, which combines the charge distribution, PAH characteristics, and the IR emission from PAHs in a given astrophysical environment. In Section~\ref{sec:results}, we present the results of our model for five different PAH molecules, namely, tetracene ($\text{C}_{18}\text{H}_{12}$), pentacene ($\text{C}_{22}\text{H}_{14}$), ovalene ($\text{C}_{32}\text{H}_{14}$), circumcoronene ($\text{C}_{54}\text{H}_{18}$), and circumcircumcoronene ($\text{C}_{96}\text{H}_{24}$), over a range of physical conditions. We discuss the application of the model in five different environments, NGC~2023, NGC~7023, the Orion Bar, the Horsehead nebula, and the diffuse interstellar medium (ISM), in Section~\ref{sec:application}. Finally, we provide a summary of this work in Section~\ref{sec:summary}.

\section{PAH emission Model}
\label{sec:emission_model}
Following \citet{Bakes:2001} we model the PAH emission taking into account the charge distribution of PAHs in astrophysical settings by adopting realistic properties of PAHs measured experimentally or calculated theoretically as available. In this section, we present a comprehensive overview of the three main components of our model: the calculation of the PAH charge distribution, the calculation of IR emission from PAHs, and the PAH characteristics, as well as a strategy for combining these to model PAH emission in astrophysical environments. We also compare our PAH emission model to other models used in previous studies.

\subsection{The PAH charge distribution model}
\label{subsec:Charge_dist_model}
The charge distribution of molecular species in astronomical environments is set by the balance between the processes of ionization and electron recombination. \citet{Bakes:1994} developed a model using the principle of ionization equilibrium to determine the charge distribution of PAHs and very small graphitic grains in a variety of physical conditions pertaining to interstellar environments. In this work, we adopt the \citet{Bakes:1994} model solely for PAHs by using the ionization, electron recombination, and electron attachment rates derived from recent theoretically calculated or experimentally measured molecular characteristics of PAHs. In this section, we describe the framework of our charge distribution model.

In our model, we assume that a PAH molecule in a given charge state, $Z$, can get ionized by absorbing a photon of energy $h\nu$
\begin{equation}
    \text{PAH}^{Z} \xrightarrow{h\nu} \text{PAH}^{Z+1} + e^{-}
\end{equation}

The ionized PAH can further undergo electron recombination 
\begin{equation}
    \text{PAH}^{Z+1} + e^{-} \to \text{PAH}^{Z} 
\end{equation}
and the neutral PAH ($Z=0$) can undergo electron attachment
\begin{equation}
    \text{PAH} + e^{-} \to \text{PAH}^{-}
\end{equation}
    
The fraction of PAHs in a given charge state, $f(Z)$, is then determined by considering the ionization balance which yields 
\begin{equation}
    f(Z) = \frac{k_{e}(Z+1)}{k_{\rm ion}(Z)}f(Z+1)
    \label{eq:frac_PAHs}
\end{equation}

\noindent where $k_{\rm ion}(Z)$ and $k_{e}(Z)$ are the photo-ionization and electron recombination rates of PAH molecule in a charge state $Z$ in units of s$^{-1}$. For $Z = -1$, $k_{e}(Z+1)$ is the electron attachment rate, $k_{ea}(Z)$.

We determine the charge distribution of PAHs by solving the set of equations given by equation~(\ref{eq:frac_PAHs}) for each charge state in conjunction with the normalization condition
\begin{equation}
    \sum_{Z\,=\,-1}^{Z\,=\,N_{\rm max}} f(Z) = 1
\end{equation}
where $N_{\rm max}$ is the highest charge state accessible to the PAH molecule.

\subsubsection{Photo-ionization rate}
\label{subsec:photo_ionization_rate}
We estimate the photo-ionization rate for a given PAH molecule in a charge state $Z$ using the following expression from \citet{Bakes:1994}:
 \begin{equation}
    k_{\rm ion}(Z) =  \pi W\int_{\nu_{Z}}^{\nu_{H}} Y_{\rm ion}(Z, \nu) \sigma_{\rm abs}(Z, \nu) \frac{B_{\nu}(T_{\rm eff})}{h\nu} d\nu\;\;\;\;\; (\text{s}^{-1})
    \label{eq:photo_ion_rate}
\end{equation}
where
    %%begin novalidate
\[
  \begin{array}{lp{0.8\linewidth}}
      
         Y_{\rm ion}(Z, \nu)  & the ionization yield of a PAH molecule in a charge state $Z$    \\
         \sigma_{\rm abs}(Z, \nu)               & the photo-absorption cross-section of a PAH molecule in a charge state $Z$ \newline Units: cm$^{2}$ \\
         B_{\nu}(T_{\rm eff})  & the Planck function for the incident blackbody radiation field at the effective temperature $T_{\rm eff}$ of the exciting star \newline
         Units: erg cm$^{-2}$ s$^{-1}$ Hz$^{-1}$ sr$^{-1}$\\
         W            & the FUV radiation field dilution factor            \\
         \nu_{Z} & the frequency of the photon corresponding to the ionization potential (IP) for a PAH molecule in charge state $Z$ \newline Units: Hz \\
         \nu_{H} & the frequency of the photon corresponding to photon energy of 13.6 eV \newline Units: Hz \\
         h & the Planck constant i.e. 6.6261$\times10^{-27}$ erg s
 \end{array}
 \]
    %%end novalidate
The Planck function is calculated as
\begin{equation}
    B_{\nu}(T_{\rm eff}) = \frac{2h\nu^{3}}{c^{2}}\frac{1}{e^{h\nu/kT_{\rm eff}}-1}
\end{equation}
where
    %%begin novalidate
\[
  \begin{array}{lp{0.8\linewidth}}
         k & the Boltzmann constant i.e. 1.3807$\times10^{-16}$ erg $\text{K}^{-1}$ \\
         c & the speed of light i.e. 2.99792458$\times10^{10}$ cm $\text{s}^{-1}$\\
 \end{array}
 \]
The FUV radiation field dilution factor is calculated as 
\begin{equation}
    W = \frac{1.6 \times 10^{-3} G_{0}}{\sigma T_{\rm eff}^{4} f_{\rm FUV}}
\end{equation}
where
    %%begin novalidate
\[
  \begin{array}{lp{0.8\linewidth}}
      
         G_{0}  & the FUV radiation field strength in Habing Field \newline 1 $G_{0}$ = 1.6 $\times 10^{-3}$ erg cm$^{-2}$ s$^{-1}$ \\
         \sigma               & the Stefan-Boltzmann constant \newline 5.6704 $\times 10^{-5}$ erg s$^{-1}$ cm$^{-2}$ K$^{-4}$\\
         f_{\rm FUV}  & the fraction of the FUV flux in the radiation field of the star with an effective temperature $T_{\rm eff}$\\
 \end{array}
 \]
    %%end novalidate
    
The fraction of the FUV flux, $f_{\rm FUV}$, is further calculated as
\begin{equation}
    f_{\rm FUV} = \frac{\pi \int_{\nu_{6}}^{\nu_{H}}B_{\nu}(T_{\rm eff})d\nu}{\sigma T_{\rm eff}^{4}}
\end{equation}
where $\nu_{6}$ is the frequency of the photon corresponding to a photon energy of 6 eV. In the calculation of W, we use the fraction of the FUV flux because the ratio of $G_{0}$ and the FUV radiation provides a measure of the radiation field dilution. Since, by definition, $G_{0}$ is the FUV radiation field, we calculate the incident blackbody FUV radiation field by multiplying $f_{\rm FUV}$ with $\sigma T_{\rm eff}^{4}$.

\subsubsection{Electron-recombination rate}
\label{subsec:electron_recombination_rate}
Measurements of electron recombination rates exist only for a few small PAHs. Therefore, we estimate the electron recombination rates theoretically by adopting the following formula proposed by \citet{Tielens:2005}:
\begin{equation}
    k_{e}(Z) = 1.3 \times 10^{-6} Z \,\, \text{N}_{C}^{1/2}\left (\frac{300}{T_{\rm gas}}\right )^{1/2}n_{e}\;\;\;\;\; (\text{s}^{-1})
    \label{eq:recombination_rate}
\end{equation}
where $\text{N}_{C}$ is the number of C atoms in a PAH molecule, $T_{\rm gas}$ is the gas temperature in K, and $n_{e}$ is the electron density in a given environment in units of cm$^{-3}$. This expression is based on the collisional rates of an electron's interaction with an ionized PAH molecule represented by a conductive disk \citep{Bakes:1994}. We note that \citet{Biennier:2006} compared the experimentally measured electron-recombination rates of small PAHs to theoretically calculated recombination rates (equation~\ref{eq:recombination_rate}) and found that the electron recombination rates increase with size and approach the theoretical estimate for larger sized PAHs.    

\subsubsection{Electron attachment rate}
\label{subsec:electron_attachment_rate}
We adopt the theoretical expression for electron attachment rates derived by \citet{Tielens:2005}:
\begin{equation}
    k_{ea}(Z=0) = 1.3 \times 10^{-7} s_{e} \,\, \text{N}_{C}^{1/2} \,\, n_{e}\;\;\;\;\; (\text{s}^{-1}) 
\label{eq:electronattachmentrate}
\end{equation}
where $s_{e}$ is the dimensionless sticking coefficient for electron attachment.

\subsection{Theoretical calculation of IR emission from PAHs}
\label{subsec:spec_calc}
In this section, we describe the procedure we use to calculate the IR emission from a PAH molecule with a charge state $Z$. First, we estimate the average photon energy, $E_{\rm avg}(Z)$, that a PAH molecule in a charge state $Z$ will absorb in a given environment that will lead to the IR emission using the following expression:
    
\begin{equation}
    E_{\rm avg}(Z) = \frac{\int_{0}^{\nu_{H}}(1 - Y_{\rm ion}(Z,\nu)) \sigma_{\rm abs}(Z, \nu) h \nu \frac{B_{\nu}(T_{\rm eff})}{h\nu} d\nu}{\int_{0}^{\nu_{H}} (1 - Y_{\rm ion}(Z, \nu))\sigma_{\rm abs}(Z, \nu) \frac{B_{\nu}(T_{\rm eff})}{h\nu} d\nu }
    \label{eq:avg_energy}
\end{equation}
where $E_{\rm avg}(Z)$ is in units of erg. The average photon energy depends on the physical conditions of the environment and the photo-absorption cross-section of the PAH molecule itself. Moreover, for a photon that is absorbed by a PAH molecule, there is always a competition between the process of IR emission and ionization. Therefore, we include a factor of $1-Y_{\rm ion}(Z,\nu)$ in the calculation of the average photon energy to account for this competition between the IR emission and photo-ionization. The factor $1-Y_{\rm ion}(Z,\nu)$ is derived from the probability, $P_{\rm emission}(Z)$, that a PAH molecule in a charge state $Z$ will absorb a photon that leads to emission as opposed to ionization:

\begin{equation}
        P_{\rm emission}(Z) = \frac{k_{\rm abs}(Z) - k_{\rm ion}(Z)}{k_{\rm abs}(Z)}
    \label{eq:P_emission}
\end{equation}
where $k_{\rm abs}(Z)$ is the photo-absorption rate given by

\begin{equation}
    k_{\rm abs}(Z) =  \pi W\int_{0}^{\nu_{H}} \sigma_{\rm abs}(Z, \nu) \frac{B_{\nu}(T_{\rm eff})}{h\nu} d\nu\;\;\;\;\; (\text{photons } \text{s}^{-1})
    \label{eq:photo_abs_rate}
\end{equation}
and $k_{\rm ion}(Z)$ is the photo-ionization rate for a PAH molecule in a charge state $Z$ in a given environment.

Substituting the expression for $k_{\rm abs}(Z)$ from equation~\ref{eq:photo_abs_rate} and $k_{\rm ion}(Z)$ from equation~\ref{eq:photo_ion_rate} in equation~\ref{eq:P_emission}, shows that $P_{\rm emission}(Z) \propto$ an averaged $\overline{1-Y_{\rm ion}(Z,\nu)}$;

\begin{equation}
        P_{\rm emission}(Z) = \frac{\int_{0}^{\nu_{H}}\sigma_{\rm abs}(Z, \nu)(1-Y_{\rm ion}(Z,\nu))\frac{B_{\nu}(T_{\rm eff})}{h\nu}d\nu}{\int_{0}^{\nu_{H}}\sigma_{\rm abs}(Z,\nu)\frac{B_{\nu}(T_{\rm eff})}{h\nu}d\nu}
\end{equation}

Once the molecule absorbs the photon of energy, $E_{\rm avg}(Z)$, it redistributes the absorbed energy over its various vibrational modes. A molecule with $N$ number of atoms will have $3N-6$ number of vibrational modes called the fundamental vibrational modes. Out of these $3N-6$ modes, only a few will be IR active and have intrinsic strengths, $\sigma_{\nu}$.
     
Upon absorption of a photon, the temperature of the molecule rises immediately. We calculate the temperature, $T_{\rm max}$, the molecule will attain upon absorption of a single photon using the following expression from \citet{Bakes:2001}: 
\begin{equation}
            E_{\rm avg}(Z) = \int_{2.7 K}^{T_{\rm max}} C_{v}(T, Z) dT
    \label{eq:max_temp_calc}
\end{equation}
where $C_{v}(T, Z)$ is the specific heat of a PAH molecule in a charge state $Z$ as a function of the temperature of the molecule in units of erg/K. We note that equation~\ref{eq:max_temp_calc} is used to calculate $T_{\rm max}$ using the $E_{\rm avg}(Z)$ determined from equation~\ref{eq:avg_energy}. In our calculation, we assume that initially before the absorption of a photon, the molecule has a temperature of 2.7 K and after the absorption of a photon of energy $E_{\rm avg}(Z)$ it reaches a temperature $T_{\rm max}$. Considering each fundamental vibrational mode of a molecule as a harmonic oscillator, we calculate the specific heat of a molecule, $C_{v}(T, Z)$, as follows \citep[equation 6.33 in][]{Tielens:2005}:
    
\begin{equation}
    C_{v}(T, Z) = k \sum_{i=1}^{3N-6} \left (\frac{h\nu_{i, Z}}{kT}\right )^{2} \frac{exp \left (\frac{h\nu_{i, Z}}{kT}\right )}{\left [exp(\frac{h\nu_{i, Z}}{kT})-1\right ]^{2}}
\end{equation}
Here, the $\nu_{i}$'s correspond to the frequencies of the fundamental vibrational modes of the PAH with charge $Z$.

In our model, we assume that the energy is quickly distributed over all available vibrational modes – the ergodic approximation. The molecule cools down via emission of IR photons through its IR active vibrational modes. We recognize that the PAH molecule can absorb yet another photon at any time during the cooling process and reach a higher temperature than $T_{\rm max}$. In other words, we consider the absorption of multiple photons by a PAH molecule before it completely cools down. We emphasize that in the case of multiple photon absorptions, each absorbed photon by a PAH molecule in a charge state $Z$ has an energy equal to $E_{\rm avg}(Z)$. We account for the effect of the absorption of multiple photons on the temperature of a PAH molecule in a charge state $Z$ and hence on its IR emission, by calculating a temperature distribution function, $G(T, Z)$. For the first photon absorption, $G_{1}(T, Z)$ is defined as follows assuming that the temperature distribution of a PAH molecule is a Poisson process \citep[see equation 6 in][]{Bakes:2001}: 

\begin{equation}
    G_{1}(T, Z) = \frac{k_{\rm abs}(Z)}{dT/dt}exp[-k_{\rm abs}(Z)\tau_{\rm min}(T)]
\end{equation}
where
    %%begin novalidate
\[
  \begin{array}{lp{0.8\linewidth}}
      
         k_{\rm abs}(Z)  & the photo-absorption rate for a PAH molecule in a charge state $Z$ 
         \newline Units: s$^{-1}$\\
         dT/dt               & the cooling rate of a PAH molecule \newline Units: K s$^{-1}$ \\
         \tau_{\rm min}(T)  & the time taken by the molecule to cool down from maximum temperature reached after one photon absorption, $T_{\rm max}$, to some temperature $T$ \newline Units: s \\
         
 \end{array}
 \]
    %%end novalidate

We calculate the rate, $dT/dt$, at which the molecule cools down following equations 3 and 4 from \citet{Bakes:2001}:
    
\begin{equation}
            \frac{dT}{dt} = \frac{4 \pi}{C_{v}(T, Z)} \sum_{i} \sigma_{\nu_{i, Z}} B_{\nu_{i}}(T) 
    \label{eq:cooling_rate}
\end{equation}
emphasizing that in equation~\ref{eq:cooling_rate}, the summation is taken over IR active vibrational modes only. Here $\sigma_{\nu_{i}, Z}$ is the intrinsic intensity of the IR active mode of a PAH molecule in a charge state $Z$ expressed as IR cross-section in units of cm$^{2}$ Hz. To calculate $\tau_{\rm min}$, we then use the following expression \citep[see equation 6.31 in][]{Tielens:2005}:
\begin{equation}
    \tau_{\rm min}(T) = \int_{T}^{T_{\rm max}} \frac{1}{dT/dt} dT
\end{equation}

For $n$ photon absorptions, we calculate the temperature distribution function, $G_{n}(T, Z)$, in an iterative fashion as follows:
\begin{equation}
    G_{n}(T,Z) = \int_{2.7}^{T} G_{n-1, T_{n-1}\to T^{'}}(T_{n-1},Z)G_{T \gets T^{'}}(T,Z) dT_{n-1}
\end{equation}

\noindent where $G_{n-1,T_{n-1}\to T^{'}}(T_{n-1}, Z)$ is the probability of finding a PAH molecule at temperature $T_{n-1}$ after $n-1$ photon absorptions that can absorb the $n^{\text{th}}$ photon and reach the temperature $T^{'}$ and $G_{T \gets T^{'}}(T, Z)$ is the probability of finding the molecule at temperature $T$ after it began to cool down from the temperature $T^{'}$ it attained after the absorption of the $n^{\text{th}}$ photon. We calculate the temperatures $T^{'}$ using equation~\ref{eq:max_temp_calc} by replacing the left hand side with $E_{\rm avg} + E_{T_{n-1}}$, sum of the energy $E_{\rm avg}$ of a new photon absorbed and the energy $E_{T_{n-1}}$ remaining in a molecule from previous $n-1$ photon absorptions, and the upper limit of the integral with the temperature $T^{'}$ that the molecule will attain after the $n^{\text{th}}$ photon absorption.

After determining the temperature distribution, $G_{n}(T,Z)$, for a PAH molecule after absorption of $n$ photons, we calculate the IR emission of a molecule through its IR active vibrational modes as follows:
\begin{equation}
            I_{\nu_{i}}(Z) = \sigma_{\nu_{i, Z}} \int_{2.7}^{T_{max, n}} B_{\nu_{i}}(T) G_{n}(T, Z) dT
    \label{eq:IR_emission_IR_active}
\end{equation}
where $T_{max,n}$ is the highest possible temperature that a PAH molecule can attain after $n$ photon absorptions. 

\subsection{Resulting PAH emission in astrophysical environments}
\label{subsec:final_model}
In order to model the PAH emission in a given environment, we combine the calculation of the PAH charge distribution described in Section~\ref{subsec:Charge_dist_model} with the theoretical calculation of IR emission from a PAH molecule described in Section~\ref{subsec:spec_calc}. In other words, we multiply the intensity of each IR active mode of a molecule in a charge state $Z$, obtained from  equation~\ref{eq:IR_emission_IR_active}, with the fraction of PAHs having the charge state $Z$, $f(Z)$, obtained from equation~\ref{eq:frac_PAHs}, and sum the product over all the charge states accessible to a PAH molecule in that environment as follows:       

\begin{equation}
        I_{\nu_{i}} =  \sum_{Z\,=\,-1}^{Z\,=\,N_{\rm max}} f(Z) I_{\nu_{i}}(Z)
\end{equation}

\subsection{PAH characteristics}
\label{subsec:characteristics_PAHs}
\begin{figure*}
    \centering
    \includegraphics[scale=0.20]{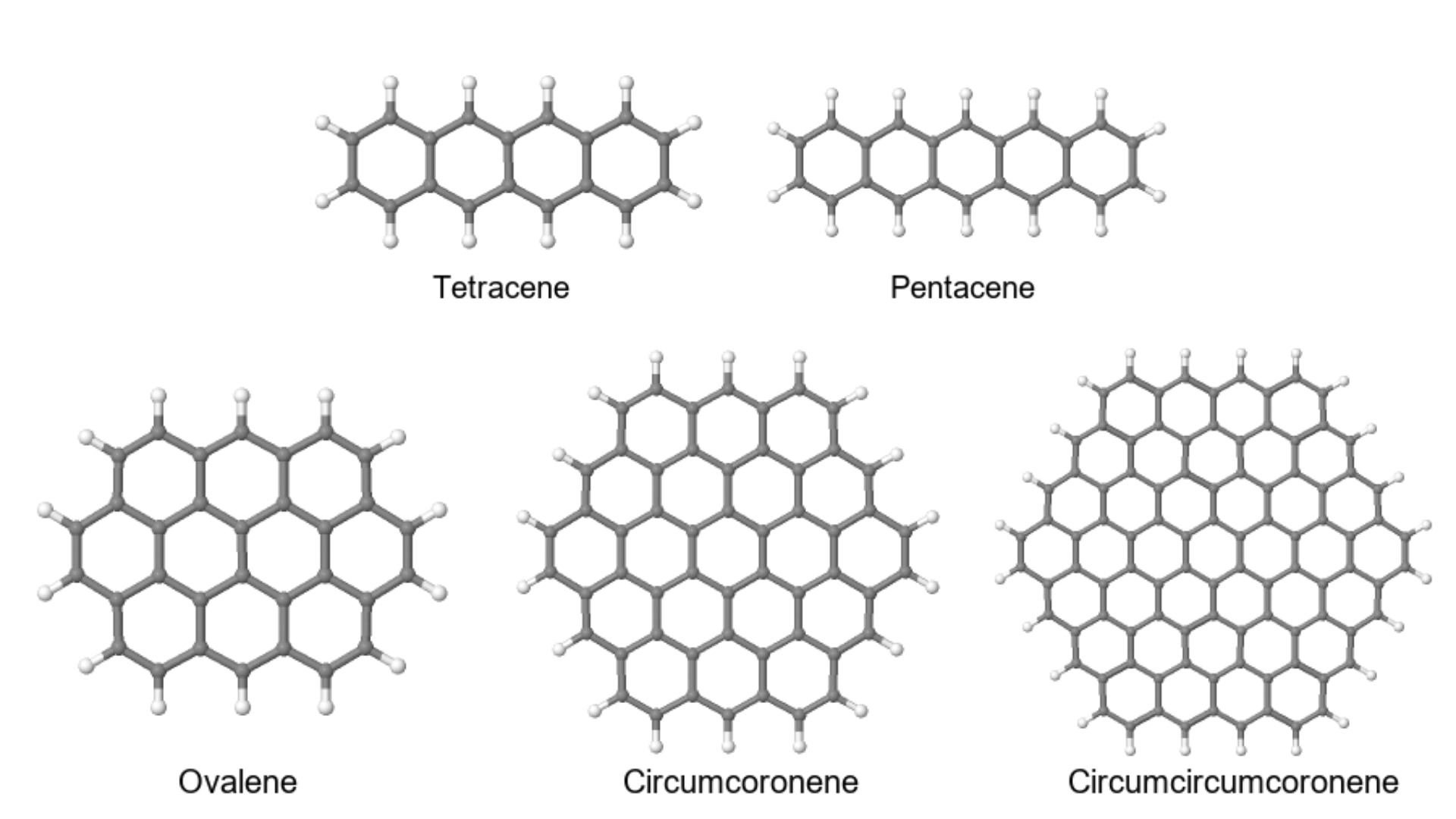}
    \caption{Molecular structure of acenes (top row) and compact PAHs (bottom row) considered in this work. The structure of each molecule is taken from version 3.20 of the NASA Ames PAH database \citep{Bauschlicher:10, Boersma:14, Bauschlicher:2018, Mattioda:2020}. }
    \label{fig:structure_PAHs}
\end{figure*}
In this paper, we study the IR emission from five representative PAHs, namely, tetracene ($\text{C}_{18}\text{H}_{12}$), pentacene ($\text{C}_{22}\text{H}_{14}$), ovalene ($\text{C}_{32}\text{H}_{14}$), circumcoronene ($\text{C}_{54}\text{H}_{18}$), and circumcircumcoronene ($\text{C}_{96}\text{H}_{24}$) covering a large range in size. The tetracene and pentacene molecules belong to the acene family, whereas the ovalene, circumcoronene, and circumcircumcoronene belong to the compact/pericondensed PAH family. Fig.~\ref{fig:structure_PAHs} shows the structure of the PAH molecules studied in this work. We chose compact PAHs in our study due to their potential relevance to the interstellar PAH family. Compact PAHs are highly stable and can withstand harsh conditions in the ISM \citep[][]{Ricca:2012}. Their spectra in the 15-20 $\mu$m range are simple in comparison to those of non-compact PAHs, highlighting their potential as candidates for the interstellar PAH family, which also exhibit relatively constant spectra in this wavelength range \citep{Boersma:2010, Ricca:2012, Andrews:2015}. We include acenes in our study as examples of catacondensed and small PAHs with a large number of solo C-H modes. The strong 11.2 $\mu$m PAH feature is associated with the out-of-plane bending of solo C-H mode and, due to their zig-zag edges, acenes  have a large number of solo Hs thereby making them relevant for the 11.2 $\mu$m PAH feature \citep{Hony:2001, Bauschlicher:2008, Candian:2015}. Though not much explored in the context of the ultimate interstellar PAH family, acenes are studied in the context of $\text{H}_{2}$, formation on PAHs owing to their large reactivity \citep[e.g.][]{Campisi:2020}. In this section, we describe the molecular characteristics we adopt to model the IR emission from PAHs.

The PAH emission model model takes into account the following molecular characteristics of PAHs: the ionization potential (IP), the ionization yield, the photo-absorption cross-section, the sticking coefficient, and the frequencies and corresponding intensities of the vibrational modes. 

\begin{table*}
\caption{Ionization potential of PAHs in various charge states.}
    \centering
    \begin{threeparttable}[t]
    \begin{tabular}{l c c c c c}
    
    \hline
    \multirow{2}{*}{Molecule} & \multicolumn{5}{c}{IP($Z$) [eV]}\\
        & $Z$ = -1 & $Z$ = 0 & $Z$ = 1 & $Z$ = 2 & $Z$ = 3\\
        \hline
        Tetracene ($\text{C}_{18}\text{H}_{12}$)  & 1.06 \tnote{1} & 6.96 \tnote{3} & 11.64 \tnote{3} & & \\ 
        Pentacene ($\text{C}_{22}\text{H}_{14}$) & 1.39 \tnote{1} & 6.61 \tnote{3} & 10.79 \tnote{3} & & \\ 
        Ovalene ($\text{C}_{32}\text{H}_{14}$)  & 1.17 \tnote{4} & 6.71 \tnote{2} & 9.82 \tnote{4} &  & \\
        Circumcoronene ($\text{C}_{54}\text{H}_{18}$)  & 1.44 \tnote{4} & 6.14 \tnote{4} & 8.81 \tnote{4} & 12.9 \tnote{5} & \\
        Circumcircumcoronene ($\text{C}_{96}\text{H}_{24}$) & 3.1 \tnote{5} & 5.7 \tnote{5} & 8.2 \tnote{5} & 10.8 \tnote{5} & 13.4 \tnote{5} \\
    
    \hline
    \end{tabular}
    \begin{tablenotes}[para]
     \item[1] \citet{Mitsui:2007}
     \item[2] \citet{Clar:1981}
     \item[3] \citet{Tobita:1994}
     \item[4] \citet{Malloci:database}
     \item[5] \citet{Bakes:1994}
   \end{tablenotes}
    \end{threeparttable}

\label{tab:Ionization_potential}
\end{table*}
    
\textbf{Ionization Potential of PAHs:} The number of charge states that are accessible to a PAH molecule depends on their IP. Since, in this work, we calculate the charge distribution of PAHs in the PDR environments where H ionizing photons are absent, the charge state $Z+1$ will be accessible to a PAH molecule if its $\text{IP}(Z)$ is less than 13.6 eV. Table \ref{tab:Ionization_potential} lists the IPs of the PAH molecules considered in this work for charge states with an IP of less than 13.6 eV. We adopt the experimentally measured IP values where available. Where experimental data was unavailable, we adopt the IP values estimated from quantum chemical calculations \citep{Malloci:database} or a conducting disk formula \citep{Bakes:1994}. Recently \citet{Wenzel:2020} modified an empirical relationship given by \citet{Weingartner:2001} to calculate the IP of PAHs. We employ the \citet{Bakes:1994} formalism for circumcoronene $Z=2$ and all the charge states for circumcircumcoronene. For these molecules, we compared the \% difference between the IPs determined using the \citet{Bakes:1994} with the IPs determined using the \citet{Wenzel:2020} formalism. We found that the difference is less than 10\% in all cases (see Appendix~\ref{sec:IPs_comparison}), demonstrating that the two formalisms produce comparable results for the large PAH molecules used in this study. Based on the IPs in Table \ref{tab:Ionization_potential}, the maximum accessible charge state for tetracene, pentacene, and ovalene is $Z = 2$, for circumcoronene $Z = 3$, and for circumcircumcoronene $Z = 4$.

\textbf{Ionization yield:} For the PAH molecules considered in this work, the ionization yields have not been measured experimentally with the exception of the singly charged ($Z = 1$) cation of ovalene \citep{Wenzel:2020}. So we employ the semi-empirical relation proposed by \citet{Jochims:1996} to estimate the ionization yields:
 \begin{equation}
          Y_{\rm ion}(Z) \approx
      \begin{cases}
            1, & h\nu \geq \text{IP(Z) + 9.2 eV} \\
            \frac{h\nu - \text{IP(Z)}}{9.2}, & h\nu < \text{IP(Z) + 9.2 eV}
      \end{cases}
    \label{eq:yield}
\end{equation}
      
We note that \citet{Jochims:1996} derived this relationship for ionization yields of neutral PAHs only. However, due to the lack of available data and formalisms to calculate ionization yields for higher charge states, we adopt the \citet{Jochims:1996} relationship for higher charge states by choosing the appropriate IPs. It is worth pointing out that \citet{Wenzel:2020} proposed a relationship to estimate ionization yields of singly charged cations of PAHs with sizes N$_{C} > 32$. We compared their ionization yields of singly charged circumcoronene and circumcircumcoronene to the yields adopted in this work and found they agree well (see Appendix~\ref{sec:yieldcomparison}), lending further support for its general use in our analysis.

%These authors compared their ionization yields of singly charged PAHs to the yields of their neutral counterparts obtained from \citet{Jochims:1996}. They conclude that equation~\ref{eq:yield} describes well the experimentally measured ionization yields for PAH cations, lending further support for its general use in our analysis.

\textbf{Photo-absorption cross-section:}
We obtain the data for the photo-absorption cross-sections from the online database of quantum chemically calculated molecular properties of PAH molecules \citep{Malloci:database}\footnote[1]{\url{https://astrochemistry.oa-cagliari.inaf.it/database/pahs.html}}. The database holds absorption cross-sections of all the PAH molecules considered in this work except circumcircumcoronene in the anionic ($Z = -1$), neutral ($Z = 0$), and cationic ($Z = 1, Z = 2$, and $Z = 3$) states. We estimate the photo-absorption cross-sections for circumcircumcoronene in the desired charge state from the photo-absorption cross-sections of circumcoronene in a corresponding charge state by scaling it with the number of carbon atoms of circumcircumcoronene, i.e. multiplying the photo-absorption cross-sections of circumcoronene with a factor of $96$/$54$ = $1.8$.

\textbf{Sticking coefficient}: The sticking coefficients, $s_{e}$, approach unity for PAHs with electron affinity $>$ 1 eV. None of the PAHs considered in this work have electron affinity, i.e. IP(-1) below 1 eV, therefore we adopt $s_{e} = 1$ following \citet{Tielens:2005}.

\textbf{Frequencies and corresponding intensities of the vibrational modes}: We obtain the frequencies corresponding to the fundamental and the IR active vibrational modes, and the intrinsic intensities corresponding to the IR active vibrational modes from the the NASA Ames PAH IR Spectroscopic database \citep[PAHdb, version 3.20;][]{Bauschlicher:2018, Boersma:14, Bauschlicher:10, Mattioda:2020} except for the tetracene dication and the pentacene anion for which the database has no data. In Appendix~\ref{sec:UID}, we provide the UIDs of the molecules in PAHdb. For the tetracene dication and the pentacene anion, we obtain the relevant data from the \citet{Malloci:database} database. In both the databases, the intrinsic intensities of the IR active vibrational modes are expressed in terms of A-value, $A$ (km/mol). The A-value is related to the laboratory measured absorbance (see equation 6 in \citet{Mattioda:2020}). For the purpose of our calculation, we convert A-value in the units of cross-section integrated over frequency following \citet{Mattioda:2020};
\begin{equation}
        \sigma_{\nu_{i}} = \frac{10^{5}}{N_{A}}\times c \times A_{i}
        \label{eq:IR_cross-section}
\end{equation}   
where $N_{A}$ is Avogadro's number i.e.  $6.02 \times 10^{23}$.

\subsection{Comparison with other PAH emission models}
\label{subsec:comparison_PAH_models}

The PAH emission has been modelled in the past following various approaches. The emission model presented in this paper is based on the work of \citet{Bakes:01:a} but differs in the following ways. Firstly, the \citet{Bakes:01:a} model does not account for the possibility of ionization when calculating the average energy absorbed by PAHs which is used to calculate the IR emission spectra. In contrast, in the current model, we include the possibility of ionization after photon absorption. Secondly, while \citet{Bakes:01:a} provided a framework to calculate the cooling rate for PAHs which we adopt in this work (see equation~\ref{eq:cooling_rate}), they used a simplified single analytical expression for calculating the cooling rate of different PAHs. Lastly, in this paper, we have adopted more recent calculations of molecular parameters of PAHs (see Section~\ref{subsec:characteristics_PAHs}).

It is worth pointing out that \citet{Verstraete:2001} also presented a  PAH emission model similar to the one presented in this work and in \citet{Bakes:01:a} but additionally incorporated the effect of the size distribution of PAHs in the calculation of the total PAH spectra in astrophysical environments. These authors adopted IR cross-sections for different PAH sizes by scaling the Einstein-A coefficients of the major PAH bands with their size. Since, in this work, we adopt the experimentally measured or theoretically calculated IR cross-sections of individual PAHs, we calculate the spectra of individual PAH molecules spanning a range from 18 - 96 carbon atoms.

Other PAH emission models use a Monte Carlo based approach to model the IR emission from PAHs \citep[e.g.][]{Mulas:1998, Joblin:2002, Malloci:2003, Mulas:2006}. We benchmark our model by comparing our model calculations with the Monte Carlo based model presented in \citet{Mulas:2006}. For the comparison, we calculate the IR emission spectra of neutral coronene in the planetary nebula IRAS~21282+5050 whose photon flux is well represented by the radiation field of $G_{0} = 5 \times 10^{5}$ in the units of Habing field and a blackbody of temperature 40,000 K (see Fig.~\ref{fig:flux_comparison_IRAS_21282}). Subsequently, we compare the flux fraction of each band of neutral coronene for both models. We present the results of the comparison and the molecular parameters of neutral coronene used to model the IR emission in Appendix~\ref{sec:comparison_Mulas}. We find that the flux fraction of the strong bands (which combined carry 88\% of the flux) differs in the two models by at most 20\%. Although there is a discrepancy in the flux fraction of weak features in the 6.69-8.24, 12.5-12.9 $\mu$m range and at 26.5 $\mu$m when compared individually, that is not the case when comparing their combined fluxes. The combined flux of weak features in the 6.69-8.24 $\mu$m and  12.5-12.9 $\mu$m range agrees between the two models within 10\% and 21\%, respectively. The discrepancies in the flux of individual weaker features could either stem from the different molecular characteristics of coronene used in the two models or due to the use of an average energy for the absorbed photon in this model versus the inclusion of the full distribution of absorbed photons in the Monte Carlo model.
        
\section{Results of the model for PAHs under astrophysical conditions}
\label{sec:results}
We use the PAH emission model to investigate the charge distribution and subsequent IR emission characteristics of the five PAHs described in Section~\ref{subsec:characteristics_PAHs} over a range of relevant astrophysical conditions. Since PAH charge distribution is set by the so-called ionization parameter, $\gamma = G_{0}\times\sqrt{T_{\rm gas}}/n_{e}$, we investigate the model results as a function of $\gamma$. 

\subsection{Charge distribution}
\label{subsec:results_charge_5_molecules}

\begin{figure}
    \centering
    \includegraphics[scale=0.40]{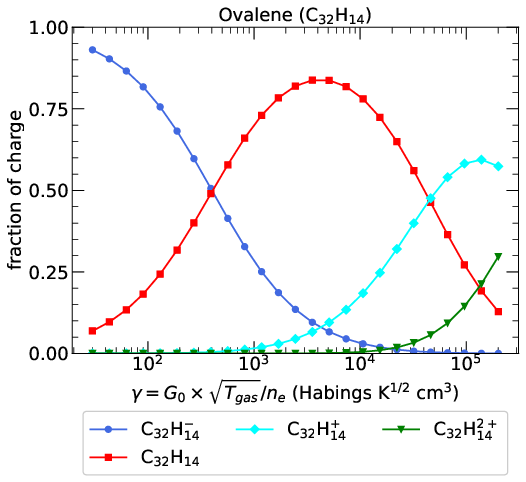} 
    \caption{Charge distribution of ovalene as a function of the ionization parameter $\gamma$, for a fixed value of $G_{0} = 2600$, $T_{\rm gas} = 300$K, and $T_{\rm eff} = 17000$K.}
    \label{fig:charge_distribution_ovalene_gamma}
\end{figure}

Adopting a fixed value of $G_{0} = 2600$ in units of the Habing field, $T_{\rm gas} = 300$ K, an electron abundance of 1.6 $\times$ 10$^{-4}$, and $10^{3} < n_{\rm gas} < 10^{6}$ cm$^{-3}$, we obtain $\gamma$ values ranging from $\sim$30--2$\,\times\, 10^{5}$ Habings K$^{1/2}$ cm$^{3}$ for which we calculate the charge distribution of the PAH molecules considered in this work. This range of $\gamma$ values cover well the variety of the physical conditions prevalent in the PDRs (see Table~\ref{tab:physical_conditions_PDRs}). We first use $T_{\rm eff}$ = 17000 K for the blackbody radiation field, which is a required parameter for determining the photo-ionization rate in the charge distribution model (see Section~\ref{subsec:Charge_dist_model}); we will consider other values later. Fig.~\ref{fig:charge_distribution_ovalene_gamma} shows the results of the calculation for ovalene, one of the PAH molecules considered in this work. The results for the remaining PAH molecules are shown in Appendix~\ref{app:chargedist} (see Fig.~\ref{fig:charge_distribution_gamma}). In comparing the results for different PAHs, it should be kept in mind that the charge distribution is controlled by the ionization rate over the recombination rate. This ratio is proportional to the ionization parameter as well as the ratio of the photo-ionization cross section to the electron-PAH interaction cross section. To first order, the former varies as $N_{C}$ while the latter varies with $N_{C}^{1/2}$. The small differences in ionization potential between the different PAHs have only a small effect given the broad energy dependence of the ionizing flux distribution. Of course, with increasing PAH size, more charge states may become accessible for $h\nu <13.6$ eV.

We find that with increasing $\gamma$ values, the dominant charge state shifts from anions through neutrals to cations. In the case of ovalene, the anions remain the dominant charge state up to $\gamma \sim$ 4$\times$ 10$^{2}$ after which the neutrals dominate up to $\gamma \sim$ 4$\times$ 10$^{4}$, beyond which the cationic charge states dominate. We note that for other PAH molecules, the general trend remains the same, but as outlined above, the precise distribution of the charge states varies somewhat with N$_{C}$; e.g., as the  size of the molecule increases, cations begin to dominate at lower $\gamma$ values. Likewise, the highest $\gamma$ value for which anions dominate decreases with increasing size. While there are differences in the precise values of the fraction of charge for individual PAH molecules, the key result that emerges from the calculations is that the anions are the dominant charge state for $\gamma$ values less than $\sim 2 \times 10^{2}$, the neutrals for $\gamma$ values in the range $\sim 10^{3}$--$10^{4}$, and the cations for $\gamma$ values greater than $\sim 5 \times 10^{5}$. These results are consistent with the charge distribution results from \citet{Bakes:2001}. We note that although the trends observed in the charge distribution of five PAHs considered in this work are similar to \citet{Bakes:01:a}, the absolute value of the fraction of PAHs in a particular charge state at a given $\gamma$ value differs due to the use of more recent experimentally, or quantum chemically determined PAH characteristics.

We now investigate how the charge distribution would change if we choose values of $G_{0}$, $T_{\rm gas}$, and $n_{e}$ other than those described above. We find that at a particular $\gamma$ value, we get the same result for the charge distribution regardless of the values of the individual parameters characterizing it. This is expected since the ratio of the photo-ionization rate and the electron-recombination rate that goes in equation~\ref{eq:frac_PAHs} is proportional to $\gamma=G_{0}\times\sqrt{T_{\rm gas}}/n_{e}$. We note that, in contrast to the electron recombination rate, the electron attachment rate (equation~(\ref{eq:electronattachmentrate})) does not depend on $T_{\rm gas}$. In any event, since any $T_{\rm gas}$ dependence scales only with $T_{\rm gas}^{1/2}$ and the relevant range in $T_{\rm gas}$ is rather limited, it has only a small effect on the charge distribution. However, we find that the results of the charge distribution of PAHs are sensitive to the excitation conditions, i.e. the effective temperature, $T_{\rm eff}$, of the radiation field (see Fig.~\ref{fig:charge_distribution_two_temperatures}). While the overall trend remains the same; that at low values of $\gamma$, anions dominate, and at high values of $\gamma$ cations dominate, the precise values of the fraction of each charge at a particular value of $\gamma$ vary with $T_{\rm eff}$.
%We emphasize that the effect of the effective temperature of the radiation field is small as the ionization rate only samples the range 6-13.6 eV and only for effective temperatures much less than 10,000 K, will the ionization rate be affected (Spaans et al).  

\subsection{PAH spectra}
\label{subsec:PAH_spectra_5_molecules}

\begin{figure}
    \centering
    \includegraphics[scale=0.37]{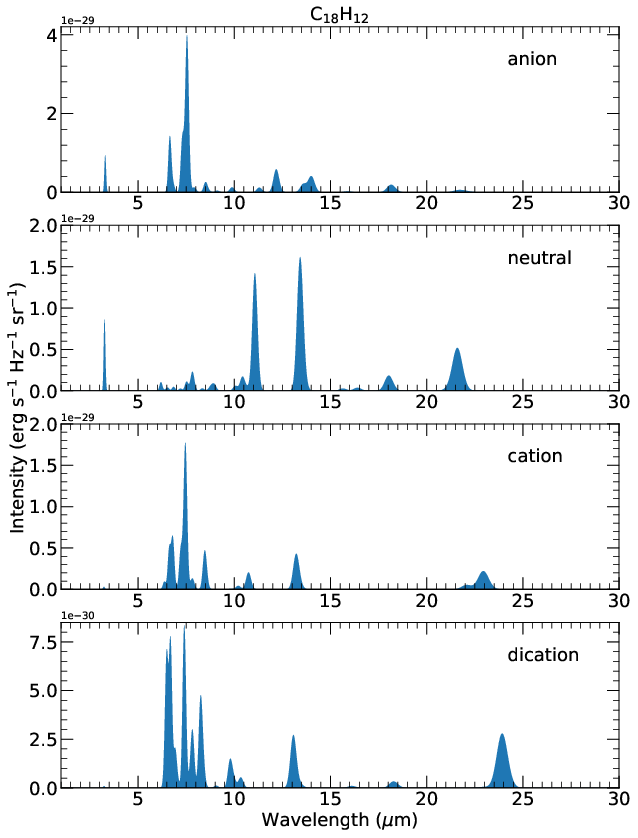}
    \caption{Calculated spectra of tetracene (C$_{18}$H$_{12}$) in the anionic, neutral, cationic, and dicationic states at excitation conditions characteristic of NGC~7023 ($G_{0}=2600$ and $T_{\rm eff}=17000$K).}
    \label{fig:spectra_tetracene}
\end{figure}

\begin{figure}
    \centering
    \includegraphics[scale=0.37]{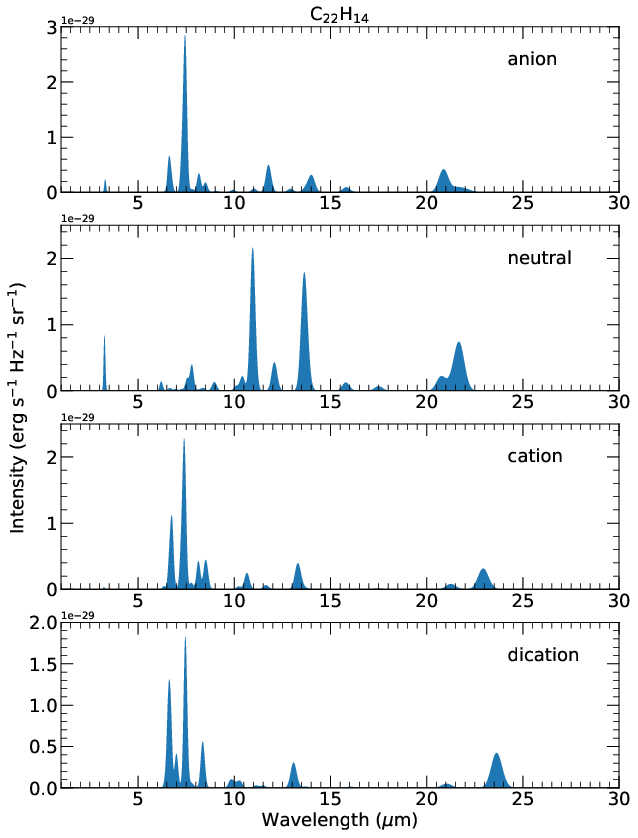}
    \caption{Calculated spectra of pentacene (C$_{22}$H$_{14}$) in the anionic, neutral, cationic, and dicationic states at excitation conditions characteristic of NGC~7023 ($G_{0}=2600$ and $T_{\rm eff}=17000$K).}
    \label{fig:spectra_pentacene}
\end{figure}

\begin{figure}
     \centering
     \includegraphics[scale=0.37]{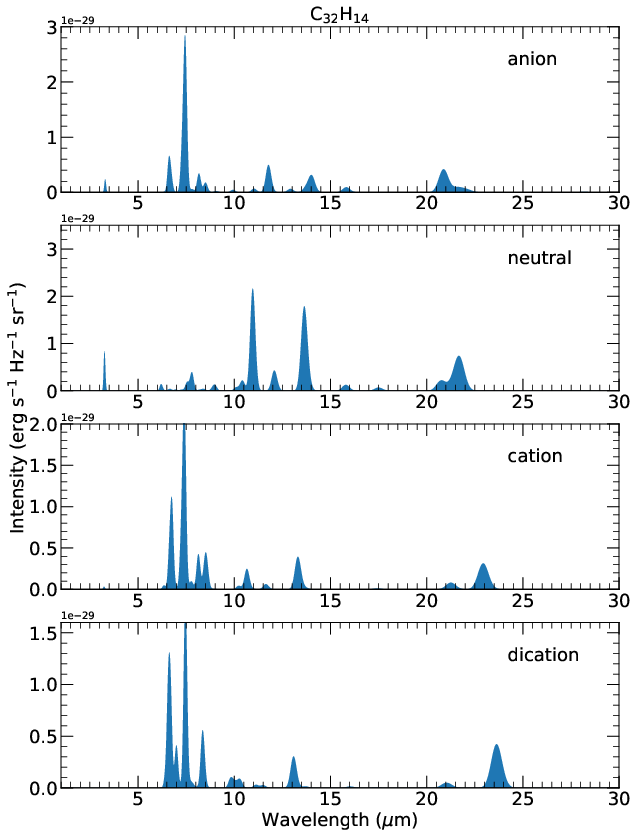}
     \caption{Calculated spectra of ovalene (C$_{32}$H$_{14}$) in the anionic, neutral, cationic, and dicationic states at excitation conditions characteristic of NGC~7023 ($G_{0}=2600$ and $T_{\rm eff}=17000$K).}
     \label{fig:spectra_ovalene}
 \end{figure}

 \begin{figure}
     \centering
     \includegraphics[scale=0.37]{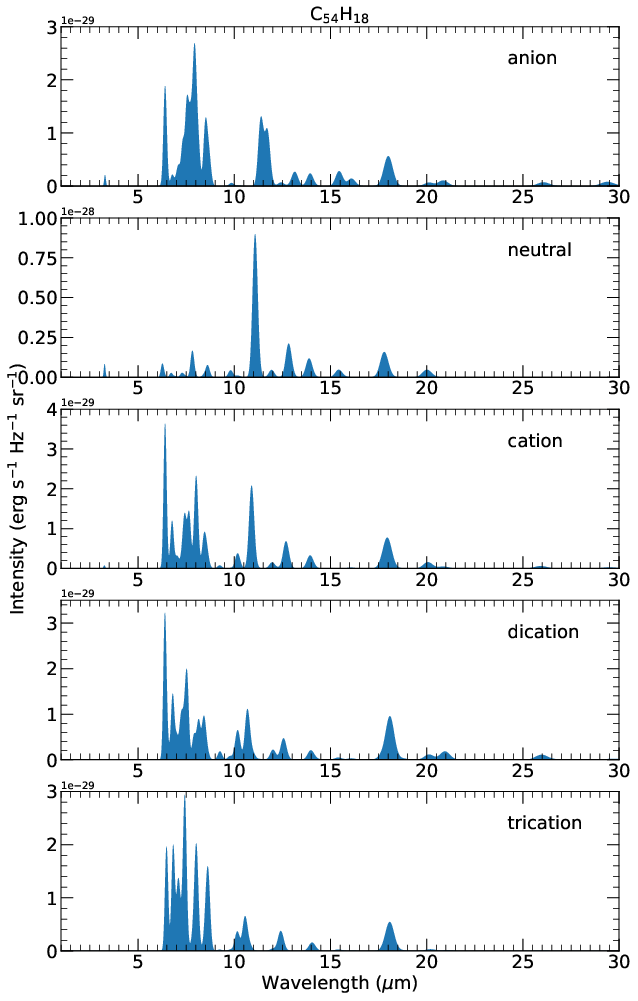}
     \caption{Calculated spectra of circumcoronene (C$_{54}$H$_{18}$) in the anionic, neutral, cationic, dicationic, and tricationic states at excitation conditions characteristic of NGC~7023 ($G_{0}=2600$ and $T_{\rm eff}=17000$K).}
     \label{fig:spectra_circumcoronene}
 \end{figure}

 \begin{figure}
     \centering
     \includegraphics[scale=0.37]{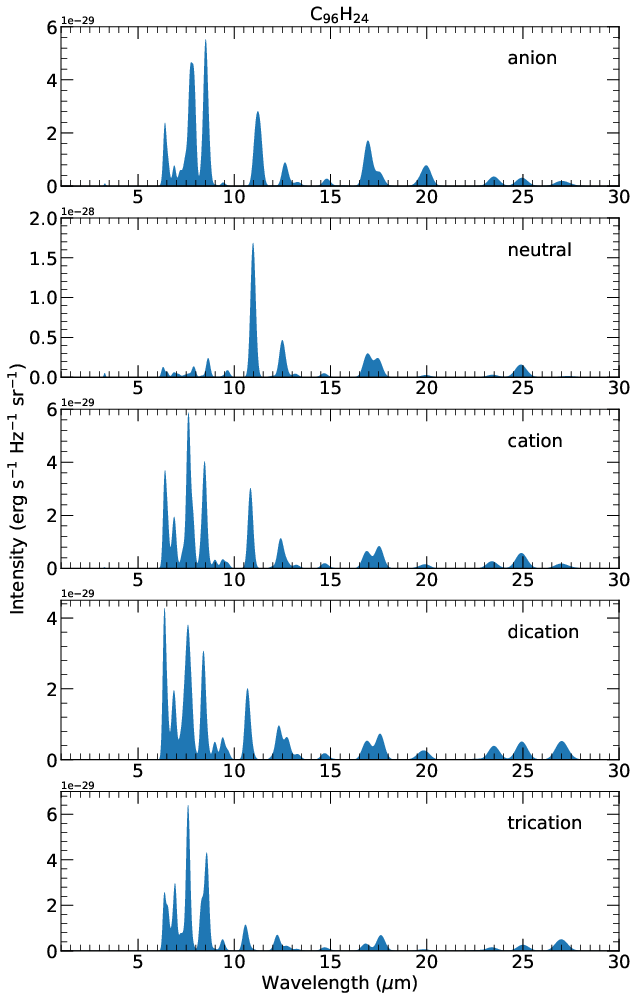}
     \caption{Calculated spectra of circumcircumcoronene (C$_{96}$H$_{24}$) in the anionic, neutral, cationic, dicationic, and tricationic states at excitation conditions characteristic of NGC~7023 ($G_{0}=2600$ and $T_{\rm eff}=17000$K).}
     \label{fig:spectra_circumcircumcoronene}
 \end{figure}

The effect of the charge state on the intrinsic strengths of the vibrational modes of a PAH molecule has been extensively analyzed previously by several authors \citep[see, e.g.][]{Langhoff:1996, Hudgins:1999, Bauschlicher:2000, Bakes:2001, Bakes:01:a}. We present the intrinsic spectra of the molecules considered here in Appendix~\ref{sec:intrinsic_spectra}. We encourage the reader to refer to the studies by, e.g. \citet{Langhoff:1996, Hudgins:1999, Bauschlicher:2000, Bakes:2001, Bakes:01:a} for a detailed discussion on the effects of the charge state on intrinsic spectra. In this section, we discuss the effects of the charge on the spectra of the PAH molecules after we apply the PAH emission model. We emphasize that after applying the emission model, the relative strength of the features will differ from the intrinsic strengths; however, the entire character of a molecular spectrum will not change.

%Figs show the intrinsic spectra of the PAH molecules for all the charge states. To facilitate the comparison, we have normalized the intrinsic spectra to the maximum intensity in each charge state. The neutrals exhibit strong features at the 3.3 $\mu$m and in the 10-15 $\mu$m range and weak features in the 6-9 $\mu$m range with 3.3 $\mu$m as the strongest feature in all the cases. The anions exhibit strong features at the 3.3 $\mu$m and in the 6-9 $\mu$m range and weak features in the 10-15 $\mu$m range. Similar to anions, the cations also exhibit strong features in the 6--9 $\mu$m region and weak features in the 10-14 $\mu$m region. The features in the 3.3 $\mu$m region, on the other hand are weaker in cations. In fact the intrinsic strength of the 3.3 $\mu$m feature is even less than that of the features in the 10-14 $\mu$m region. Longwards of 15 $\mu$m, the molecules in all the charge states exhibhit few weak features with strength of these features in neutrals slightly greater than that in other charge states.  

Figs.~\ref{fig:spectra_tetracene}-\ref{fig:spectra_circumcircumcoronene} show the spectra of PAHs considered in this work after the application of the emission model. We calculate the spectra in each charge state for the excitation conditions typical of NGC~7023 ($G_{0}$ = 2600 and $T_{\rm eff}$=17000 K) following the procedure outlined in section~\ref{subsec:spec_calc}. The average energies absorbed by the PAH molecule corresponding to the excitation conditions for which we calculate the spectra in each charge state are given in Table~\ref{tab:average_energy}. To allow for multiphoton absorptions in calculating the IR emission from PAHs, we include three-photon absorptions to calculate the spectra presented in Figs.~\ref{fig:spectra_tetracene}-\ref{fig:spectra_circumcircumcoronene} and checked that the $G(T)dT$ function and the calculated spectra have converged at that point. Essentially, we perform three iterations on the temperature distribution function, $G(T)dT$, after which our calculated intensities of each vibrational mode converge. Once we obtain the strength of each IR active mode, we convolve the intensity of each vibrational mode with a Gaussian profile with a full width at half maximum (FWHM) determined by the spectral resolving power of 200.

The anionic and cationic charge states exhibit the strongest features in the 6--9 $\mu$m region, characteristic of the C-C stretching and C-H in-plane bending modes for all of the PAH molecules considered here, while the neutral charge states exhibit the strongest features in the 10--15 $\mu$m region characteristic of the C-H out of plane bending modes. While there are differences in the relative intensities of the PAH features in the 6--9 $\mu$m region for cationic and anionic charge states, these differences are not as distinct as those that distinguish neutrals from the other charge states. The intensity of the 3.3 $\mu$m feature, on the other hand, appears to differentiate between different charge states. The 3.3 $\mu$m feature is strongest in neutral charge states and weakest in cationic charge states. The intensity of this feature varies between molecules in the anionic charge states, with tetracene exhibiting a strong 3.3 $\mu$m feature compared to the other molecules. However, for a given PAH molecule, the 3.3 $\mu$m feature is always stronger in anions than in cations. As a result, the relative intensity of the 3.3 $\mu$m and 6--9 $\mu$m features may serve as a distinguishing feature between anionic and cationic charge states.  Finally, we note that there are few strong features in the region long wards of 15 $\mu$m, but we could not identify a systematic trend among the features in this region that can distinguish between different charge states. %As an example, we observe the shifting of the strong feature towards higher wavelengths in the 20--25 $\mu$m region in tetracene, pentacene, and ovalene, but not in circumcoronene and cirumcircumcoronene.  

\subsubsection{The 6.2/11.2 band ratio}
\label{subsubsec:62_112_ratio_5_molecules}

\begin{figure}
     \centering
     \includegraphics[scale=0.37]{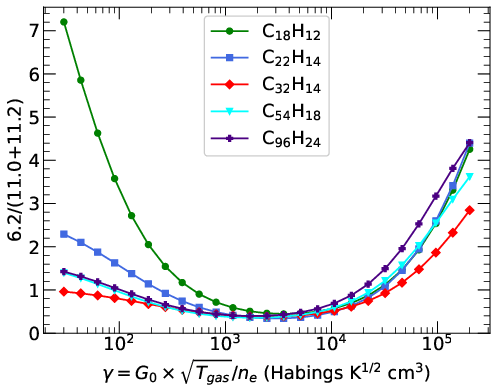}
     \caption{The 6.2/(11.0+11.2) band ratio for the five PAH molecules considered in this work as a function of the ionization parameter $\gamma$. High values of 6.2/(11.0+11.2) on the right end of the curve result from the large fraction of cations at high $\gamma$ values, whereas high 6.2/(11.0+11.2) values on the left end of the curve result from the large fraction of anions at low $\gamma$ values. See text for details of the calculation of the 6.2/(11.0+11.2) band ratio.}
     \label{fig:ratio_plot}
 \end{figure}
 
Traditionally, the ratio of the 6.2 and 11.2 $\mu$m PAH features is used as an indicator of the PAH charge state in astrophysical environments where low values imply neutrals as the dominant charge state and high values imply cations as the dominant charge state \citep[e.g.][]{Peeters:2002, Berne:2007, Boersma:2018}. In light of the results presented in Figs.~\ref{fig:spectra_tetracene}-\ref{fig:spectra_circumcircumcoronene} as discussed above, this interpretation may be unreliable for high values of this ratio. To investigate how the similarity between the features of anions and cations in the 6--9 $\mu$m region translate in the interpretation of the 6.2/11.2 band ratio, we analyze this ratio as a function of the ionization parameter $\gamma$.

We obtain the spectra of a PAH molecule in all the charge states at the conditions that are typical of NGC~7023. We note that for circumcircumcoronene in charge state $Z=4$, there is no available data on the frequencies and the intensities of the vibrational modes, and therefore, we do not include it in our calculation. However, it does not have a significant impact on the results of the PAH ratio because the contribution from the $Z = 4$ state of circumcircumcoronene for the $\gamma$ values investigated here is negligible (see Fig.~\ref{fig:charge_distribution_gamma}). For the calculation of the 6.2/11.2, we do not convolve the spectra with Gaussian functions. Instead, for each charge state, we extract the intensities corresponding to the C-C modes (6.2 $\mu$m) by adding the intensities of all the modes in the 6.1--6.9 $\mu$m range. We recognize that we use a broad wavelength range to measure the 6.2 $\mu$m band in order to include the strongest C-C stretching mode in the measurement for all the PAH molecules. In this pursuit, we occasionally miss the traditional 6.2 $\mu$m band. Therefore, when comparing these results to the astronomical observations, we analyze trends rather than any spectroscopic detail in the PAH emission from astrophysical environments. For C-H modes (11.0+11.2 $\mu$m), we add the intensities of all the modes in the 10.0--12.0 $\mu$m range. We note that, although the wavelength range used to calculate the intensities of C-H modes theoretically consists of a couple of weak and strong features, only the features at 11.0 and 11.2 $\mu$m bands are observed, with the 11.2 $\mu$m PAH band being the dominant band. Therefore, for the remainder of the paper, we will refer to the intensities of the C-H modes as the 11.0+11.2 $\mu$m band. This will also serve as a reminder to the reader that the weak 11.0 $\mu$m PAH band characteristic of cations is also included in the measurement of the C-H modes. We multiply the extracted intensities of the C-C and C-H modes for a charge state $Z$ of a molecule with its corresponding charge fraction. Finally, we add up the weighted intensities of the C-C and C-H modes for all the charge states of a molecule to obtain the 6.2 and 11.0+11.2 $\mu$m band intensities, respectively.

Fig.~\ref{fig:ratio_plot} shows the 6.2/(11.0+11.2) band ratio as a function of $\gamma$ for all the PAH molecules considered in this work. We note that high values of 6.2/(11.0+11.2) can be reached for both low and high values of $\gamma$. We recall from section~\ref{subsec:results_charge_5_molecules} that anions dominate at low values of $\gamma$ and cations at high values of $\gamma$. Therefore, the high values of 6.2/(11.0+11.2) at low values of $\gamma$ result from the contribution of anions to the 6.2 $\mu$m band, thereby illustrating that observed high values of 6.2/(11.0+11.2) do not always necessarily imply a large fraction of cations. For intermediate $\gamma$ values, where neutrals dominate, the curve of 6.2/(11.0+11.2) is relatively flat for all molecules exhibiting almost similar values. In contrast, the 6.2/(11.0+11.2) ratio varies for the different molecules at low and high $\gamma$ values, although there is no systematic trend between the molecules. Finally, we note that compared to compact PAHs, acenes exhibit relatively high 6.2/(11.0+11.2) values at the left end of the curve, where anions dominate. We discuss the implications of this result further in section~\ref{subsec:PAH_emission_PDR_environments} where we compare the results of the PAH emission model with observations of PDRs.

\subsubsection{6.2/(11.0+11.2) vs 3.3/(11.0+11.2)}
\label{subsubsec:62_112_vs_33_112}
\begin{figure*}
    \centering
    \begin{tabular}{cc}

    \includegraphics[scale=0.50]{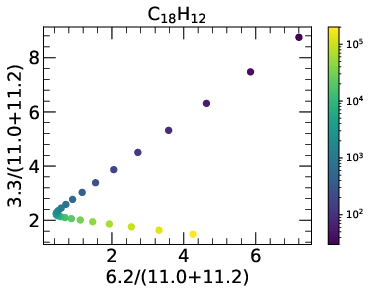} &  \includegraphics[scale=0.50]{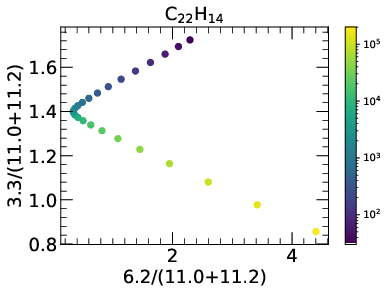}\\
    \includegraphics[scale=0.50]{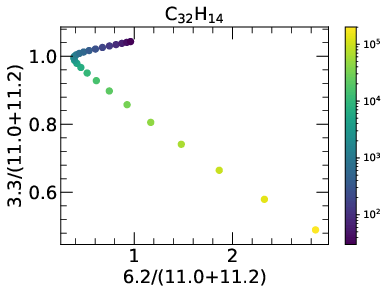} & 
    \includegraphics[scale=0.50]{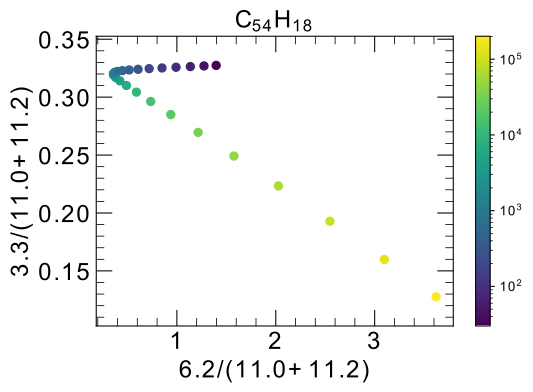} \\
    \includegraphics[scale=0.50]{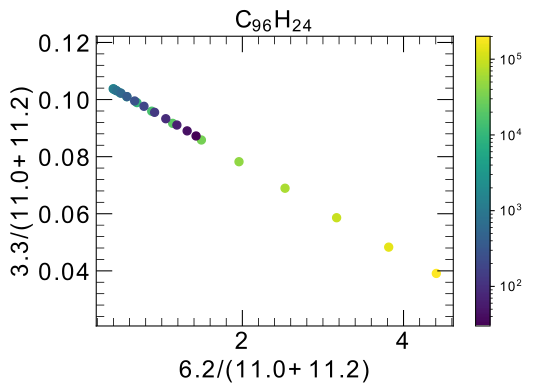}\\
    \end{tabular}
    \caption{Ratios of the 6.2/(11.0+11.2) vs 3.3/(11.0+11.2) color-coded with $\gamma$ values for the five PAH molecules considered in this work. Except for circumcircumcoronene, there are two values of 3.3/(11.0+11.2) for a given value of 6.2/(11.0+11.2), one originating from low $\gamma$ values and the other from high $\gamma$ values.  See text for details of the calculation of the 6.2/(11.0+11.2) and 3.3/(11.0+11.2) band ratios.}
    \label{fig:distinction_anions_cations}
    \end{figure*}

Given that high values of the 6.2/(11.0+11.2) ratio can arise from both anions and cations, it becomes important to have a tool to distinguish between these two charge states while interpreting astronomical observations. As noted in section~\ref{subsec:PAH_spectra_5_molecules}, the relative intensities of the features in the 3.3 and 6--9 $\mu$m range can be used to distinguish between cations and anions. To investigate this, we analyze plots of the 6.2/(11.0+11.2) vs 3.3/(11.0+11.2) for the PAH molecules considered here. To calculate the 3.3/(11.0+11.2) ratio, we extract the intensity of the 3.3 $\mu$m feature by adding up intensities of all the modes in the 3.0--4.0 $\mu$m range for each charge state before weighting them with their corresponding charge fraction. We show the resulting plots in fig.~\ref{fig:distinction_anions_cations}. There are essentially two branches for each molecule, with both branches overlapping in the case of circumcircumcoronene. In one branch, the 3.3/(11.0+11.2) values decrease with increasing 6.2/(11.0+11.2) values. This branch corresponds to $\gamma$ values $> 10^{3}$ where neutrals and cations dominate. However, in the second branch, the 3.3/(11.0+11.2) values increase with increasing 6.2/(11.0+11.2) values for the acenes and ovalene, remain almost constant with increasing 6.2/(11.0+11.2) values for circumcoronene, and decreases with increasing 6.2/(11.0+11.2) values for circumcircumcoronene. The second branch originates from $\gamma$ values $< 10^{3}$ where anions dominate. The 3.3/(11.0+11.2) vs 6.2/(11.0+11.2) plots separate cations from anions in acenes, ovalene, and circumcoronene, but not for circumcircumcoronene. We further note that the slope of the second branch originating from anions not only vary with respect to the molecule considered but also with respect to the excitation conditions (see Fig.~\ref{fig:distinction_anions_cations_Orion_bar}). For example, at $G_{0}=26000$ and $T_{\rm eff} = 40000 $K, the 3.3/(11.0+11.2) values in the second branch decrease with increasing 6.2/(11.0+11.2) values for circumcoronene whereas they remain constant for $G_{0}=2600$ and $T_{\rm eff}$ = 17000 K. Therefore, we conclude that the 3.3/(11.0+11.2) vs 6.2/(11.0+11.2) plots can not be used as a diagnostic tool for astronomical purposes. However, the plots presented here show that corresponding to a single value of 6.2/(11.0+11.2), there are two values of 3.3/(11.0+11.2). This result demonstrates the influence of charge state on 3.3/(11.0+11.2) ratio which is traditionally used to determine the PAH size \citep[e.g.][]{Allamandola:1989, Schutte:1993, Ricca:2012, Croiset:2016, Knight:2021}. It also has implications in interpreting the 3.3/(11.0+11.2) - 6.2/(11.0+11.2) diagnostic plots often used in the literature to determine PAH charge and size \citep[e.g.][]{Draine:2001, Maragkoudakis:2018, Maragkoudakis:2020}. We discuss this further in Section~\ref{subsec:grid_plots}.  

\section{Application of the model to PDRs}
\label{sec:application}
We apply the PAH emission model to the IR observations of four well-studied PDRs and the diffuse ISM that sample various physical conditions.

\subsection{Environments}
\label{subsec_environments}
In the following sections, we describe the environments and present a relevant discussion on the measurements of physical conditions ($G_{0}$, $n_{\rm gas}$, and $T_{\rm gas}$) in these environments. We emphasize that the measurement of the physical conditions strongly depends on the method adopted to measure these, and as such, there exists a range of values for some of the parameters across the literature. In such cases, we adopt an average value of the parameters over the range present in the literature. To calculate $n_{e}$ from $n_{\rm gas}$, we adopt an electron abundance, $X_{e}$, of $1.6\times10^{-4}$ from \citet{Sofia:2004} assuming that the $X_{e}$ is a result of the ionization of C. Table~\ref{tab:physical_conditions_PDRs} presents the physical conditions we adopt, the $\gamma$ values we derive from these, and the $T_{\rm eff}$ of the radiation field in the environments considered here. For the diffuse ISM, we use the expression provided by \citet{Tielens:2005} for the energy dependence of the radiation field.

\begin{table*}
\caption{Physical conditions in the environments studied in this work.}
    \centering
    \begin{threeparttable}[t]
    \begin{tabular}{c c c c c c c}
    \hline
    PDR environment & $G_{0}$ (Habings field)\tnote{a} & $n_{\rm gas}$ (cm$^{-3}$) \tnote{b} & $T_{\rm gas}$ (K) \tnote{c} & $X_{e}$ \tnote{d} & $T_{\rm eff}$ (K) \tnote{e} & $\gamma$ \\
    \hline
    NGC~7023 & 2600  & $10^{4}$   & 400   & $1.6\times 10^{-4}$  & 17000   & $3.3\times10^{4}$\\
    NGC~2023 & 17000 & $2.0 \times 10^{5}$   & 500  & $1.6\times 10^{-4}$   & 23000  & $1.2\times10^{4}$\\
    Horsehead Nebula & 100  & $2.0\times 10^{5}$   & 100   & $1.6\times 10^{-4}$  & 33000  & 30\\
    Orion Bar & 26000 & $10^{5}$  & 500  & $1.6\times 10^{-4}$ & 40000  & $3.6\times10^{4}$\\
    diffuse ISM & 1.7 & 30 & 80 &  $1.6\times 10^{-4}$ &  & $3.2\times10^{3}$\\
    
    \hline
    \end{tabular}
    \begin{tablenotes}[para]
     \item[a] NGC~7023: \citet{Chokshi:88}; NGC~2023: \citet{Sheffer:11}; Horsehead Nebula: \citet{Abergel:2003}; Orion Bar: \citet{Marconi:1998}; diffuse ISM: \citep{Habing:1968, Draine:1978} 
     \item[b] NGC~7023: \citet{Chokshi:88}, \citet{Joblin:10}, \citet{Kohler:2014}, \citet{Beranrd:2015}, \citet{Joblin:18}; NGC~2023: \citet{Steiman-Cameron:97}, \citet{Sheffer:11}; Horsehead Nebula: \citet{Habart:2005}; Orion Bar: \citet{Parmar:1991}, \citet{Tauber:1994}, \citet{Young_Owl:2000}, \citet{Bernard-Salas:2012}, \citet{Joblin:18}; diffuse ISM: \citep{Wolfire:2003}   
     \item[c] NGC~7023: \citet{Fuente:1999},  \citet{Fleming:2010}; NGC~2023: \citet{Steiman-Cameron:97}, \citet{Sheffer:11}; Horsehead Nebula: \citet{Habart:2011}; Orion Bar: \citet{Allers:2005}; diffuse ISM: \citep{Wolfire:2003}   
     \item[d] \citet{Sofia:2004}
     \item[e] NGC~7023: \citep{Racine:1968, Witt:1980, van_den:1997}; NGC~2023: \citep{Racine:1968, Mookerjea:09}; Horsehead Nebula: \citep{Edwards:1976, Abergel:2003}; Orion bar: \citep{Kraus:2007}
    \end{tablenotes}
    \end{threeparttable}
    
    \label{tab:physical_conditions_PDRs}
\end{table*}

\subsubsection{NGC~7023}
\label{subsubsec:NGC_7023}
NGC~7023 is a reflection nebula illuminated by HD~200775, a spectroscopic binary \citep[Herbig B3Ve - B5;][]{Racine:1968, Witt:1980, van_den:1997}. It is situated at a distance of 361 $\pm$ 6 pc \citep{gaia16,gaia18b} and is a well-studied reflection nebula due to its high surface brightness and proximity to the Earth. There are three PDRs in NGC~7023 towards the North West (NW), South West (SW), and East of the central star. In this work, we focus on the NW PDR. The structure of NGC~7023 towards the NW PDR is easy to visualize with stratified layers of a cavity carved out by the illuminating star, followed by the almost edge-on PDR at the edges of the cavity. The PDR is made of diffuse gas embedded with filaments of dense gas. In this work, we adopt the values of the physical conditions estimated previously by several authors. We take the estimate of $G_{0}=2600$ in units of Habing Field from \citet{Chokshi:88} who estimated this from the far-infrared (FIR) continuum intensity assuming that the UV radiation from the star is absorbed and reradiated in the IR by the dust. There are several estimates for $n_{\rm gas}$ in the literature. Using observations of O\,I (63 $\mu$m), C\,II (158 $\mu$m), and low J $^{12}$CO and $^{13}$CO lines, \citet{Chokshi:88} estimated $n_{\rm gas}\sim4\times 10^{3}$ cm$^{-3}$ in dense clumps and 500 cm$^{-3}$ outside of the clumps in the NW PDR. \citet{Joblin:10} estimated $n_{\rm gas}\sim7\times 10^{3}$ cm$^{-3}$ from the Herschel measurements of the C\,II (158 $\mu$m) line alone. \citet{Kohler:2014} modelled intermediate- to high-J CO lines in NGC~7023 and estimated $n_{\rm gas}\sim5\times 10^{4} - 5\times 10^{6}$ cm$^{-3}$ from high-J CO lines and $10^{4}-10^{5}$ cm$^{-3}$ from low-J CO lines. \citet{Beranrd:2015} estimated densities of $1.7\times 10^{4} - 1.8\times 10^{5}$ cm$^{-3}$ from the measurements of the O\,I (145 $\mu$m) line. Recently, \citet{Joblin:18} modelled high-J CO lines and estimated $n_{\rm gas}$ of $4\times10^{4}$ cm$^{-3}$ at the edge of the PDR. These estimates of $n_{\rm gas}$ range from values as low as 500 cm$^{-3}$ in the diffuse gas to as high as $10^{6}$ cm$^{-3}$ in the dense filaments. We note that the estimates of $n_{\rm gas}$ strongly depend on the line being used for the estimation that probe regions of different densities within the PDR. Here, we adopt an average value of $\sim 10^{4}$ cm$^{-3}$. For $T_{\rm gas}$, we adopt a value of $\sim$400 K measured using pure H$_{2}$ rotational lines by \citet{Fuente:1999} based on the Infrared Space Observatory (ISO) observations. This value of $T_{\rm gas}$ is in agreement with the calculation of $T_{\rm gas}$ by \citet{Fleming:2010} based on the Spitzer observations.

\subsubsection{NGC~2023}
\label{subsubsec:NGC_2023}
NGC~2023 is a reflection nebula in the Orion constellation illuminated by B1.5V star HD~37903 \citep{Racine:1968, Mookerjea:09}. It is at a distance of 403 $\pm$ 4 pc from the Earth \citep{Kounkel:18}. The illuminating star has carved out a cavity in the molecular gas, the edges of which form the PDR. In this work, we focus on the bright ridge in the PDR known as the South ridge. We adopt the physical conditions in the ridge from \citet{Sheffer:11}. They determined a value of radiation field $\chi = 10^{4}$ in terms of Draine interstellar radiation field, which translates to a $G_{0} = 1.7\times 10^{4}$ in terms of the Habing field and $n_{\rm gas}$ of $2\times 10^{5}$ cm$^{-3}$ using pure rotational $\text{H}_{2}$ emission lines. They also measured excitation temperatures, $T_{\rm ex}$, and found that $T_{\rm ex}$ ranges from 240--700 K. Since transitions between low-J levels of the pure rotational $\text{H}_{2}$ probe $T_{\rm gas}$ owing to their low critical densities, we adopt an average $T_{\rm gas}$ value of $\sim 500$ K. These estimates for the physical conditions are consistent with the estimates obtained using measurements of [\textrm{O~{\textsc{i}}}] (63, 145 $\mu$m), [\textrm{C~{\textsc{ii}}}] (158 $\mu$m), [\textrm{Si~{\textsc{ii}}}] (35 $\mu$m) and CO lines \citep{Steiman-Cameron:97}.

\subsubsection{Horsehead Nebula}
\label{subsubsec:Horsehead}
The Horsehead nebula is a dark nebula situated due west of NGC~2023 in the Orion constellation. The Horsehead nebula lies at the edge of the L1630 molecular cloud and is illuminated by the O9.5V binary system $\sigma$ Orionis star \citep{Edwards:1976, Abergel:2003}. The illuminating star and the horsehead nebula are almost in the same plane perpendicular to our line of sight, presenting an almost edge-on view of the PDR. It is also a prototypical example of a low illumination PDR. In this work, we focus on the narrow filament at the edge of the PDR. We adopt a $G_{0}$ of 100 Habings field from \citet{Abergel:2003}. These authors estimated $G_{0}$ based on dilution assuming a distance of $\sim$ 3.5 pc between $\sigma$ Orionis and the PDR. We take $n_{\rm gas}=2\times 10^{5}$ cm$^{-3}$ estimated from the H$_{2}$ 1-0 S(1) line, low-J CO lines, and dust continuum emission by \citet{Habart:2005}. \citet{Habart:2011} estimated rotational temperatures for low-J pure rotational lines of H$_{2}$ ranging from 250 -- 400 K. However, a quick look at the model parameters in the PhotoDissociation Region Toolbox, an online toolbox to estimate physical conditions from observations \citep{Kaufman:2006, Pound:2008}, shows that at $G_{0}$ = 100 and $n_{\rm gas}$ = $2\times 10^{5}$, the $T_{\rm gas}$ is $\sim 100$ K. Therefore, in this work, we adopt a $T_{\rm gas}$ of 100 K.

\subsubsection{Orion Bar}
\label{subsubsec:Orion_bar}
The Orion Bar is a PDR situated in the Orion nebula at a distance of 414 $\pm$ 7 pc \citep{men07}. It is illuminated by four stars of type O-B in the trapezium cluster with $\theta^{1}$ Orionis C being the most massive and luminous star \citep{Kraus:2007}. It is one of the brightest and the most studied PDR. The Orion Bar lies at the edge of the \textrm{H~{\textsc{ii}}} region created by the star. For this work, we adopt a $G_{0}$ of $2.6 \times 10^{4}$ Habings field determined by \citet{Marconi:1998} using near-IR observations of an O\,I fluorescent line. The generally accepted picture of the Orion Bar entails dense clumps of density $10^{6}$ cm$^{-3}$ embedded in diffuse gas of density $5\times10^{4}$ cm$^{-3}$. These estimates for the density have been derived consistently by several authors using H$_{2}$, CO, HCN, and HCO$^{+}$ observations \citep{Parmar:1991, Tauber:1994, Young_Owl:2000, Bernard-Salas:2012}. \citet{Joblin:18} derived a gas density of $10^{5}$ cm$^{-3}$ at the PDR front by modelling the high-J CO lines and attributed the origin of this emission to small, high thermal pressure structures within the PDR rather than the interclump material. In this work, we adopt an average $n_{\rm gas}$ of $10^{5}$ cm$^{-3}$. For $T_{\rm gas}$, we use the value of 500 K estimated from pure rotational H$_{2}$ lines by \citet{Allers:2005}.   

\subsubsection{Diffuse ISM}
\label{subsubsec:diffuse_ISM}
In this work, we adopt the physical conditions characteristic of the Cold Neutral Medium. For $G_{0}$, we take a value of 1.7 in the units of Habing field \citep{Habing:1968, Draine:1978}. For $n_{\rm gas}$ and $T_{\rm gas}$, we take the values estimated by \citet{Wolfire:2003}. These authors estimated average $n_{\rm gas}$ and $T_{\rm gas}$ from detailed modelling of the gas heating rate and the gas-phase abundances of interstellar gas and found average $n_{\rm gas}$ of $\sim$ 30cm$^{-3}$, and $T_{\rm gas}$ of $\sim$ 80 K.

\subsection{Charge distribution in PDR environments}
\label{subsec:charge_distribution_PDRs}

\begin{figure*}
    \centering
    \begin{tabular}{cc}

    \includegraphics[scale=0.40]{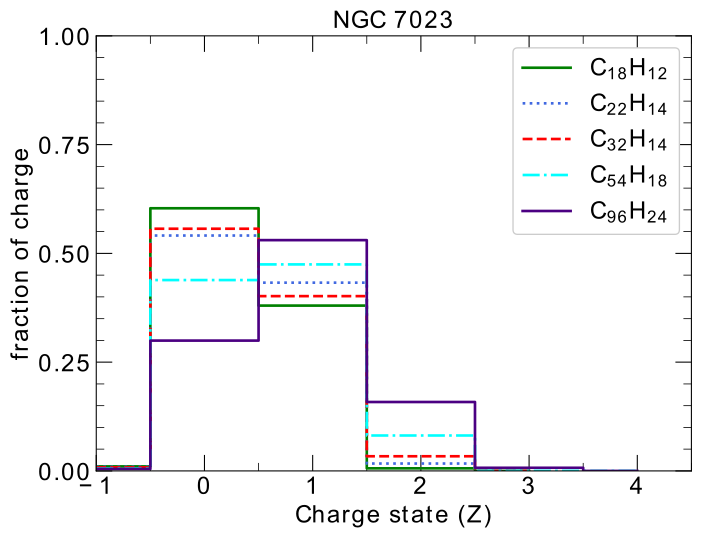} &  \includegraphics[scale=0.40]{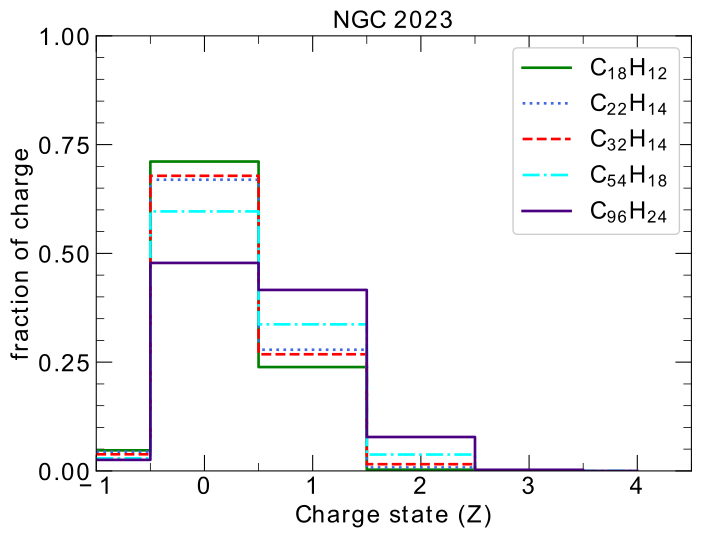}\\
    \includegraphics[scale=0.40]{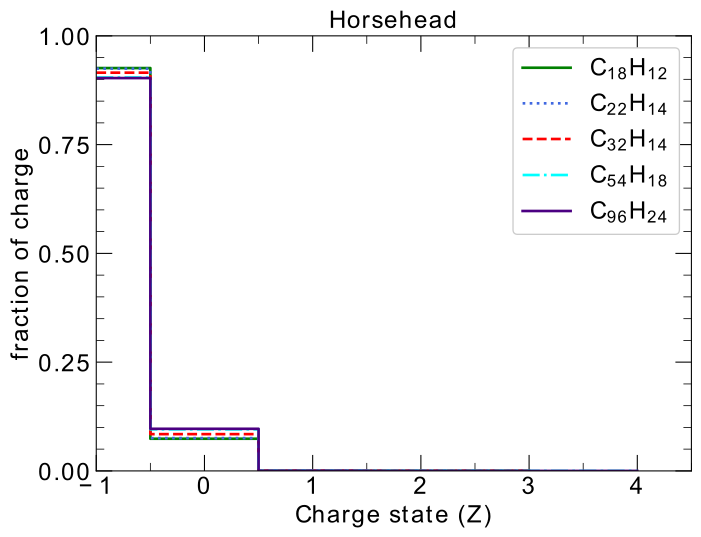} &
    \includegraphics[scale=0.40]{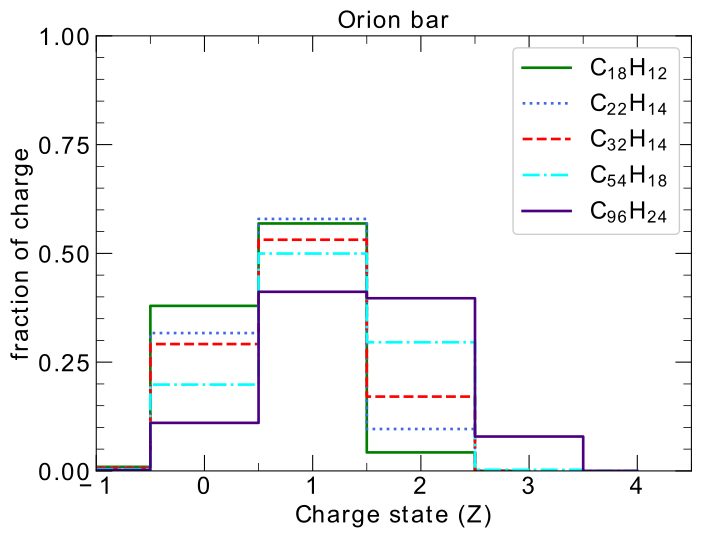} \\
    \includegraphics[scale=0.40]{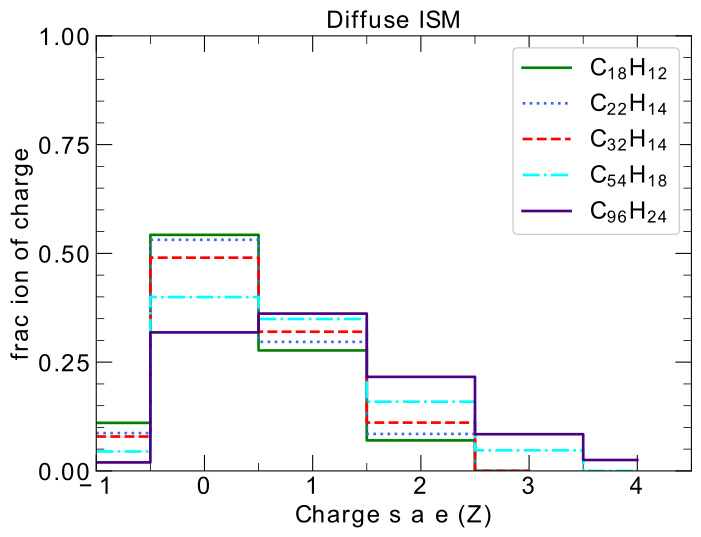} & \\
    \end{tabular}
    \caption{Charge distribution of the five PAHs considered in this work in NGC~7023, NGC~2023, the Horsehead nebula, the Orion Bar, and the diffuse ISM.}
    \label{fig:charge_PDRs}
    \end{figure*}
    
In Fig.~\ref{fig:charge_PDRs} we show the charge distribution predicted for the five PAH molecules, C$_{18}$H$_{12}$, C$_{22}$H$_{14}$, C$_{32}$H$_{14}$, C$_{54}$H$_{18}$, and C$_{96}$H$_{24}$, by the charge distribution model (see Section~\ref{subsec:Charge_dist_model}) in NGC~7023, NGC~2023, the Horsehead nebula, the Orion Bar, and the diffuse ISM. Here, we briefly summarize the model predictions for each of the environment. In NGC~7023, neutrals emerge as the dominant charge state for the small-sized PAHs (e.g., C$_{18}$H$_{12}$, C$_{22}$H$_{14}$, C$_{32}$H$_{14}$) and cations for the large-sized PAHs (e.g., C$_{96}$H$_{24}$). For the intermediate-sized PAHs (e.g., C$_{54}$H$_{18}$), both neutrals and cations have similar charge fractions of roughly 50\%. In NGC~2023, the neutrals emerge as the dominant charge state for all the  PAHs considered here, though the fraction of neutrals decreases as the size of the PAH molecule increases. The Horsehead nebula presents an interesting case where the dominant charge state is anions for all the PAHs with a minimal decrease in the fraction of anions with increasing size of the PAH molecule. In the Orion Bar, the dominant charge state is cations with singly charged cations as the dominant charge state in small and intermediate-sized PAHs and singly and doubly charged cations as the dominant charge states in large-sized PAHs. Finally, in the diffuse ISM, neutrals are the dominant charge state for small sized PAHs and cations for the large sized PAHs.

The charge distribution of PAHs in NGC~7023 has previously been modelled by \citet{Montillaud:2013} and \citet{Andrews:2016}. Our results are comparable with \citet{Andrews:2016} who also predicted a significant fraction of PAH cations in the NGC~7023 PDR. On the contrary, \citet{Montillaud:2013} predicts a higher fraction of PAH neutrals than the fraction of cations at the PDR front. This stark difference in the calculated charge fractions arises from the differences in the adopted values of the recombination rate coefficients, gas temperature, and electron density in the two studies. \citet{Montillaud:2013} used a recombination rate that was $\sim$50\% higher than the one adopted in the current work, resulting in a higher fraction of PAH neutrals than cations. 

\subsection{Validating the model}
\label{subsec:PAH_emission_PDR_environments}
\begin{table}
\caption{Average energy absorbed by each charge state of PAH molecules considered in this work in NGC~7023, NGC~2023, the Horsehead nebula, the Orion Bar, and the diffuse ISM.}
    \centering
    \begin{tabular}{l c c c c c}

        \hline
        \multicolumn{6}{c}{\textbf{
        NGC~7023}}\\
        \multirow{2}{*}{Molecule} & \multicolumn{5}{c}{$E_{\rm avg}$(Z) [eV]}\\
        & Z = -1 & Z = 0 & Z = 1 & Z = 2 & Z = 3 \\
        \hline
        Tetracene   & 4.30 & 5.27 & 5.17 & 4.73 & \\ 
        Pentacene  & 4.07 & 4.89 & 4.81 & 4.44 & \\
        Ovalene  & 4.70 & 5.36 & 5.49 & 5.35 & \\
        Circumcoronene   & 4.51 & 5.05 & 5.15 & 5.15 & 5.16  \\
        Circumcircumcoronene  & 4.68 & 5.01 & 5.12 & 5.12 & 5.16  \\
        \hline
        \multicolumn{6}{c}{\textbf{NGC~2023}}\\
        \multirow{2}{*}{Molecule} & \multicolumn{5}{c}{$E_{\rm avg}$(Z) [eV]}\\
        & Z = -1 & Z = 0 & Z = 1 & Z = 2 & Z = 3 \\
        \hline
        Tetracene   & 4.66 & 6.03 & 6.32 & 5.91 & \\ 
        Pentacene  & 4.45 & 5.66 & 5.95 & 5.70 & \\
        Ovalene  & 5.04 & 6.00 & 6.36 & 6.34 & \\
        Circumcoronene   & 4.85 & 5.61 & 5.91 & 6.09 & 6.09  \\
        Circumcircumcoronene  & 5.10 & 5.55 & 5.85 & 5.98 & 6.09  \\
        \hline
        \multicolumn{6}{c}{\textbf{Horsehead nebula}}\\
        \multirow{2}{*}{Molecule} & \multicolumn{5}{c}{$E_{\rm avg}$(Z) [eV]}\\
        & Z = -1 & Z = 0 & Z = 1 & Z = 2 & Z = 3 \\
        \hline
        Tetracene   & 5.02 & 6.97 & 7.70 & 7.44 & \\ 
        Pentacene  & 4.84 & 6.64 & 7.37 & 7.35 & \\
        Ovalene  & 5.32 & 6.78 & 7.40 & 7.59 & \\
        Circumcoronene   & 5.14 & 6.32 & 6.88 & 7.30 & 7.31  \\
        Circumcircumcoronene  & 5.51 & 6.20 & 6.76 & 7.10 & 7.31  \\
        \hline
        \multicolumn{6}{c}{\textbf{Orion Bar}}\\
        \multirow{2}{*}{Molecule} & \multicolumn{5}{c}{$E_{\rm avg}$(Z) [eV]}\\
        & Z = -1 & Z = 0 & Z = 1 & Z = 2 & Z = 3 \\
        \hline
        Tetracene   & 5.16 & 7.39 & 8.29 & 8.11 & \\ 
        Pentacene  & 5.01 & 7.09 & 7.99 & 8.07 & \\
        Ovalene  & 5.43 & 7.15 & 7.88 & 8.16 & \\
        Circumcoronene   & 5.25 & 6.06 & 7.34 & 7.87 & 7.88 \\
        Circumcircumcoronene  & 5.69 & 6.53 & 7.21 & 7.63 & 7.88 \\
        \hline
        
         \multicolumn{6}{c}{\textbf{diffuse ISM}}\\
        \multirow{2}{*}{Molecule} & \multicolumn{5}{c}{$E_{\rm avg}$(Z) [eV]}\\
        & Z = -1 & Z = 0 & Z = 1 & Z = 2 & Z = 3 \\
        \hline
        Tetracene   & 5.07 & 6.60 & 6.98 & 6.63 & \\ 
        Pentacene  & 4.90 & 6.30 & 6.71 & 6.50 & \\
        Ovalene  & 5.36 & 6.44 & 6.84 & 6.80 & \\
        Circumcoronene   & 5.19 & 6.05 & 6.40 & 6.57 & 6.57  \\
        Circumcircumcoronene  & 5.51 & 5.97 & 6.33 & 6.47 & 6.57  \\
        \hline
    \end{tabular}

    \label{tab:average_energy}
\end{table}

\begin{figure*}
    \centering
    \begin{tabular}{cc}

    \includegraphics[scale=0.40]{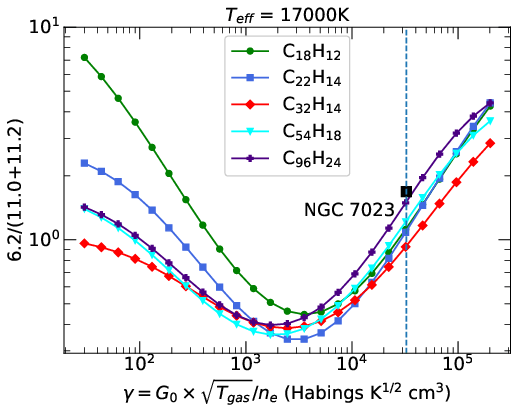} &  \includegraphics[scale=0.40]{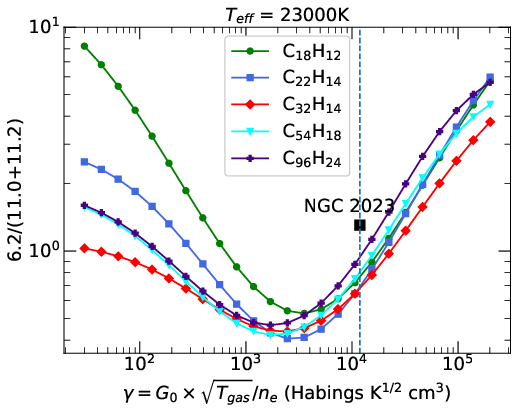}\\
    \includegraphics[scale=0.40]{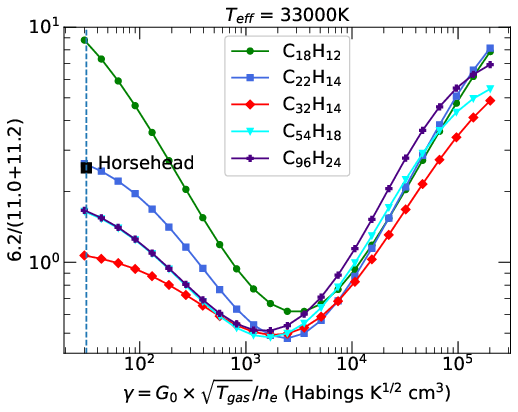} &
    \includegraphics[scale=0.40]{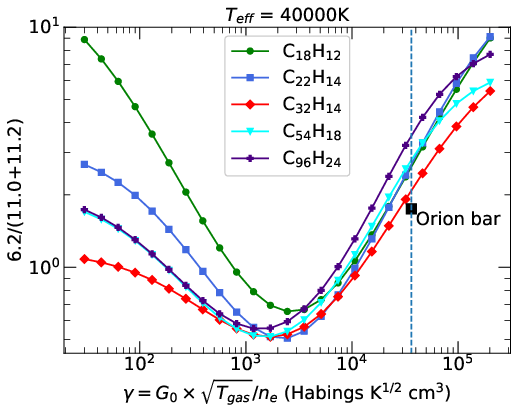} \\
    \includegraphics[scale=0.40]{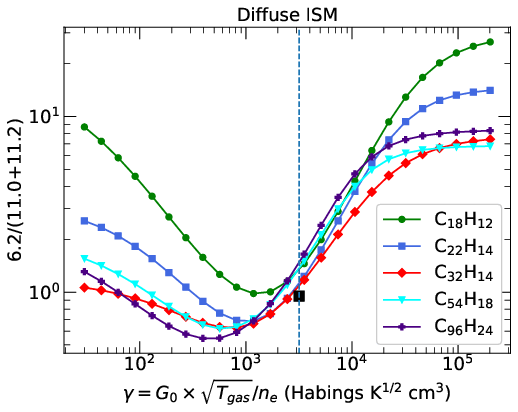} &
    \\
    \end{tabular}
    \caption{The 6.2/(11.0+11.2) band ratio as a function of the ionization parameter $\gamma$. Individual panels correspond to the 6.2/(11.0+11.2) ratio determined from the PAH spectra calculated at excitation conditions characteristic for NGC~7023 ($G_{0}=2600$ and $T_{\rm eff}=17000$K), NGC~2023 ($G_{0}=17000$ and $T_{\rm eff}=23000$K), the Horsehead nebula ($G_{0}=100$ and $T_{\rm eff}=33000$K), the Orion Bar  ($G_{0}=26000$ and $T_{\rm eff}=40000$K), and the diffuse ISM. See text for details about the method of calculation of the 6.2/(11.0+11.2) band ratio. The observational data points are also shown in the corresponding panels (black square). The dashed line shows the adopted $\gamma$ value for the environments considered here.}
    \label{fig:ratio_62_112_PDRs}
    \end{figure*}    
    
\begin{figure*}
    \centering
    \begin{tabular}{cc}

    \includegraphics[scale=0.40]{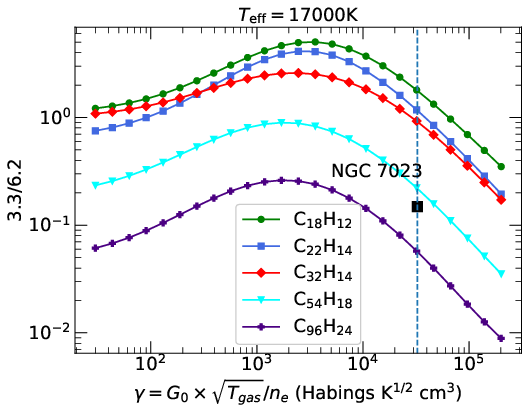} &  \includegraphics[scale=0.40]{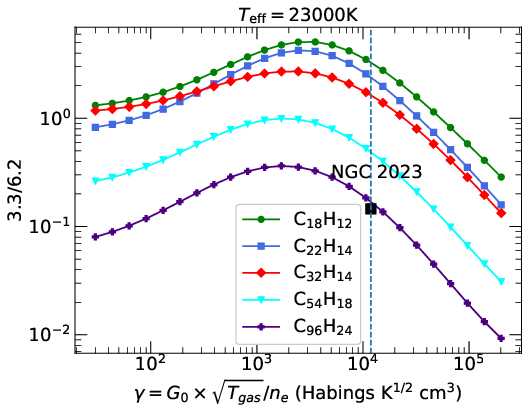}\\
    \includegraphics[scale=0.40]{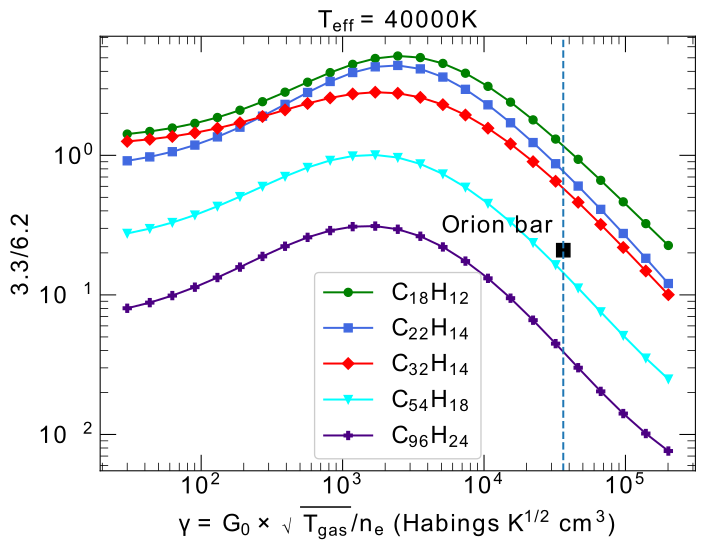} & \includegraphics[scale=0.40]{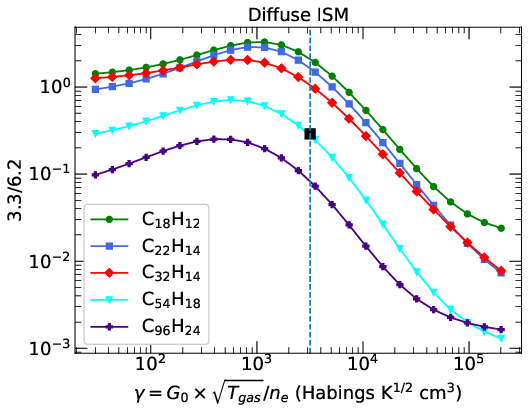}\\
    
    \end{tabular}
    \caption{The 3.3/6.2 band ratio as a function of the ionization parameter $\gamma$. Individual panels correspond to the 3.3/6.2 ratio determined from the PAH spectra calculated at excitation conditions characteristic for NGC~7023 ($G_{0}=2600$ and $T_{\rm eff}=17000$K), NGC~2023 ($G_{0}=17000$ and $T_{\rm eff}=23000$K), the Orion Bar  ($G_{0}=26000$ and $T_{\rm eff}=40000$K), and the diffuse ISM. See text for details about the method of calculation of the 3.3/6.2 band ratio. The observational data points are also shown in the corresponding panels (black square). The dashed line shows the adopted $\gamma$ value for the environments considered here.}
    \label{fig:ratio_33_62_PDRs}
    \end{figure*}

In order to test the PAH emission model, we compare the relative intensities of the PAH features, in particular the 6.2/(11.0+11.2) and 3.3/6.2 band ratios, predicted by the model with the observations of the PDRs and the diffuse ISM. In the PDRs, we take the Spitzer Infrared Spectrograph (IRS) Short Low (SL) observations for the 6.2/(11.0+11.2) band ratio, and ISO - Short Wavelength Spectrometer (SWS) observations for the 3.3/6.2 band ratio. For the 6.2/(11.0+11.2) band ratios, we take the flux measurements at the NW PDR of NGC~7023 from \citet{Stock:16}, at the South ridge in NGC~2023 from \citet{Peeters:2017}, at the Horsehead PDR from the Spitzer archive (published previously by \citet{Ochsendorf:2015}), and the Orion Bar from \citet{Knight:2021:a}. For the 3.3/6.2 band ratio, we take the flux measurements in all the environments from \citet{Peeters:2002, van:2004}. We note that the ISO-SWS apertures are larger than the Spitzer IRS SL apertures and are not centered on the same positions, therefore the observations of 6.2/(11.0+11.2) and 3.3/6.2 are not necessarily at the same location. Moreover, because of the lack of the ISO-SWS observations in the Horsehead nebula, we do not calculate the 3.3/6.2 band ratios from the model for the excitation conditions corresponding to the Horesehead nebula. In the diffuse ISM, we take the observations for the 6.2/(11.0+11.2) and 3.3/6.2 from AROME observations from \citet{Giard:1994} and ISO camera (ISOCAM) from \citet{Flagey:2006}.

To determine the 6.2/(11.0+11.2) and 3.3/6.2 band ratios from the model, we compute the spectra for each charge state of a PAH molecule at the average energy set by $T_{\rm eff}$ and the PAH characteristics (see equation~\ref{eq:avg_energy}). In Table~\ref{tab:average_energy}, we present the average energy absorbed by a specific PAH molecule in each charge state in the environments considered here. To validate our calculation of $E_{avg}(Z)$, we compare our calculated values for pentacene and ovalene with the values presented in Table~3 of \citet{Mulas:2006} (see Appendix~\ref{sec:validating_abs_energy_calc} for details) and conclude that the $E_{avg}(Z)$ values agree to better than 1\%. Except for the diffuse ISM, where a single photon absorption suffices, we used three photon absorptions to calculate the spectra. From the calculated spectra, we first determine the strengths of the 3.3, 6.2, and 11.0+11.2 $\mu$m bands in each charge state of a molecule following the procedure described in section~\ref{subsec:PAH_spectra_5_molecules} and weight the strengths with the fraction of charge states before adding the band strengths in all the charge states to determine the 3.3, 6.2, and 11.0+11.2 $\mu$m bands for a PAH molecule. Since the charge distribution is a sensitive function of $\gamma$, whose value for the PDRs is not strictly constrained (see discussion in section~\ref{subsec_environments}), we calculate the 3.3/6.2 and 6.2/(11.0+11.2) values as a function of $\gamma$.

Fig.~\ref{fig:ratio_62_112_PDRs} shows the predicted values of the 6.2/(11.0+11.2) as a function of $\gamma$ for the excitation conditions characteristic for NGC~2023, NGC~7023, the Horsehead nebula, the Orion Bar and the diffuse ISM. The corresponding observed values of the 6.2/(11.0+11.2) band ratio are also shown in the figure. For NGC~7023 and NGC~2023, the predicted value of 6.2/(11.0+11.2) ratio for C$_{96}$H$_{24}$ compares well with the observational value, for the Horsehead nebula, the predicted value for C$_{22}$H$_{14}$ compares well the observational value. For the Orion Bar and the diffuse ISM, the predicted value for C$_{32}$H$_{14}$ compares well with the observational value. We emphasize that since the $\gamma$ of these sources is not strictly constrained, the observational data point can shift in the 6.2/(11.0+11.2) vs $\gamma$ plane and, as a consequence, the size of the molecule for which there will be a good match between the model prediction and the observation may change. The 6.2/(11.0+11.2) vs $\gamma$ plots shown here do not predict the size of the PAH molecule but rather shows that the PAH emission model based on the charge distribution can explain the observed PAH emission. Moreover, it shows that compact PAHs, which are considered potential candidates for grandPAHs - a set of few stable PAHs comprising the astronomical PAH population \citep{Andrews:2015} - can effectively produce the observed 6.2/(11.0+11.2) ratios in NGC~7023, NGC~2023, the Orion Bar, and the diffuse ISM. In contrast, in the Horsehead nebula, an acene shows a good comparison with the observed value. Thus, while the grandPAH hypothesis remains plausible in high illumination PDRs like NGC~7023, NGC~2023, and the Orion Bar, it is not the case in low illumination PDRs like the Horsehead nebula. We emphasize that a good match between the predicted 6.2/(11.0+11.2) for C$_{22}$H$_{14}$ and the observed value in the Horsehead nebula does not necessarily mean that acenes are present in the PDR environment of Horsehead nebula. It merely shows that either C$_{22}$H$_{14}$ or a molecule behaving like C$_{22}$H$_{14}$ (e.g., molecules with a high ratio for the intrinsic strength of the 6.2/(11.0+11.2) bands in anions) is responsible for the observed emission in the Horsehead nebula. Finally, we reiterate that a high value of 6.2/(11.0+11.2) does not always indicate the presence of a large fraction of cations in the astronomical environment. As shown in Fig.~\ref{fig:ratio_62_112_PDRs}, while the high 6.2/(11.0+11.2) ratio in the Orion Bar is a result of the large fraction of cations, the high 6.2/(11.0+11.2) ratio in the Horsehead nebula is a result of the large fraction of anions.

Fig.~\ref{fig:ratio_33_62_PDRs} shows the plots of predicted and observed values of 3.3/6.2 as a function of $\gamma$ for NGC~7023, NGC~2023, the Orion Bar, and the diffuse ISM. The observed values of NGC~7023 compare well with predicted values for the compact PAHs C$_{54}$H$_{18}$ and C$_{96}$H$_{24}$. For the NGC~2023, the observed value shows a good match with C$_{96}$H$_{24}$, for the Orion bar and the diffuse ISM, it matches with C$_{54}$H$_{18}$. These results demonstrate that the PAH emission model based on the charge distribution presented here can effectively explain the observed PAH emission. However, as discussed above, a good match between the observed value and model prediction does not predict the precise size of the molecule as the observed value can shift in the plane of the 3.3/6.2 vs $\gamma$ due to uncertainty in $\gamma$ in the PDRs and the uncertainty in the electron recombination and attachment cross-sections (c.f., section~\ref{subsec:Caveat}).

\subsection{Revisiting the 6.2/(11.0+11.2) vs 3.3/(11.0+11.2) as a diagnostic tool}
\label{subsec:grid_plots}
\begin{figure*}
    \centering
    \begin{tabular}{cc}
    \includegraphics[scale=0.44]{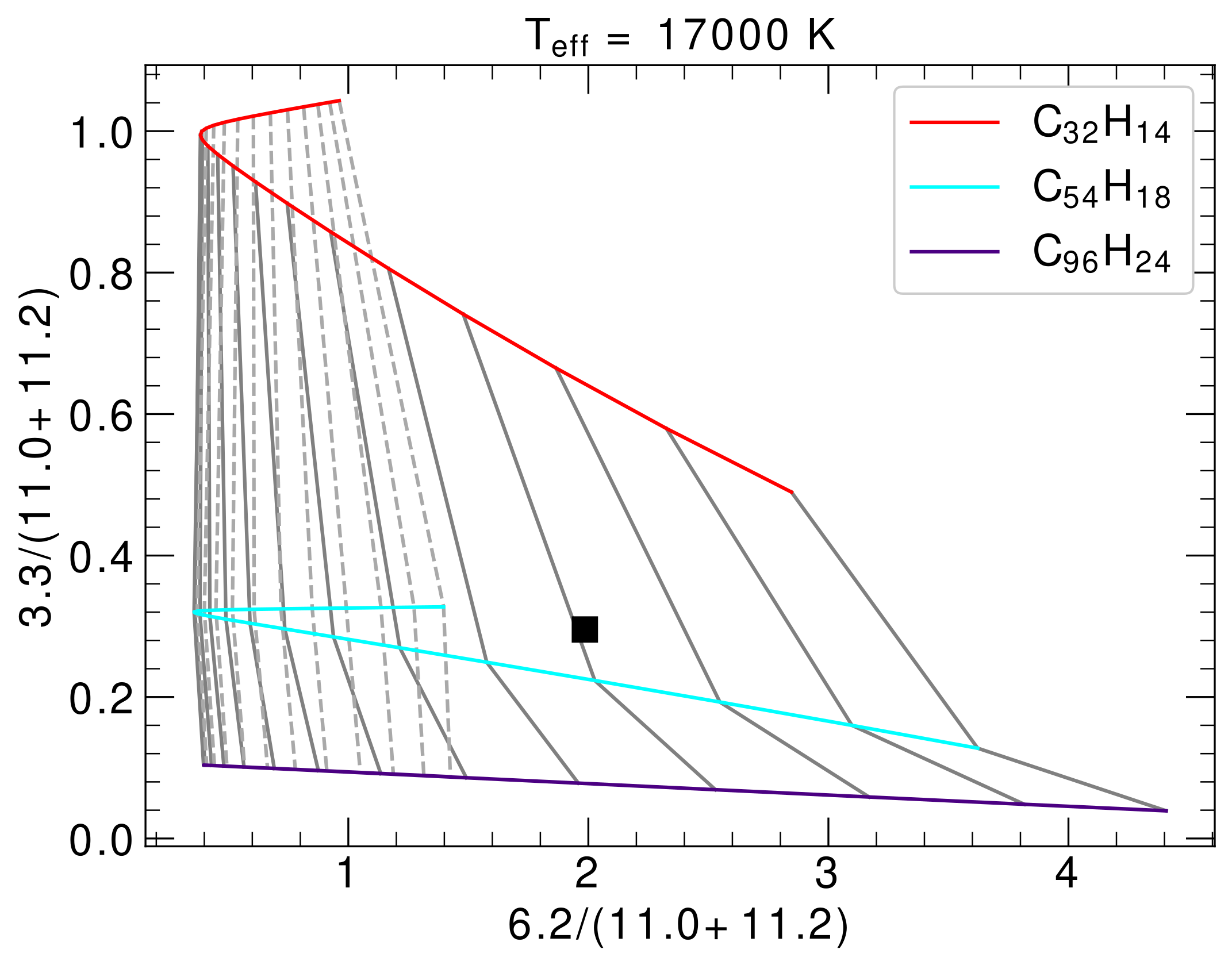} &
    \includegraphics[scale=0.44]{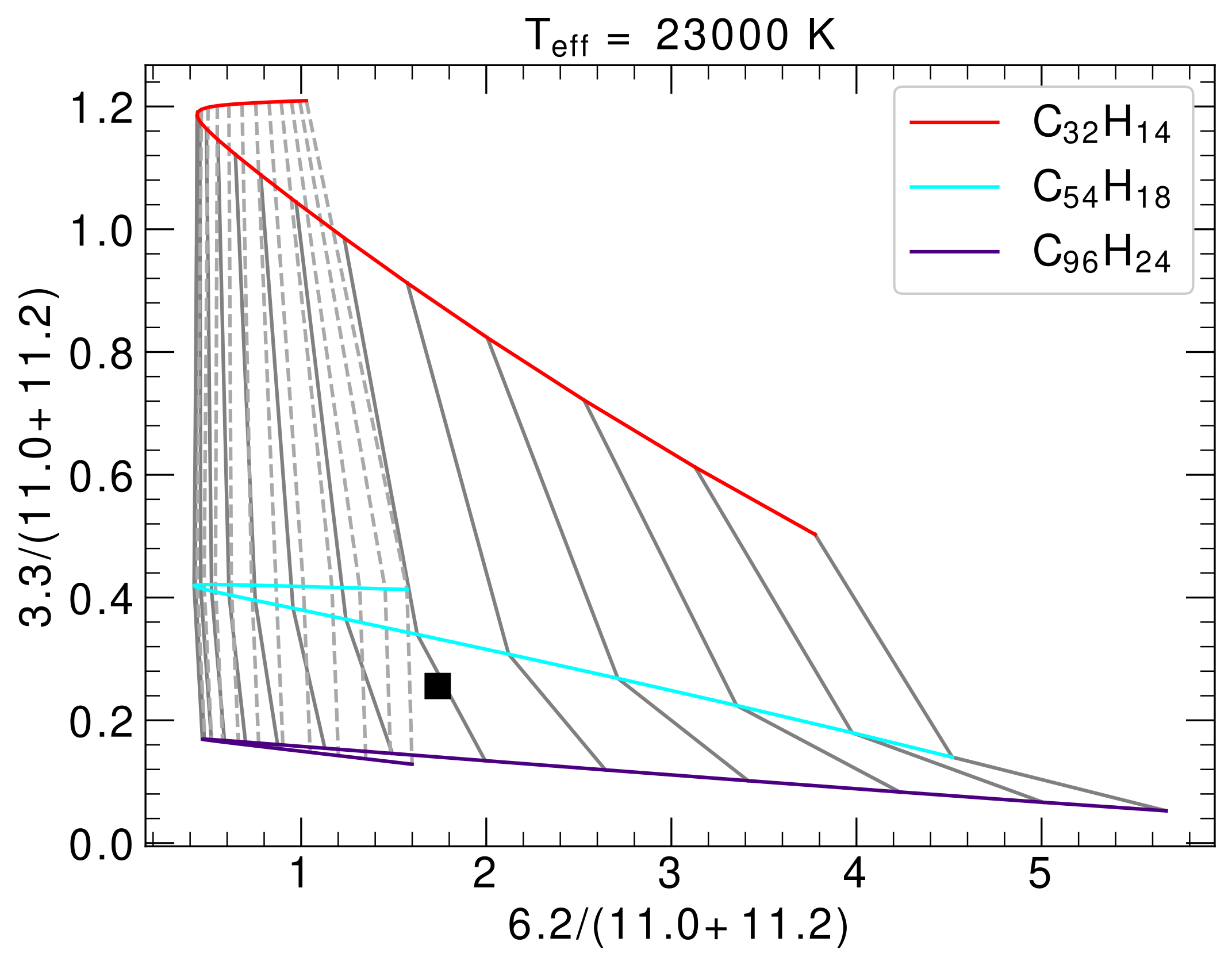} \\
      \includegraphics[scale=0.44]{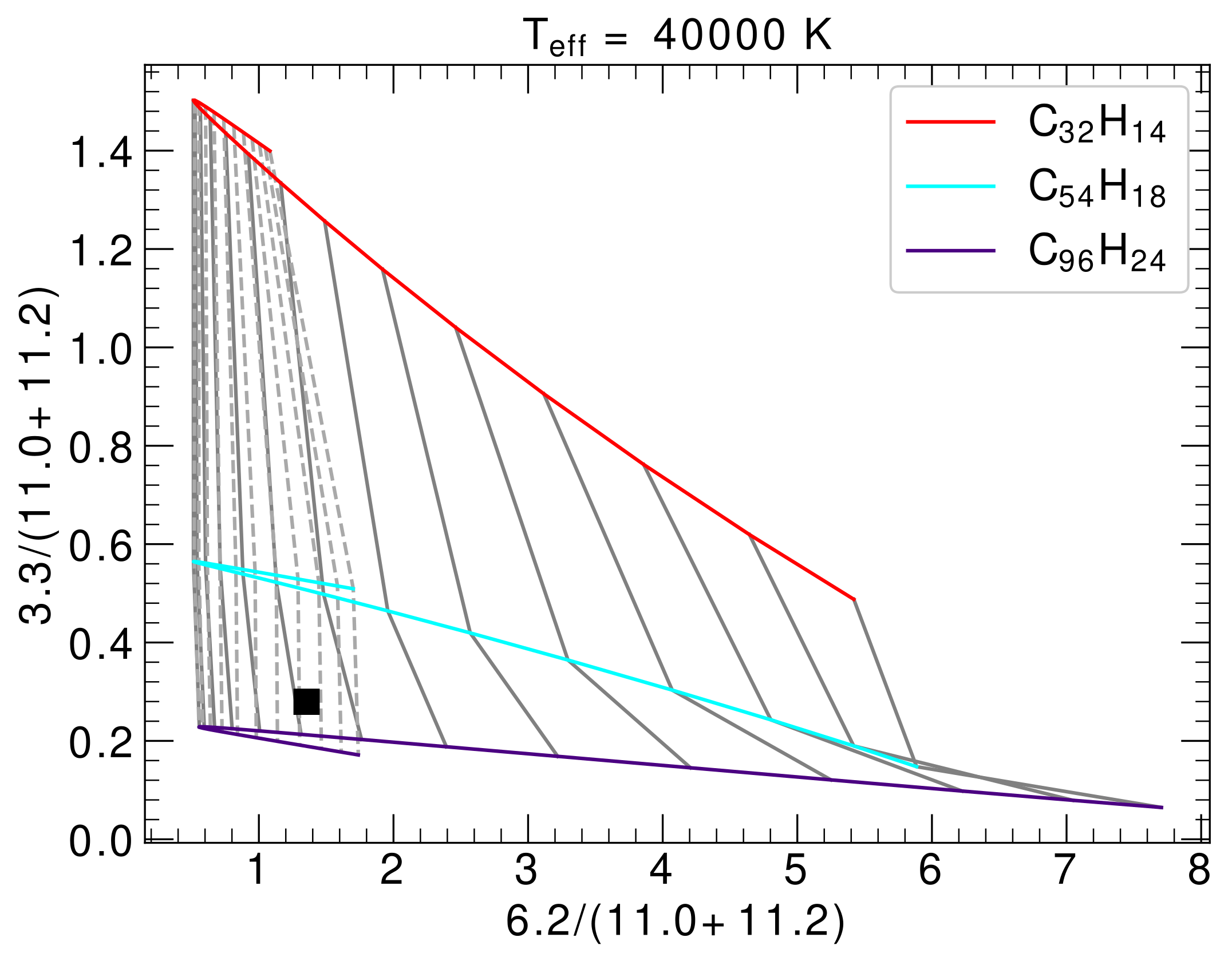} &
    \includegraphics[scale=0.44]{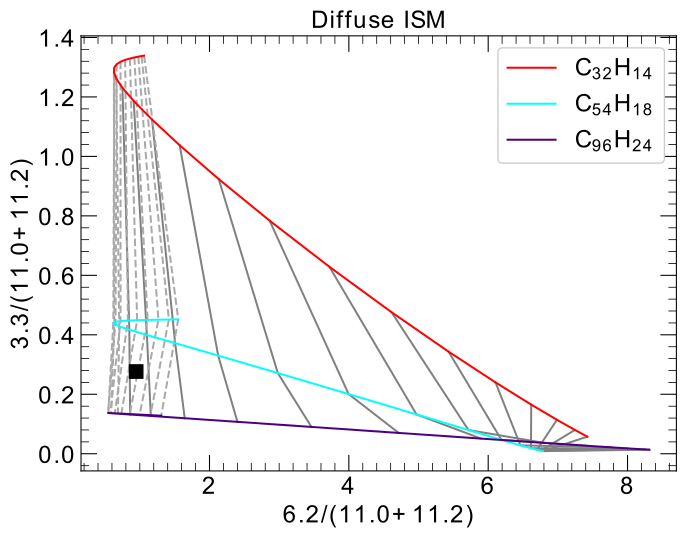} \\
    \end{tabular}
    \caption{The 6.2/(11.0+11.2) vs 3.3/(11.0+11.2) plots of compact PAHs considered in this work over $\gamma$ values ranging from 30 - 2$\times 10^{5}$. The points of constant $\gamma$ values for each PAH molecule are joined together. In each panel, the high $\gamma$ values are at the right end of the plot and decrease towards the left. Each panel corresponds to the ratios determined from the PAH spectra calculated for different excitation conditions. $T_{\rm eff}=17000$K corresponds to conditions characteristic of NGC~7023, $T_{\rm eff}=23000$K to NGC~2023, and $T_{\rm eff}=40000$K to the Orion bar. The corresponding observational data point is also shown in each panel.   }
    \label{fig:diagnostc_tool_entire_gamma_range}
\end{figure*}

\begin{figure*}
    \centering
    \begin{tabular}{cc}
    \includegraphics[scale=0.44]{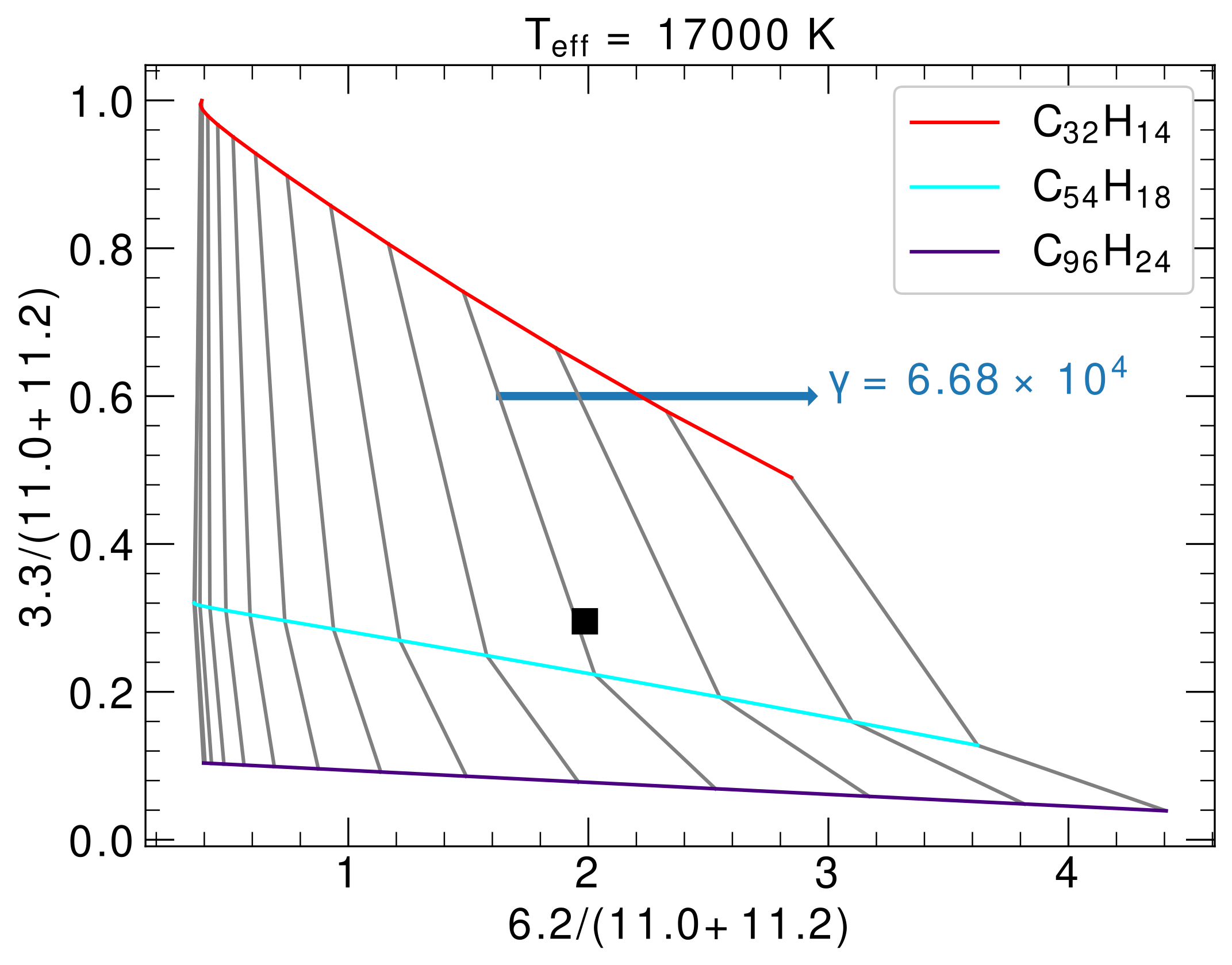} &
    \includegraphics[scale=0.44]{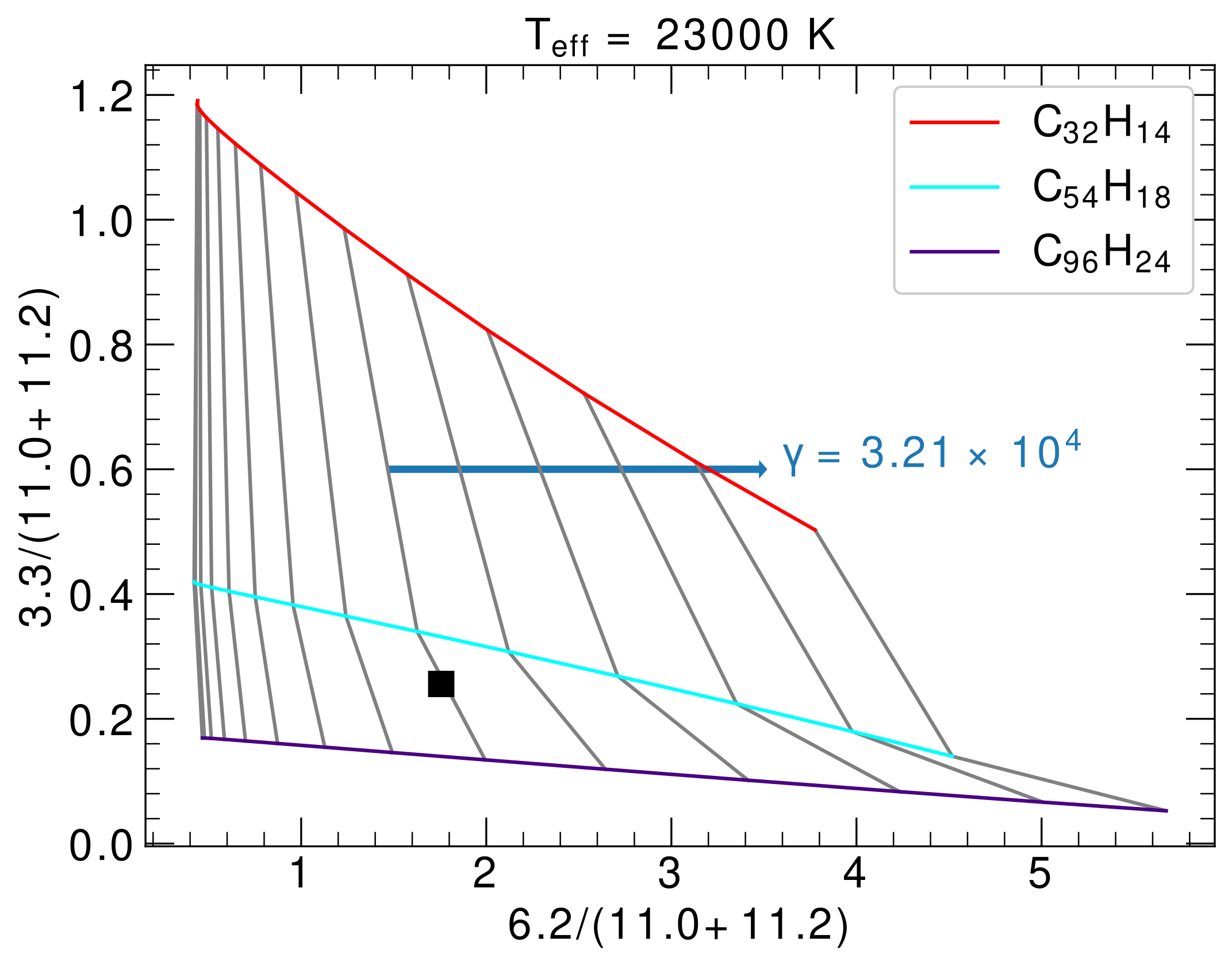} \\
      \includegraphics[scale=0.44]{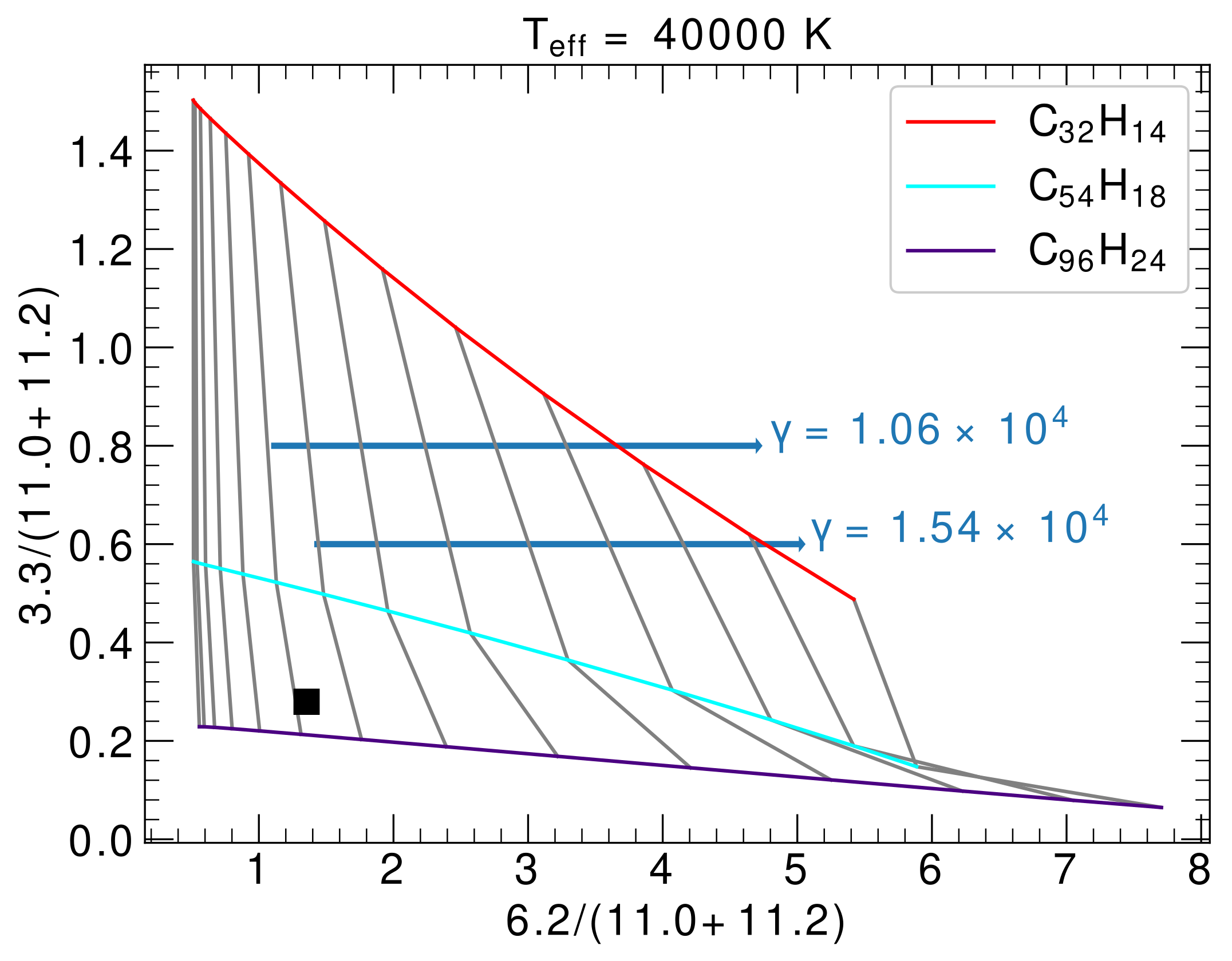} &
    \includegraphics[scale=0.44]{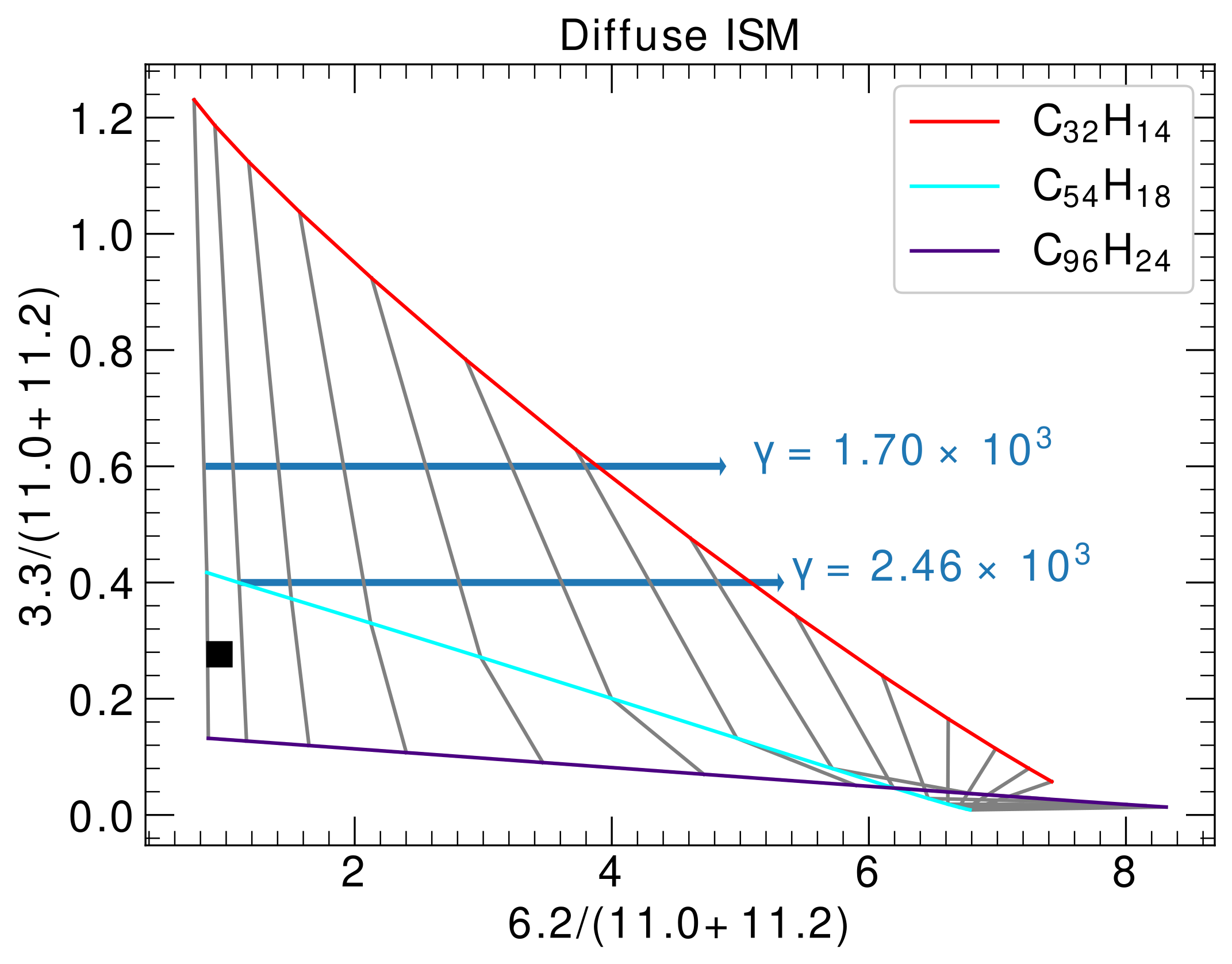} \\
    \end{tabular}
    \caption{A section of the 6.2/(11.0+11.2) vs 3.3/(11.0+11.2) plots shown in Fig.~\ref{fig:diagnostc_tool_entire_gamma_range} demonstrating that the 6.2/(11.0+11.2) vs 3.3/(11.0+11.2) can be used as a tool to determine the precise $\gamma$ value and PAH size in astrophysical environments if we have a rough estimation of the range of $\gamma$ values. }
    \label{fig:diagnostc_tool_half_gamma_range}
\end{figure*}

The 6.2/11.2 and 3.3/11.2 band ratios are often employed to determine the charge and the size of the emitting PAH population in astrophysical environments \citep[e.g.][]{Draine:2001, Croiset:2016, Boersma:16, Maragkoudakis:2020, Knight:2021}. Furthermore, the 6.2/11.2 band ratio has been related to the ionization parameter $\gamma$ \citep[e.g.][]{Galliano:2008, Fleming:2010, Boersma:16, Stock:16}. However, the results of the PAH emission model show that the interpretation of the 6.2/(11.0+11.2) ratio is ambiguous in the low and high regimes of $\gamma$. Moreover, we find that for a PAH molecule, two values for the 3.3/(11.0+11.2) may correspond to one 6.2/(11.0+11.2) value (section~\ref{subsubsec:62_112_vs_33_112}); one for high $\gamma$ values and the other for low $\gamma$ values. This adds a layer of complexity to the interpretation of PAH size from observations of the 3.3/11.2 ratio in astronomical environments. Given these results, we revisit the diagnostic potential of the 6.2/(11.0+11.2) vs 3.3/(11.0+11.2) ratios and investigate whether these are reliable indicators of $\gamma$ and the average size of the astronomical PAH population.

We analyze the 6.2/(11.0+11.2) vs 3.3/(11.0+11.2) plots for the excitation conditions typical of NGC~7023, NGC~2023, the Orion bar, and the diffuse ISM (see Fig.~\ref{fig:diagnostc_tool_entire_gamma_range}). Essentially we plot the 6.2/(11.0+11.2) vs 3.3/(11.0+11.2) curves for compact PAHs in the same plane and join the curves for different PAH molecules corresponding to the same $\gamma$ values forming a grid. We note that the grid corresponding to each excitation temperature overlaps for a certain range of $\gamma$ values.  We emphasize that this overlapping results from the fact that two values of 3.3/(11.0+11.2) corresponds to a single value of 6.2/(11.0+11.2).

To understand the diagnostic potential of 6.2/(11.0+11.2) vs 3.3/(11.0+11.2), we investigate Fig.~\ref{fig:diagnostc_tool_entire_gamma_range} by considering two scenarios. In one scenario, we assume no prior knowledge of the $\gamma$ value of the astrophysical region, whereas, in the other scenario, we assume a rough estimate for the range of $\gamma$. In the first scenario, where we have no idea about the $\gamma$ value, it becomes unfeasible to use these plots as indicators of $\gamma$. For NGC~7023 and NGC~2023, the observed data point is in the region of the grid where there is no overlap, implying a unique solution for $\gamma$ and PAH size. In contrast, for the Orion bar and the diffuse ISM, the observed data point is in the overlapping region of the grid, implying a unique solution for PAH size but not for $\gamma$. As a result, in the first scenario, the 3.3/(11.0+11.2) ratio can be used to predict PAH size, but the 6.2/(11.0+11.2) ratio cannot be used to predict $\gamma$. However, in the second scenario, where we assume prior knowledge about the range of $\gamma$ values in an astrophysical environment, these diagnostic plots have enormous potential because now we are essentially sampling only a portion of the 6.2/(11.0+11.2) vs 3.3/(11.0+11.2) grid, avoiding the grid-overlap problem (see Fig.~\ref{fig:diagnostc_tool_half_gamma_range}). For example, assuming that the $\gamma$ values for NGC~7023, NGC~2023, and the Orion bar range from $10^{3}$--$10^{5}$, we show the part of the 6.2/(11.0+11.2) vs 3.3/(11.0+11.2) plot from Fig.~\ref{fig:diagnostc_tool_entire_gamma_range} corresponding to this range in Fig.~\ref{fig:diagnostc_tool_half_gamma_range}. From  Fig.~\ref{fig:diagnostc_tool_half_gamma_range} we can now infer that in NGC~7023 the size of the emitting PAH population is $\sim$ 50 N$_{C}$, and the $\gamma$ value is $\sim$7$\times 10^{4}$, in NGC~2023, size is $\sim$ 70 N$_{C}$ and $\gamma$ is $\sim 3 \times 10^{4}$, in the Orion Bar, the size is $\sim$ 80 N$_{C}$ and $\gamma$ is $\sim 1.5 \times 10^{4}$, and in the diffuse ISM, the size is $\sim$ 70 N$_{C}$ and $\gamma$ is $\sim 2 \times 10^{3}$. We note that the inferred $\gamma$ values differ from the ones we adopt in this work (see Table~\ref{tab:physical_conditions_PDRs}) by about a factor of 2. This difference could be due to uncertainty in the $\gamma$ values we adopt (see section~\ref{subsec_environments}) or due to uncertainty in the electron recombination rates that we discuss in section~\ref{subsec:Caveat}. The key point of this discussion, however, is not to infer the precise values of PAH size and $\gamma$ values but to illustrate that the 6.2/(11.0+11.2) vs 3.3/(11.0+11.2) grids are only useful if we know an approximate range of $\gamma$ values for the astrophysical environment considered. If we have no prior knowledge of $\gamma$, only the 3.3/(11.0+11.2) can be used to predict the PAH size but the 6.2/(11.0+11.2) cannot predict the $\gamma$ values.

\subsection{Caveat in the model}
\label{subsec:Caveat}

In this section, we highlight some of the caveats in the PAH model presented in this paper. Firstly, we acknowledge that the charge distribution based PAH emission model is not a complete model as it does not account for other changes in the molecular properties of PAHs, such as changes in the size distribution, hydrogenation, or molecular structure. While changes in the PAH charge state are the primary drivers of observed PAH variation \citep[e.g.][]{Joblin:1996, Galliano:2008, Rosenberg:11, Sidhu:2021}, a complete PAH emission model must also account for other changes in the molecular properties of PAHs in order to simulate the astronomical PAH observations fully. Secondly, to model the PAH emission, we have approximated the radiation fields of the astrophysical environments using black bodies with effective temperatures corresponding to the illuminating star in those environments. Such an approximation results in an uncertainty in the calculated photo-ionization rate and the average energy absorbed by a PAH molecule for which we calculate the IR emission spectrum. As excitation occurs through a broad range of energies, any changes introduced by this will be small and we elected to neglect this but a full model could account for this. Thirdly, due to uncertainty in the measurements of the physical conditions in astrophysical environments, there is an uncertainty in the $\gamma$ values that we adopt in this work which then influences the comparison of the predicted values of the 6.2/(11.0+11.2) and 3.3/6.2 with the observed values (see section~\ref{subsec:PAH_emission_PDR_environments}). Lastly, we calculate the band ratios using a model based on the PAH charge distribution, which is determined by the ratio of the photo-ionization and electron recombination/attachment rates.
In our model, we use analytical relations involving PAH characteristics to calculate these rates. While the photo-ionization rate calculation involves PAH characteristics, the ionization yield, and absorption cross-sections that have been measured experimentally or calculated theoretically for astrophysically relevant PAHs, the electron recombination and attachment rates calculations involve no such experimentally measured PAH characteristics (see section~\ref{subsec:Charge_dist_model}). This lack of laboratory data on electron recombination and attachment rates results in uncertainty in the charge distribution calculations, which then propagates into calculations of PAH band ratios. Therefore, laboratory experiments to determine electron recombination and attachment rates of astrophysically relevant PAHs are highly desirable.

\section{Conclusions}
\label{sec:summary}
We present a PAH emission model that calculates the IR emission from PAHs in PDRs taking into account the PAH charge distribution. Following \citet{Bakes:01:a}, the model first calculates the charge distribution of PAHs in the PDRs based on the physical conditions, $G_{0}$, $n_{\rm gas}$, $T_{\rm gas}$, $X_{e}$. The model then computes the IR emission from a specific PAH molecule by adding the emissions from all relevant charge states and weighing them according to their charge distribution. We model the IR emission from five different PAH molecules, tetracene ($\text{C}_{18}\text{H}_{12}$), pentacene ($\text{C}_{22}\text{H}_{14}$), ovalene ($\text{C}_{32}\text{H}_{14}$), circumcoronene ($\text{C}_{54}\text{H}_{18}$), and circumcircumcoronene ($\text{C}_{96}\text{H}_{24}$), adopting their recent experimentally measured or quantum chemically calculated data on PAH characteristics. The selected PAHs span a wide range in physical and chemical properties and are therefore a suitable set of species for the analysis of observations.

We show that anions are the dominant charge state for $\gamma < 2 \times 10^{2}$, neutrals for $10^{3} < \gamma < 10^{4}$, and cations for $\gamma > 5 \times 10^{5}$. Based on the analysis of the PAH spectra of the molecules considered in this work in all the charge states, we show that anionic and cationic charge states exhibit similar spectral characteristics with strong features in the 6--9 $\mu$m region and weak features in the 10--15 $\mu$m region. We investigated the implications of this similarity between cationic and anionic charge states on the 6.2/(11.0+11.2) band ratio and discovered that the large contribution from either anions and cations can result in high values of the 6.2/(11.0+11.2) band ratio. We further found that the charge state of PAHs also influences the 3.3/(11.0+11.2) band ratio beyond the well-known dependence on the PAH size.

We also model the PAH emission in five astrophysical environments (NGC~7023, NGC~2023, the Horsehead nebula, the Orion Bar, and the diffuse ISM) for the five PAH molecules considered and show that changes in the charge distribution can account for the observed variations in the IR emission in these environments. We further find that anions, which have previously been overlooked, are the dominant charge carriers in low illumination PDRs such as the Horsehead nebula. In light of the similarity between the spectral characteristics of cations and anions, we revisited the diagnostic potential of the 6.2/(11.0+11.2) vs 3.3/(11.0+11.2) plots to determine the $\gamma$ of astrophysical environments and PAH size, respectively. We find that even in the absence of prior knowledge about an approximate value of $\gamma$, the 3.3/(11.0+11.2) can be used to infer PAH size, but the 6.2/(11.0+11.2) cannot be used to infer $\gamma$. Comparing the model predictions of PAH emission with observations highlights the need for more experiments to determine the electron recombination and attachment cross-sections of astrophysically relevant PAHs as the lack of this experimental data results in an uncertainty in the charge distribution calculation which then propagates into PAH emission calculation.

\section*{Acknowledgements}
The authors thank the referee for providing valuable comments which led to the improvement of this paper. AS acknowledges support from the Mitacs Globalink Research Award. EP and JC acknowledge support from an NSERC Discovery Grant.

\section*{Data Availability}
The data underlying this article will be shared on reasonable request to the corresponding author.
%%%%%%%%%%%%%%%%%%%%%%%%%%%%%%%%%%%%%%%%%%%%%%%%%%



\bibliographystyle{apj}
\bibliography{main}

%%%%%%%%%%%%%%%%%%%%%%%%%%%%%%%%%%%%%%%%%%%%%%%%%%

%%%%%%%%%%%%%%%%% APPENDICES %%%%%%%%%%%%%%%%%%%%%
\appendix
\appendixpage
\addappheadtotoc

\begin{appendices}
\section{Comparison between different formalisms to calculate the ionization potential}
\label{sec:IPs_comparison}
In order to access the quantitative impact of the choice of the \citet{Bakes:1994} formalism used to calculate IPs, we compared our calculated IPs with an another formalism presented in \citet{Wenzel:2020}. We present the results of the comparison in Table~\ref{tab:IP_comparison}. In all cases, the differences are small and given that the species interact with a broad range of UV photon energies, ionization and excitation rates are not very sensitive to these differences. 
\begin{table*}
\caption{Comparison between the IPs calculated using the \citet{Bakes:1994} and \citet{Wenzel:2020} formalisms for the PAH molecules for which we employed the \citet{Bakes:1994} formalism.}
\label{tab:IP_comparison}
    \centering
    
    \begin{tabular}{c c c c}
    
    \hline
    \multicolumn{4}{c}{Circumcoronene}\\
    \multirow{2}{*}{Charge state (Z)} & \multicolumn{2}{c}{IP(Z) [eV]} & \multirow{2}{*}{\% difference}\\
        & \citet{Wenzel:2020} & \citet{Bakes:1994} &  \\
    \hline
    2 & 12.0 & 12.9 & 7.4\\
    \hline
    \multicolumn{4}{c}{Circumcircumcoronene}\\
    \multirow{2}{*}{Charge state (Z)} & \multicolumn{2}{c}{IP(Z) [eV]} & \multirow{2}{*}{\% difference}\\
        & \citet{Wenzel:2020}  & \citet{Bakes:1994} &  \\
    \hline
    -1 & 2.8 & 3.1 & 9.6 \\
    0 & 5.4 & 5.7 & 5.3 \\
    1 & 7.9 & 8.2 & 3.7 \\
    2 & 10.5 & 10.8 & 2.8 \\
    3 & 13.1 & 13.4 & 2.2 \\
    \hline
    \end{tabular}
\end{table*}

\section{Comparison between the ionization yields determined using different formalisms}
\label{sec:yieldcomparison}
\begin{figure}
    \centering
    \begin{tabular}{c}
    \includegraphics[scale=0.44]{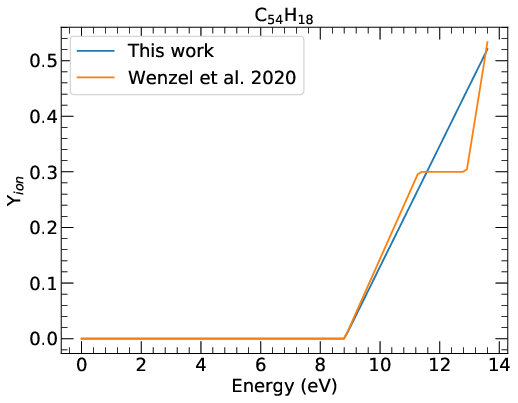} \\
    \includegraphics[scale=0.44]{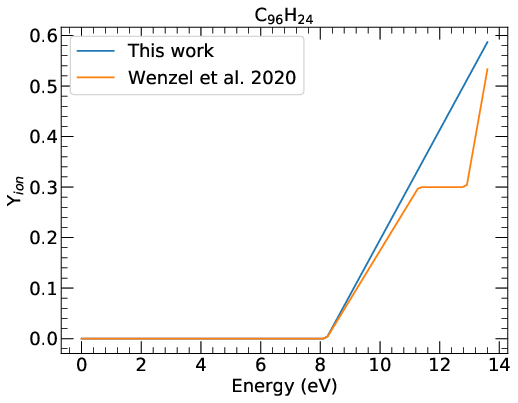} \\
      
    \end{tabular}
    \caption{Comparison of the ionization yields of singly charged circumcoronene and circumcircumcoronene determined using the formalism adopted in this work to that presented in \citet{Wenzel:2020}.}
    \label{fig:yield_comparison}
\end{figure}

In Fig.~\ref{fig:yield_comparison}, we present the comparison between the ionization yields calculated using the formalism adopted in this work and the one presented in \citet{Wenzel:2020} for singly charged circumcoronene and circumcircumcoronene. This comparison reveals that the simple expression pioneered by \citet{Jochims:1996} agress well with the more detailed expression derived by \citet{Wenzel:2020}.

\section{PAH database UIDs}
\label{sec:UID}
In Table~\ref{tab:UID}, we present the UIDs of the PAH molecules studied in this work in the PAHdb. 
\begin{table*}
\caption{UIDs of the molecules in the PAHdb.}
\label{tab:UID}
    \centering
    \begin{threeparttable}[t]
    \begin{tabular}{l c c c c c}
    
    \hline
    \multirow{2}{*}{Molecule} & \multicolumn{5}{c}{UID}\\
        & $Z$ = -1 & $Z$ = 0 & $Z$ = 1 & $Z$ = 2 & $Z$ = 3\\
    \hline
    Tetracene ($\text{C}_{18}\text{H}_{12}$)  & 210  & 282  & 283 & -- & \\ 
    Pentacene ($\text{C}_{22}\text{H}_{14}$) &  -- & 307  & 308  & 672 & \\ 
    Ovalene ($\text{C}_{32}\text{H}_{14}$)  & 11  & 4  & 5  & 10 & \\
    Circumcoronene ($\text{C}_{54}\text{H}_{18}$)  & 46  & 37  & 38  & 44  & \\
    Circumcircumcoronene ($\text{C}_{96}\text{H}_{24}$) & 114  & 108  & 111  & 112  & 113 \\
    
    \hline
    \end{tabular}
    
    \end{threeparttable}
\end{table*}

\section{Comparison of the IR emission calculations with another model}
\label{sec:comparison_Mulas}
\begin{figure}
    \centering
    
    \includegraphics[scale=0.44]{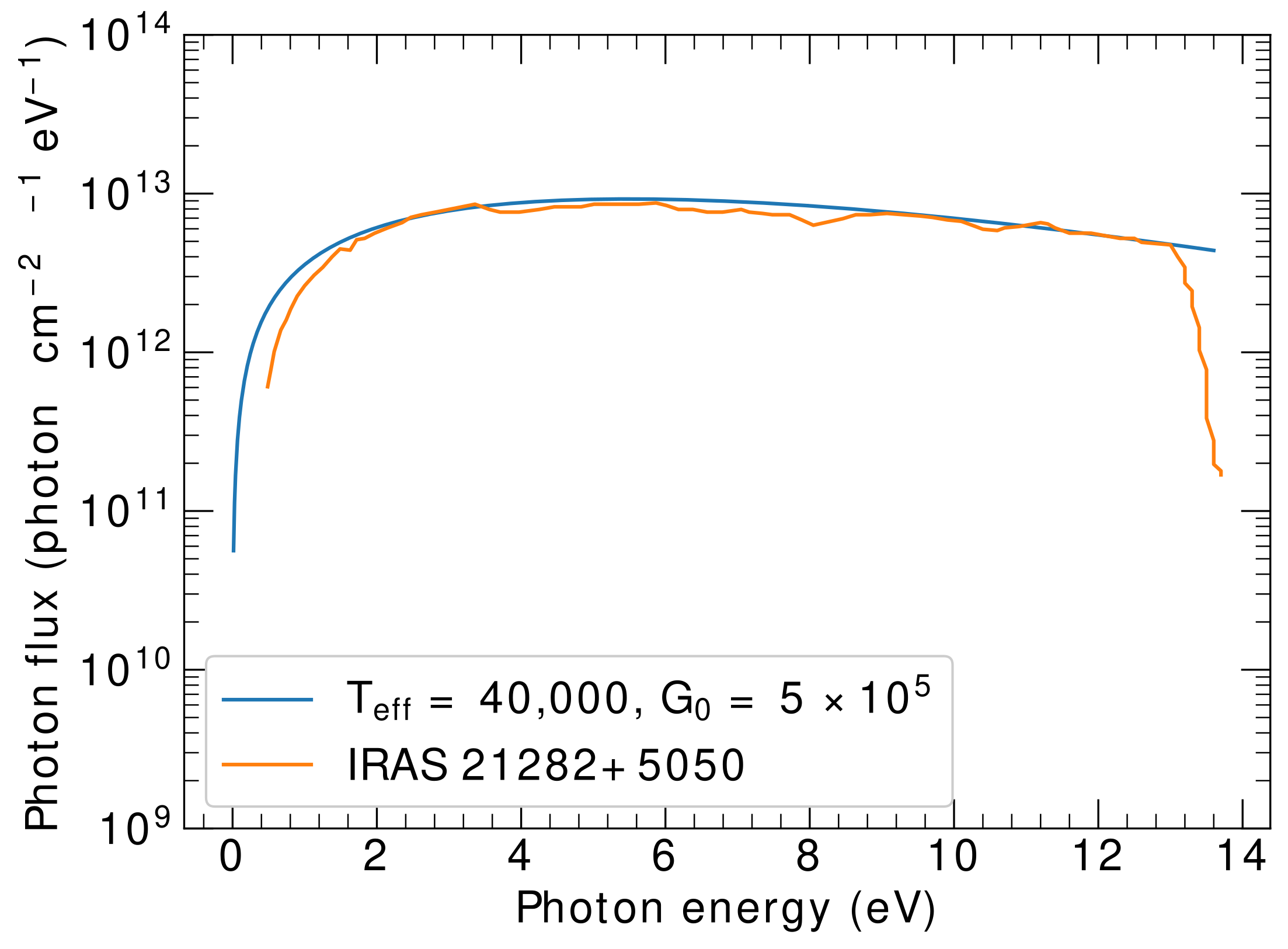} 
    
    \caption{The photon flux of the planetray nebula IRAS~21282+5050 taken from \citet{Mulas:2006} compares well with the photon flux estimated using a blackbody of temperature = 40,000 K and a radiation field strength of $G_{0} = 5 \times 10^{5}$.}
    \label{fig:flux_comparison_IRAS_21282}
\end{figure}

To benchmark our calculations of IR emission, we compared our calculations for neutral coronene (C$_{24}$H$_{12}$) in the planetary nebula IRAS~21282+5050 with the results presented in Table~2 of \citet{Mulas:2006}. Fig.~\ref{fig:flux_comparison_IRAS_21282} shows that the radiation field of IRAS~21282+5050 is well characterized by a blackbody of temperature 40,000 K and a radiation field strength of $G_{0} = 5 \times 10^{5}$ in the units of Habing field, which we will use as input for the radiation field in our model. We note that the calculations in Table~2 of \citet{Mulas:2006} do not include the possibility of ionization. Therefore, we also calculated the IR emission spectra excluding the possibility of ionization after photon absorption for this comparison. We adopted the photo-absorption cross-section from the \citet{Malloci:database} online database and the frequencies and corresponding intensities of the vibrational modes of neutral coronene from the NASA Ames PAH database. In Table~\ref{tab:model_comparison_Mulas}, we present the comparison of the flux fraction determined using the model presented in this work to that in \citet{Mulas:2006}. The two models agree well. 

\begin{table*}
\caption{Comparison of the flux fraction carried by each band of neutral coronene in the planetray nebula IRAS~21282+50 calculated using the model presented in this work to that in \citet{Mulas:2006} (see Appendix~\ref{sec:comparison_Mulas} for further details).}
\label{tab:model_comparison_Mulas}
    \centering
    \begin{threeparttable}[t]
    \begin{tabular}{c c c c}
    
    \hline
    \multicolumn{2}{c}{This work} & \multicolumn{2}{c}{\citet{Mulas:2006}} \\ 
    Peak ($\mu$m) & Flux fraction (\%) & Peak ($\mu$m) & Flux fraction (\%) \\
    \hline
    3.26 & 46.62 & 3.26 & 46.68\\
    3.28 & 2.67 & 3.29 & 3.18\\
    6.28 & 5.31 & 6.24 & 5.01\\
    6.73 & 0.54 & 6.69 & 0.09\\
    7.25 & 0.33 & 7.21 & 0.99\\
    7.67 & 9.23 & 7.63 & 6.05\\
    8.29 & 0.37 & 8.25 & 2.63\\
    8.82 & 2.36 & 8.78 & 2.75\\
    11.6 & 27.18 & 11.57 & 27.89\\
    12.5 & 0.04 & 12.43 & 0.09\\
    12.9 & 1.68 & 12.90 & 1.29\\
    18.2 & 3.08 & 18.21 & 3.04\\
    26.5 & 0.49 & 26.20 & 0.19\\
    81.6 & 0.09 & 80.60 & 0.11\\
    
    \hline
    \end{tabular}
    
    \end{threeparttable}
\end{table*}

\section{Charge Distribution}
\label{app:chargedist}
\begin{figure*}
    \centering
    \begin{tabular}{cc}

    \includegraphics[scale=0.40]{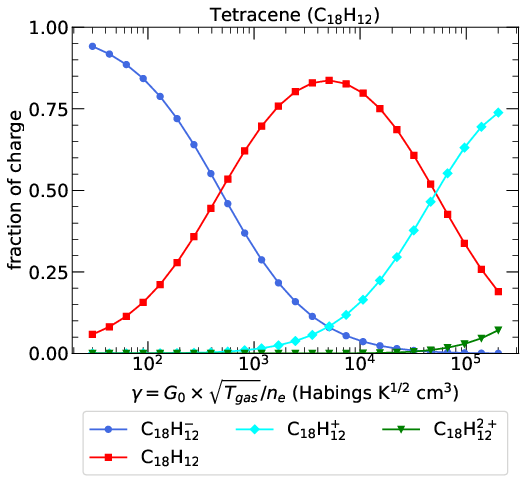} &  \includegraphics[scale=0.40]{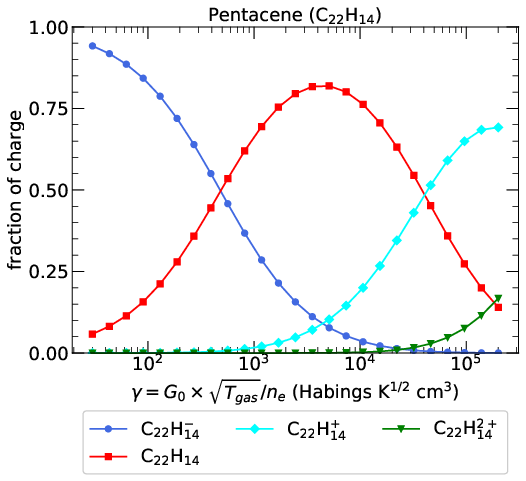}\\
    \includegraphics[scale=0.40]{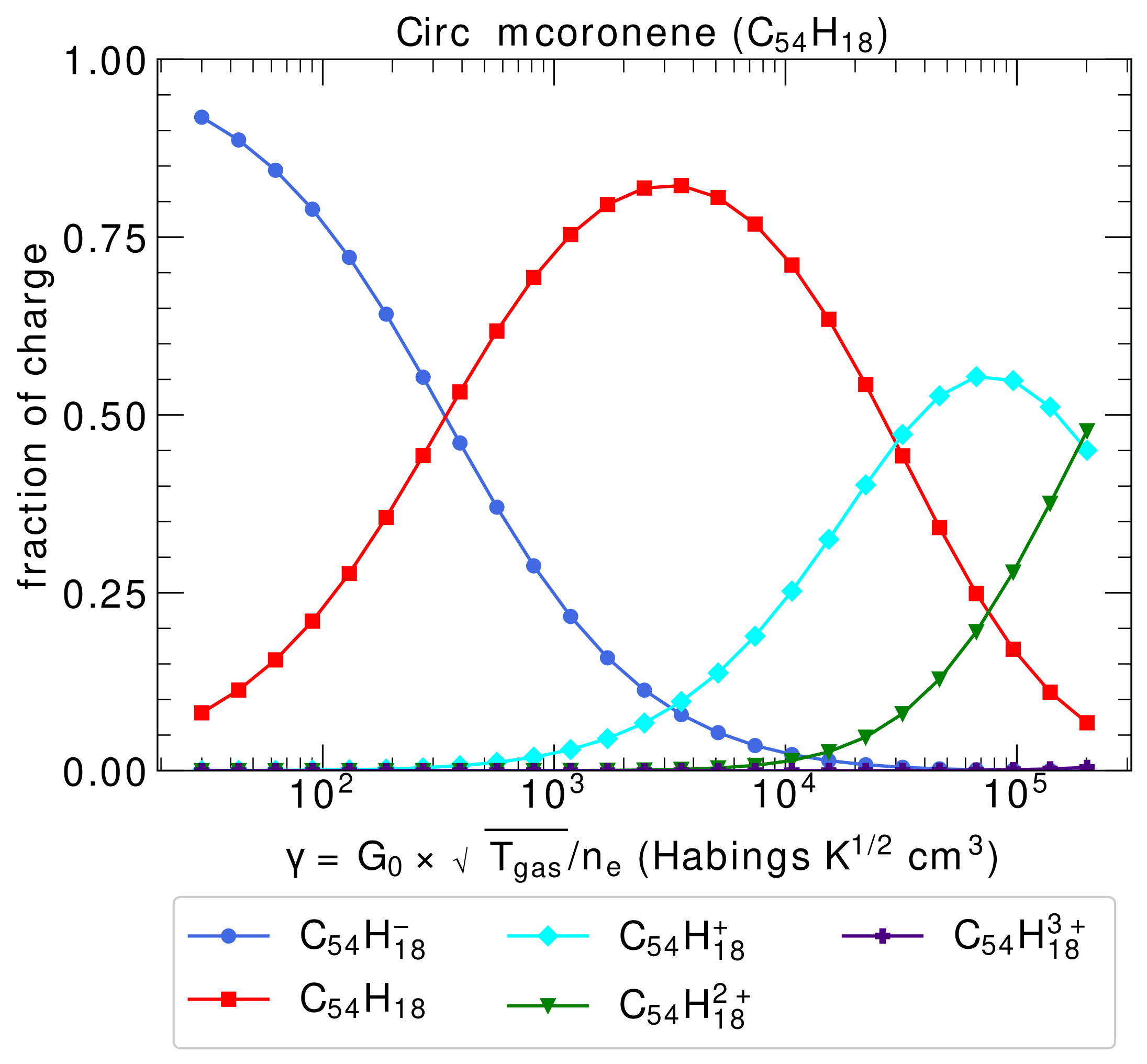} &
    \includegraphics[scale=0.40]{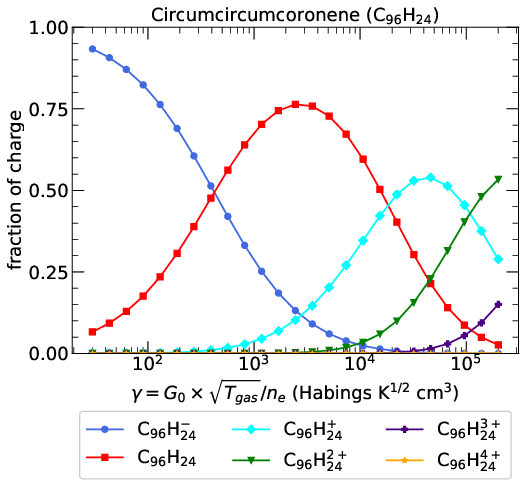}\\
    \end{tabular}
    \caption{Charge distribution of tetracene, pentacene, circumcoronene, and cirumcircumcoronene as a function of the ionization parameter $\gamma$, for a fixed value of $G_{0} = 2600$, $T_{\rm gas} = 300$K, and $T_{\rm eff} = 17000$K.}
    \label{fig:charge_distribution_gamma}
    \end{figure*}

In Fig.~\ref{fig:charge_distribution_gamma}, we present the charge distribution of tetracene ($\text{C}_{18}\text{H}_{12}$), pentacene ($\text{C}_{22}\text{H}_{14}$), circumcoronene ($\text{C}_{54}\text{H}_{18}$), and circumcircumcoronene ($\text{C}_{96}\text{H}_{24}$), as a function of $\gamma$, for a fixed value of $G_{0}$=2600, $T_{\rm gas} = 300$K, and $T_{\rm eff}=17000$K. 

\section{Effect of the change in the excitation conditions on the charge distribution of PAHs}
\begin{figure*}
    \centering
     \begin{tabular}{cc}

    \includegraphics[scale=0.40]{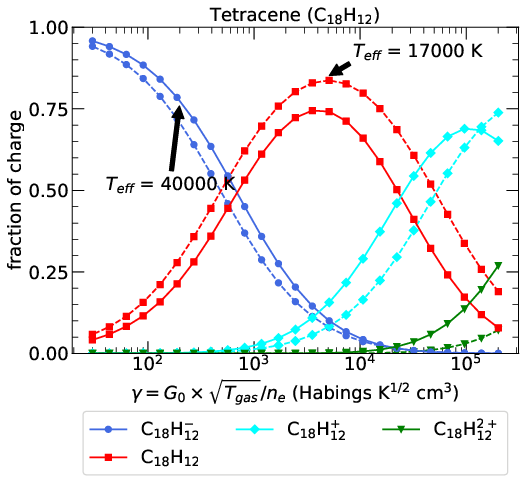} &  \includegraphics[scale=0.40]{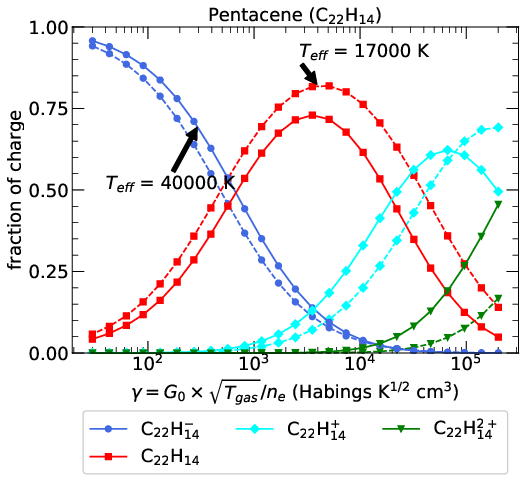}\\
     \includegraphics[scale=0.40]{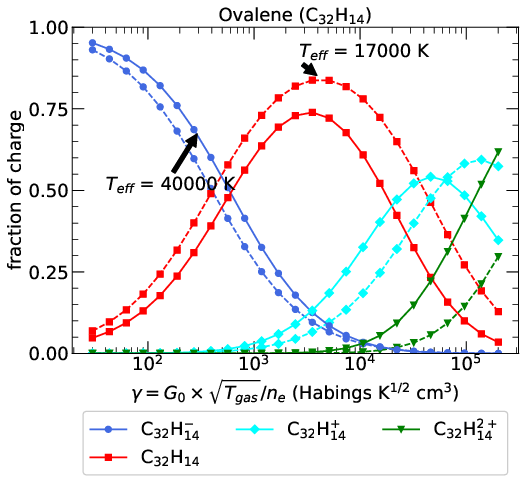} & 
     \includegraphics[scale=0.40]{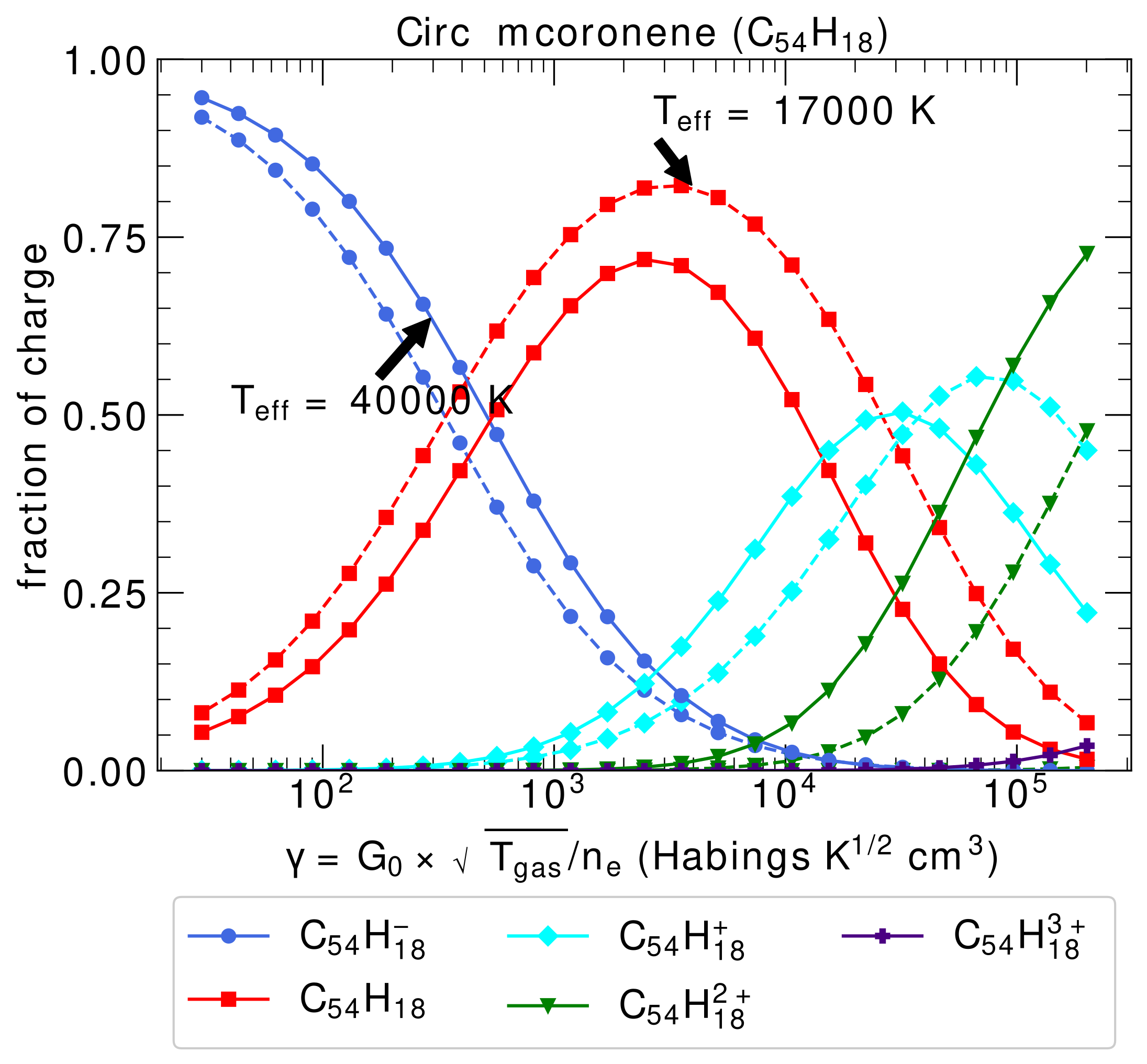} \\
     \includegraphics[scale=0.40]{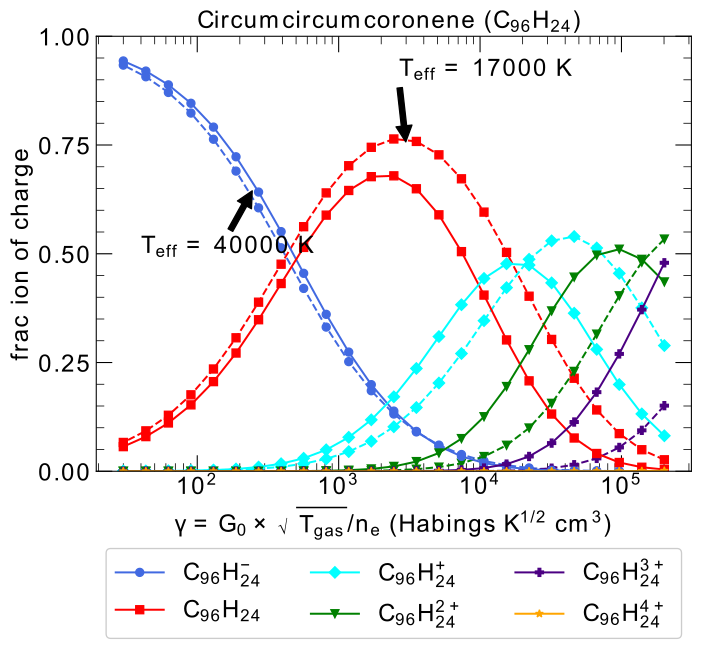}\\
     \end{tabular}
     \caption{Charge distribution of the five PAHs considered in this work as a function of the ionization parameter $\gamma$ for two different effective temperatures at which we calculate the blackbody radiation field in order to calculate charge distribution (see text for details).}
    \label{fig:charge_distribution_two_temperatures}
     \end{figure*}

In order to investigate the influence of the change in excitation conditions on the charge distribution of PAHs, we modelled the charge distribution at $T_{\rm eff} = 17000$K and $T_{\rm eff} = 40000$K for the blackbody radiation field and present the results in Fig.~\ref{fig:charge_distribution_two_temperatures}.  
     
\section{Intrinsic spectra of PAHs}
\label{sec:intrinsic_spectra}
\begin{figure}
     \centering
     \includegraphics[scale=0.37]{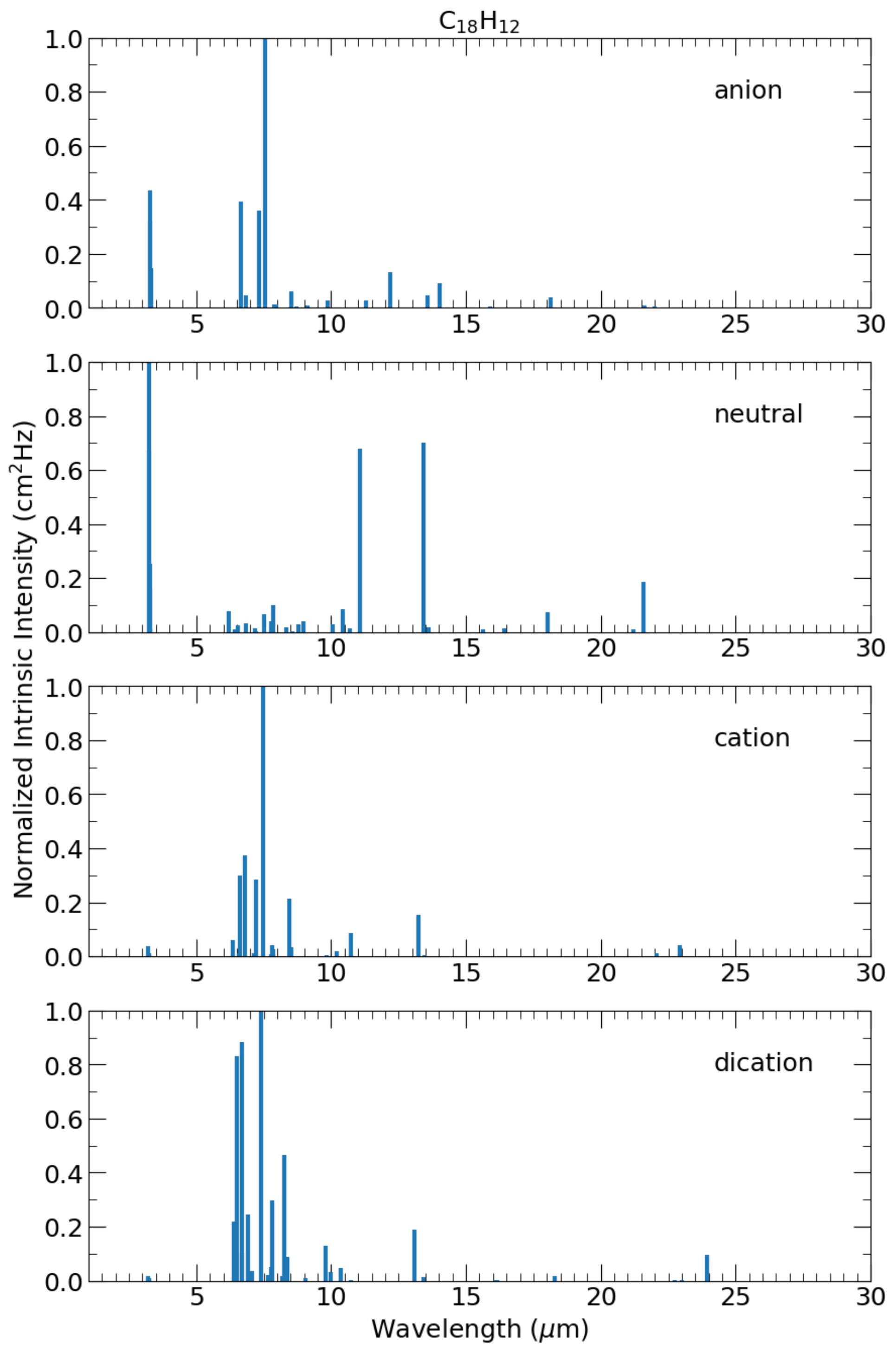}
     \caption{Intrinsic stick spectra of tetracene normalized to the maximum intensity in anionic, neutral, cationic, and dicationic charge states. }
     \label{fig:intrinsic_spectra_tetracene}
 \end{figure}

\begin{figure}
     \centering
     \includegraphics[scale=0.37]{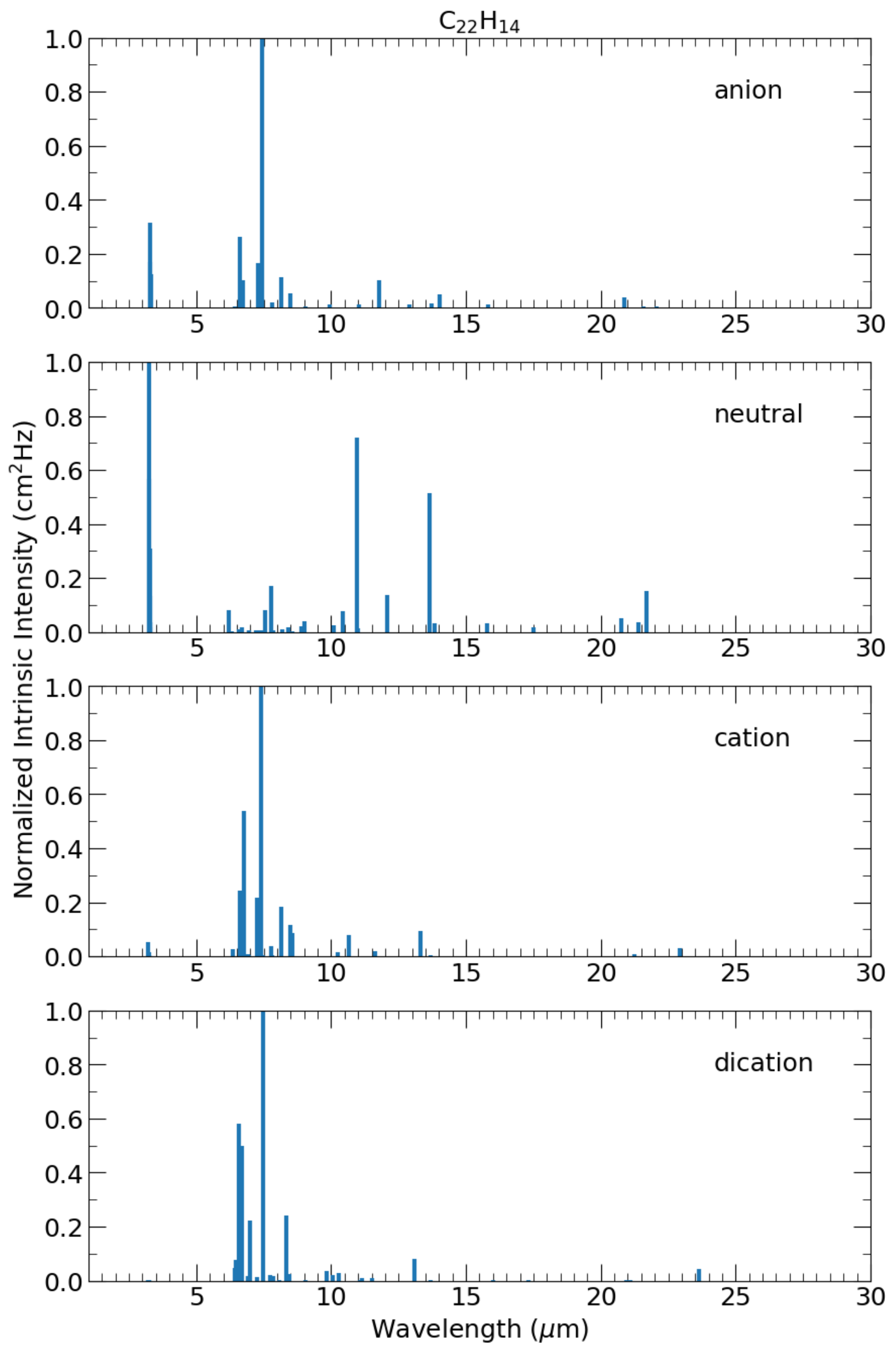}
     \caption{Intrinsic stick spectra of pentacene normalized to the maximum intensity in anionic, neutral, cationic, and dicationic charge states.}
     \label{fig:intrinsic_spectra_pentacene}
 \end{figure}

\begin{figure}
     \centering
     \includegraphics[scale=0.37]{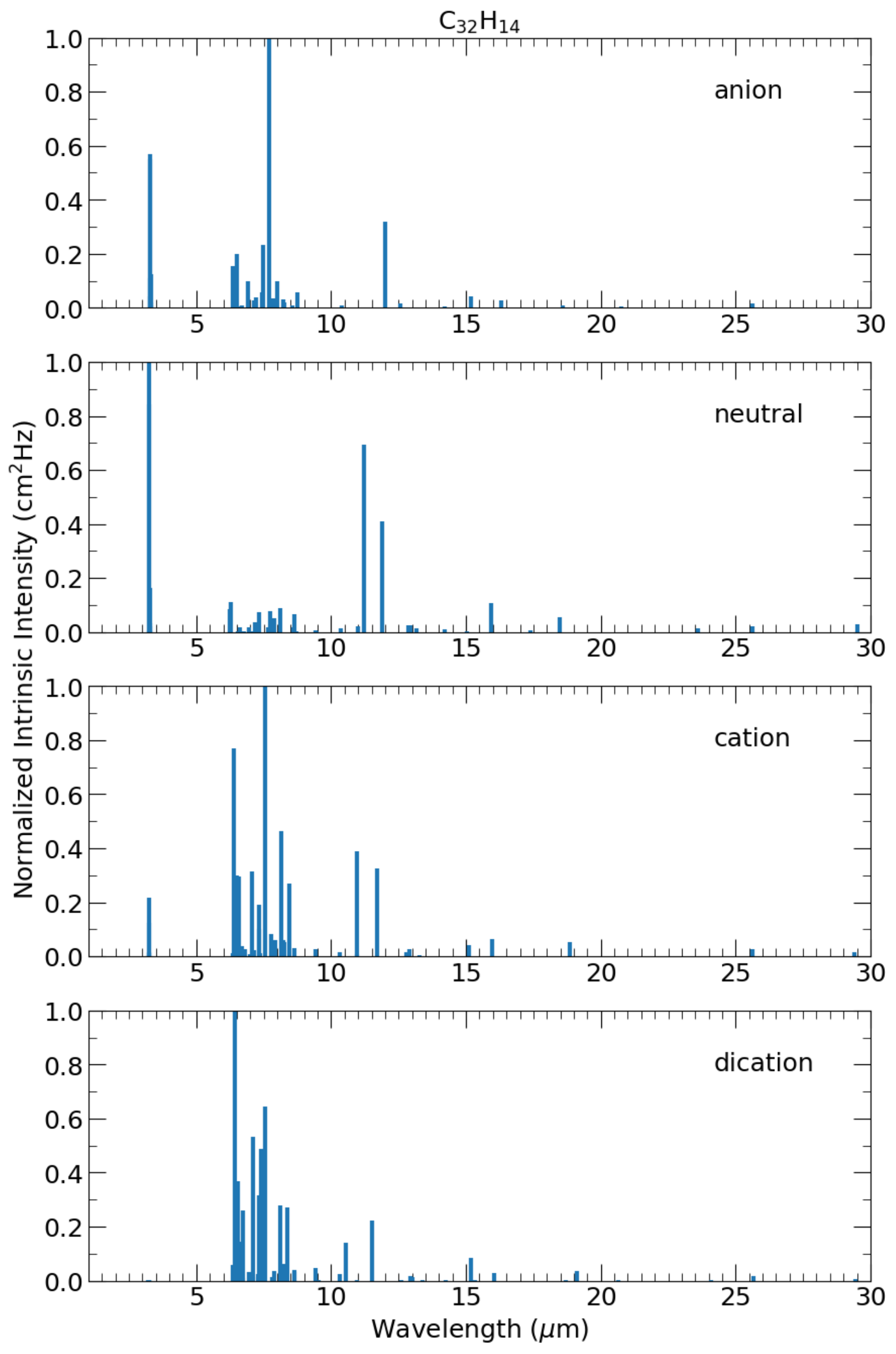}
     \caption{Intrinsic stick spectra of ovalene normalized to the maximum intensity in anionic, neutral, cationic, and dicationic charge states.}
     \label{fig:intrinsic_spectra_ovalene}
 \end{figure}

\begin{figure}
     \centering
     \includegraphics[scale=0.37]{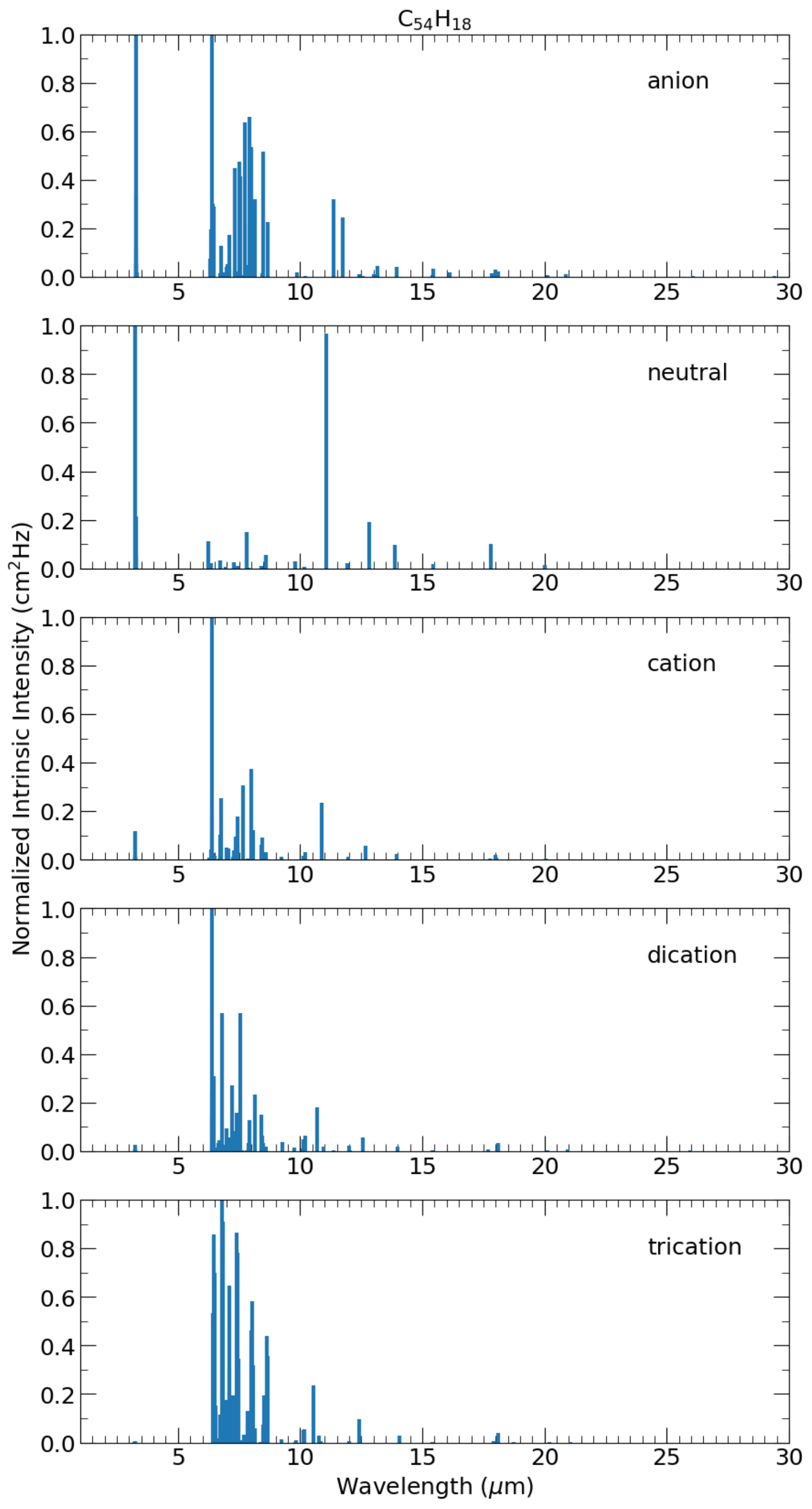}
     \caption{Intrinsic stick spectra of circumcoronene normalized to the maximum intensity in anionic, neutral, cationic, dicationic, and tricationic charge states.}
     \label{fig:intrinsic_spectra_circumcoronene}
 \end{figure}

\begin{figure}
     \centering
     \includegraphics[scale=0.37]{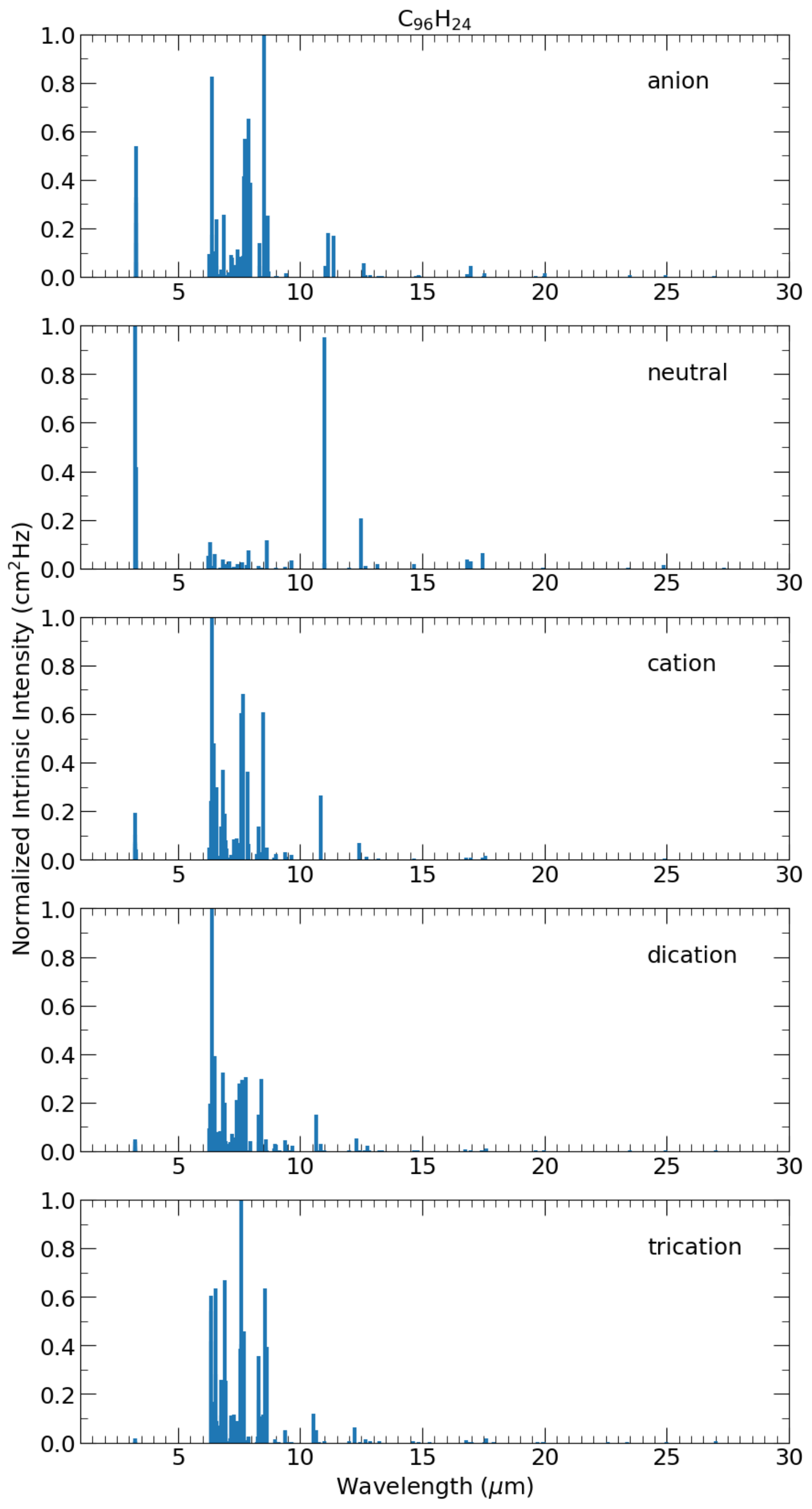}
     \caption{Intrinsic stick spectra of circumcircumcoronene normalized to the maximum intensity in anionic, neutral, cationic, dicationic, and tricationic charge states.}
     \label{fig:intrinsic_spectra_circumcircumcoronene}
 \end{figure}

Figs.~\ref{fig:intrinsic_spectra_tetracene}-\ref{fig:intrinsic_spectra_circumcircumcoronene}, show the intrinsic spectra of the PAH molecules for all the charge states. To facilitate the comparison, we have normalized the intrinsic spectra to the maximum intensity in each charge state. The neutrals exhibit strong features at the 3.3 $\mu$m and in the 10-15 $\mu$m range and weak features in the 6-9 $\mu$m range with 3.3 $\mu$m as the strongest feature in all the cases. The anions exhibit strong features at the 3.3 $\mu$m and in the 6-9 $\mu$m range and weak features in the 10-15 $\mu$m range. Similar to anions, the cations also exhibit strong features in the 6--9 $\mu$m region and weak features in the 10-14 $\mu$m region. The features in the 3.3 $\mu$m region, on the other hand are weaker in cations. In fact the intrinsic strength of the 3.3 $\mu$m feature is even less than that of the features in the 10-14 $\mu$m region. Longwards of 15 $\mu$m, the molecules in all the charge states exhibhit few weak features with strength of these features in neutrals slightly greater than that in other charge states.

\section{6.2/(11.0+11.2) vs 3.3/(11.0+11.2)}
\begin{figure*}
    \centering
    \begin{tabular}{cc}

    \includegraphics[scale=0.50]{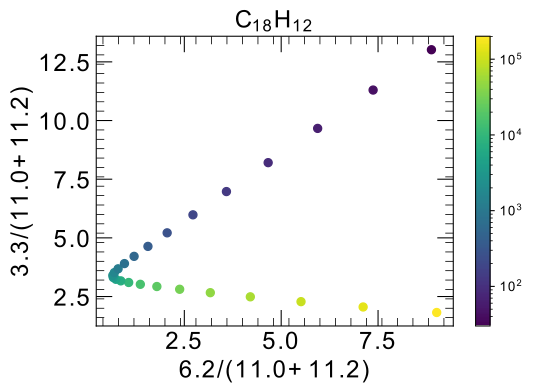} &  \includegraphics[scale=0.50]{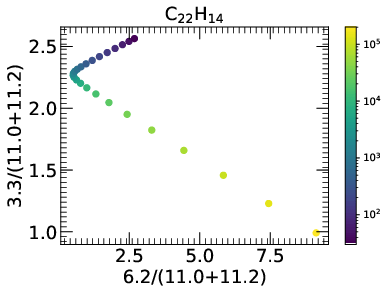}\\
    \includegraphics[scale=0.50]{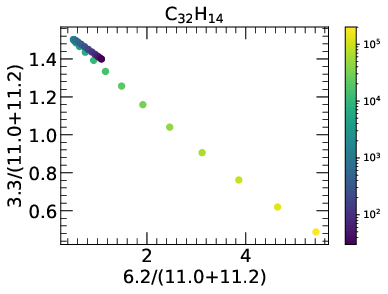} & 
    \includegraphics[scale=0.50]{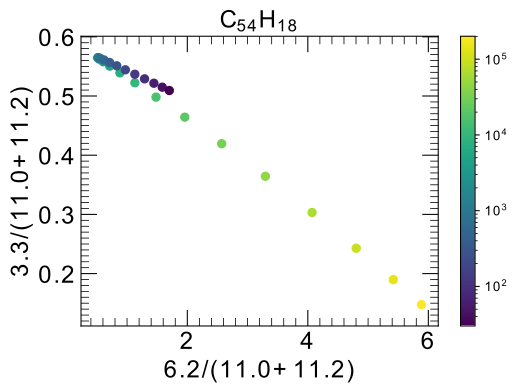} \\
    \includegraphics[scale=0.50]{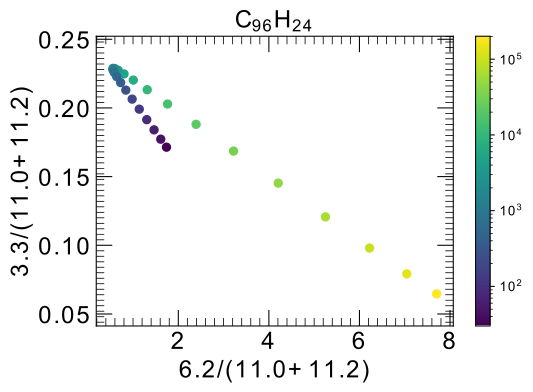}\\
    \end{tabular}
    \caption{Ratios of the 6.2/(11.0+11.2) vs 3.3/(11.0+11.2) color coded with $\gamma$ values for the five PAH molecules considered in this work at $G_{0}=2600$ and $T_{\rm eff}=40000$K. These plots demonstrate that the slope of the branch corresponding to low $\gamma$ values is sensitive to excitation conditions. }
    \label{fig:distinction_anions_cations_Orion_bar}
    \end{figure*}

In fig.~\ref{fig:distinction_anions_cations_Orion_bar}, we present the ratios of 6.2/(11.0+11.2) vs 3.3/(11.0+11.2) color coded as a function of $\gamma$ at $T_{\rm eff}$ = 40000 K, demonstrating the effect of the change in the excitation conditions on the slopes of the two  branches observed in the plots.

\section{Validating the calculation of the average energy absorbed by a PAH molecule}
\label{sec:validating_abs_energy_calc}

In our calculation of $E_{\rm avg}(Z)$, we account for the possibility of ionization by the absorbed photon. While the PAH model presented in \citet{Mulas:2006} also include the possibility of ionization by the absorbed photon, these authors account for it by including it as one of the molecular relaxation processes in the Monte Carlo modelling instead of in the calculation of $E_{\rm avg}(Z)$. To make a consistent comparison with the results of \citet{Mulas:2006}, we calculate the average energy absorbed by a PAH molecule in a charge state $Z$ excluding the possibility of ionization using the following expression:
\begin{equation}
    E_{\rm avg}(Z) = \frac{\int_{0}^{\nu_{H}} \sigma_{\rm abs}(Z, \nu) h \nu \frac{B_{\nu}(T_{\rm eff})}{h\nu} d\nu}{\int_{0}^{\nu_{H}} \sigma_{\rm abs}(Z, \nu) \frac{B_{\nu}(T_{\rm eff})}{h\nu} d\nu }
    \label{eq:avg_energy_excluding_ionization}
\end{equation}

While none of the environments studied in this work are part of the study presented in \citet{Mulas:2006}, we note that the distribution of the photon flux of the Orion bar, characterized by the $T_{eff}$ = 40,000 K, is comparable to the planetary nebula IRAS~21282+5050 (see Fig.~\ref{fig:flux_comparison_IRAS_21282}) studied by \citet{Mulas:2006}. Therefore, we calculate the $E_{avg}(Z)$ for coronene and pentacene in the Orion bar in the neutral and cationic charge state. In Table~\ref{tab:Evg_comparison} we present the results of this comparison.

\begin{table*}
\caption{$E_{\rm avg}(Z)$ values in the Orion Bar (this work) and in the planetary nebula IRAS~21282+5050 (\citet{Mulas:2006}). The Orion Bar and IRAS~21282+5050 have comparable photon flux.}
\label{tab:Evg_comparison}
    \centering
    
    \begin{tabular}{c c c c}
    
    \hline
    \multirow{2}{*}{Molecule} & \multirow{2}{*}{Charge state (Z)} & \multicolumn{2}{c}{$E_{\rm avg}(Z)$ [eV]} \\ 
      &  & IRAS~21282+5050 & Orion Bar  \\
    \hline
    Pentacene ($\text{C}_{22}\text{H}_{14}$) & 0 & 8.3 & 8.3 \\
    & 1 & 8.2 & 8.2 \\
    Ovalene ($\text{C}_{32}\text{H}_{14}$) & 0 & 8.2 & 8.2\\
    & 1 & 8.2 & 8.3 \\
    \hline
    \end{tabular}
\end{table*}

\end{appendices}
%%%%%%%%%%%%%%%%%%%%%%%%%%%%%%%%%%%%%%%%%%%%%%%%%%

% Don't change these lines
\bsp	% typesetting comment
\label{lastpage}
\end{document}